\documentclass[conference,compsoc]{IEEEtran}
\usepackage[nocompress]{cite}
\usepackage{amsmath,amssymb,graphicx,booktabs,array,longtable,placeins,xspace}
\usepackage[table]{xcolor}
\usepackage[hidelinks]{hyperref}
\usepackage[caption=false,font=footnotesize]{subfig}
\usepackage{tcolorbox}
\usepackage{fontawesome6}
\usepackage{simpleicons}
\definecolor{ebBlue}{HTML}{DCE3F0}
\definecolor{ebAmber}{HTML}{F4CC79}
\definecolor{ebBeige}{HTML}{ECD7BC}
\definecolor{ebCyan}{HTML}{CCFDFE}
\definecolor{ebCoral}{HTML}{ED8683}
\definecolor{ebPink}{HTML}{F7D3D2}
\definecolor{ebWarm}{HTML}{F2F2EE}
\definecolor{ebDeepBlue}{HTML}{B9C6E0}
\definecolor{ebInk}{HTML}{2E5BE6}

\definecolor{ghdark}{HTML}{24292F}
\definecolor{hfyellow}{HTML}{FFD21E}
\definecolor{hfink}{HTML}{1F2937}
\newcommand{\linkpill}[5]{%
  \href{#5}{\tcbox[on line,colback=#1,colframe=#1,boxrule=0pt,arc=7pt,
    left=6pt,right=7pt,top=2.5pt,bottom=2.5pt,boxsep=0pt,
    fontupper=\sffamily\upshape\small\color{#2}]{#3\hspace{0.45em}#4}}}

\newenvironment{packeditemize}{
    \begin{list}{$\bullet$}{
            \setlength{\labelwidth}{4pt}
            \setlength{\itemsep}{0pt}
            \setlength{\leftmargin}{\labelwidth}
            \addtolength{\leftmargin}{\labelsep}
            \setlength{\parindent}{0pt}
            \setlength{\listparindent}{\parindent}
            \setlength{\parsep}{0pt}
            \setlength{\topsep}{1pt}}}{\end{list}}

\newcount\soktablefitpass
\newcommand{\sokfitcolumns}[2]{%
  \soktablefitpass=0
  \loop
    \sbox0{#1}%
    \dimen0=\dimexpr\linewidth-1pt-\wd0\relax
    \divide\dimen0 by #2
    \advance\tabcolsep by \dimen0
    \ifdim\tabcolsep<1.5pt\setlength{\tabcolsep}{1.5pt}\fi
    \advance\soktablefitpass by 1
  \ifnum\soktablefitpass<8\repeat}

\newcommand{\jev}{Jev-1.13.0\xspace}
\newcommand{\sokjev}{Jev-1.13\xspace}
\newcommand{\deepseek}{DeepSeek-V4.1-Flash\xspace}
\newcommand{\semif}{SemIf-Qwen3.5-4B\xspace}
\newcommand{\glm}{GLM-5.3-Flash\xspace}
\newcommand{\sokglm}{GLM-4.7-Flash\xspace}
\newcommand{\qwenflash}{Qwen3.8-Flash\xspace}
\newcommand{\qwenjson}{Qwen3.5-4B-JSON\xspace}
\newcommand{\sokqwen}{Qwen3.5-4B\xspace}
\newcommand{\gemini}{Gemini-3.1-Flash-Lite\xspace}
\newcommand{\anyjev}{AnyJev-L0\xspace}
\newcommand{\laya}{Laya\xspace}

\graphicspath{{figures/}}

\hypersetup{pdftitle={SoK: Semantic Decision Engines in Network Control Loops},pdfauthor={Delong Li, Chen Li, Xu Wang, Haochen Gong, Rui Lang, Guangsheng Yu}}
\begin{document}
\title{SoK: Semantic Decision Engines in Network Control Loops}
\author{\IEEEauthorblockN{Delong Li, Chen Li, Xu Wang, Haochen Gong, Rui Lang, and Guangsheng Yu}
\IEEEauthorblockA{School of Electrical, Mechanical and Biomedical Engineering\\University of Technology Sydney, Sydney, Australia\\[0.35em]
\linkpill{ghdark}{white}{\faGithub}{OniReimu/SoK-JEV}{https://github.com/OniReimu/SoK-JEV}\hspace{0.7em}%
\linkpill{hfyellow}{hfink}{\simpleicon{huggingface}}{OniReimu/SoK-JEV}{https://huggingface.co/datasets/OniReimu/SoK-JEV}}}
\maketitle
\begin{abstract}
A semantic decision engine such as Jev can return a valid answer and still miss a network deadline, select an infeasible action or leave the service unverified. We systematize 139 paper families by decision interface, execution path and check ownership. Fifty families claim that their engine fits a control loop or time budget, but only four support the claim with matched measurement. Across all 139, four report deadline attainment. The gap concentrates where the decision has no deterministic computation step. Those 72 families make 22 of the claims, none supported, and name a coverage owner in only two. Bounded tests under one event model show that each gap can reverse an admission verdict. A decision that meets a 10\,s budget for every isolated request meets it for none once decisions queue ahead of replayed execution times. The same engine passes one coverage check and fails another. We derive a minimum reporting record, design rules and a research agenda for admitting decision engines to control loops.
\end{abstract}

\IEEEpeerreviewmaketitle
\section{Introduction}\label{sec:introduction}
A network controller receives a request to move a service, restore a route or change a radio policy. Before the action takes effect, the controller interprets the request, checks the available state and resources and admits work to an execution path. A quick answer helps only when this complete path meets the requested outcome and its deadline. For example, the near-real-time controller in an open radio access network (O-RAN) operates on typical 10\,ms--1\,s timescales. State collection, communication, decision waiting, control and observation all consume that budget~\cite{oran2026architecture}.

\begin{figure}[!t]
\centering
\includegraphics[width=\columnwidth]{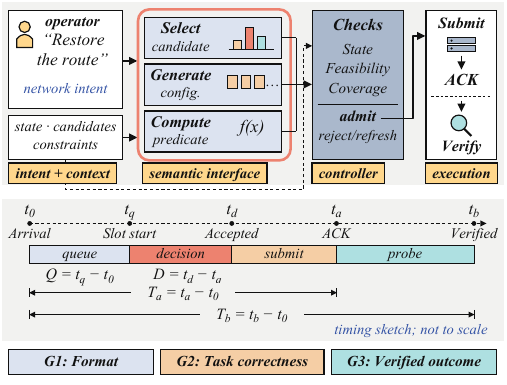}
\caption{Three alternative interfaces inside a control path. The controller checks state, feasibility and coverage before admission. Compute evaluates $f(x)$. Application programming interface (API) or configuration acknowledgement (ACK) and verified outcome are distinct endpoints. G1--G3 report format validity, task correctness and verified outcome.}
\label{fig:control-boundary}
\end{figure}

Language interfaces make such requests easier to express, but they also move part of the controller's responsibility into the interpreter. How much moves depends on the interface. The interpreter may choose among alternatives that the controller supplies, construct fields or configurations that the request leaves out, or pass a formalized request to a rule or solver that computes the answer. In each case a well-formed output can still be wrong for the network, because a schema-valid answer can exceed a shared quota, come from an incomplete catalogue or rest on contradictory observations. Conversely, a justified rejection can be the correct outcome even though no service is activated. Figure~\ref{fig:control-boundary} places these decisions in the surrounding control path.

We ask which network control loops can admit a semantic decision engine and which checks the controller must keep. The literature rarely measures the complete path that this question requires. Of the 139 paper families we review, 50 claim that their engine fits a control loop or time budget, and 4 report measurement matched to the claimed budget and boundary. Nine families report latency at the 95th percentile or above, and four report deadline attainment ($\S$\ref{sec:evidence-audit}). In the systematization, the 72 families whose decision has no deterministic computation step make 22 of the 50 loop claims, none with matched measurement. Only 23 of them name a feasibility owner and 2 a coverage owner, against 45 and 11 of the 67 families with such a step.

We test whether the missing measurements and checks change an admission verdict, using bounded measurements in one event model. The service-orchestration study~\cite{li2026replacing}, the RAN study~\cite{li2026intent} and our application tasks become instances of one framework, and fixed-action interventions on transport and edge execution stacks vary only the decision delay. Each test measures the consequence of one gap. For missing load evidence, a decision that meets a 10\,s budget for every isolated request meets it for none at 4.992 arrivals/s once decisions queue ahead of measured execution times. For unnamed check ownership, an engine that owns the coverage check requests a new candidate for 45 of 48 contracts without one but escalates 4 of 120 infeasible routes.
This systematization of knowledge (SoK) contributes:
\begin{packeditemize}
\item A three-axis systematization of 139 families by decision interface, control and execution path, and check ownership, mapped to standardized network functions and recoded blind to measure its reproducibility. It shows that check ownership and load evidence concentrate in families with a deterministic computation step.
\item An evidence audit with independent coding and explicit denominators, which finds matched support for 4 of 50 loop claims and yields a minimum reporting record.
\item Bounded tests of the gaps that the systematization exposes. One event model with four diagnostic lenses (measurement boundary, workload summary, check placement and robustness scope) compares the service, RAN and application studies without pooling their effects. Fixed-action delay interventions on transport and edge stacks quantify endpoint shares and queue amplification.
\item Design rules for admitting decision engines and a research agenda, each derived from a gap in the systematization or the audit.
\end{packeditemize}

\section{Background}\label{sec:background}
\begin{figure*}[!t]
\centering
\includegraphics[width=\textwidth]{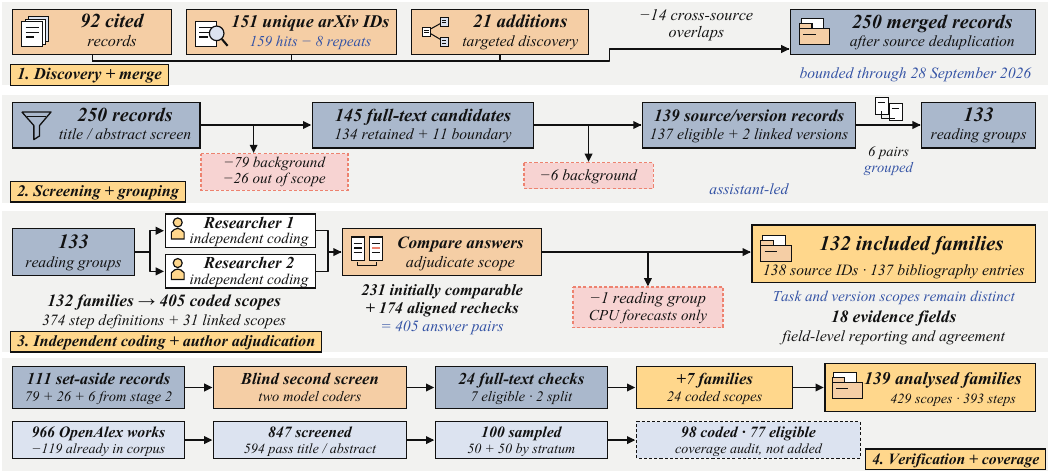}
\caption{Corpus formation and independent coding. Records, versions, groups, families and scopes are distinct units. Screening was assistant-led. Two researchers independently coded the groups, then jointly adjudicated. The 231 initially comparable pairs and 174 aligned rechecks form two strata. Stage 4 adds the blind second screening of all set-aside records, which contributes seven families, and the OpenAlex coverage audit, whose sampled records stay outside the analysed set. Search and coding definitions appear in Appendix~\ref{app:e0-protocol}, and agreement is reported in Table~\ref{tab:e0-agreement} and Appendix~\ref{app:agreement}.}
\label{fig:e0-screening}
\end{figure*}

\subsection{Control Loops and Their Time Budgets}
O-RAN distinguishes non-real-time (non-RT) radio access network (RAN) intelligent controller (RIC) functions above approximately 1\,s from near-real-time (near-RT) RIC functions at approximately 10\,ms--1\,s~\cite{oran2026architecture}. Management and orchestration can act on slower operational timescales. An interpreter shares these intervals with state collection, communication and execution. The architecture also describes faster device-internal real-time (RT) control. Distributed applications (dApps), proposed outside the standardized two-tier RIC hierarchy, extend artificial intelligence and machine learning toward this faster regime~\cite{doro2024dapps}.

Non-RT applications (rApps) supply policy or enrichment through the non-RT RIC's A1 services, while near-RT applications (xApps) consume measurements and issue control through the near-RT RIC's E2 services. E2 service models distinguish measurement reporting, including Key Performance Measurement (KPM), from control functions~\cite{oran2024e2sm}. A control acknowledgement and a later KPM change establish different milestones, and service verification requires an outcome check.

\subsection{Intent, Analytics and Closed-Loop Functions}
The 3rd Generation Partnership Project (3GPP) defines intent-driven management in Technical Specification (TS)~28.312, including feasibility checks before fulfilment~\cite{3gpp2026intent}. TS~23.288 describes the Network Data Analytics Function (NWDAF), while TS~28.104 covers Management Data Analytics (MDA), its functions and services~\cite{3gpp2026nwdaf,3gpp2025mda}. Analytics informs policy and orchestration, while installation and outcome verification occur downstream.

The European Telecommunications Standards Institute (ETSI) separates analytics, intelligent decision, orchestration and control in its Zero-touch network and Service Management (ZSM) architecture~\cite{etsi2019zsm002}. Its closed-loop automation specification adds governance and coordination of closed loops~\cite{etsi2021zsm0091}. TM Forum's TMF921 interface provides intent and report resources between an intent owner and handler, and IG1253 frames their intent-management relationship~\cite{tmforum2024tmf921,tmforum2022ig1253}. Requests for Comments (RFCs)~9315 and 9316 supply intent-based networking concepts and intent classification, respectively. Both are Informational documents from the Internet Research Task Force's Network Management Research Group~\cite{clemm2022intentbased,li2022intentclassification}.

\subsection{Selection, Generation and Computation}
Selection returns a member of a supplied alternative set, which may include rejection, clarification or escalation. Typed-choice application programming interfaces (APIs), candidate-constrained language models and discriminative classifiers can implement this interface. Generation constructs content not already represented as a complete alternative, including extracted values, programs, configuration edits and plans. Schema or grammar constraints govern output form, whereas the task predicate determines semantic correctness~\cite{geng2025jsonschemabench,beurer-kellner2024guiding,li2026xgrammar2}. Computation uses rules, solvers or verifiers to evaluate formalized conditions or synthesize a satisfying assignment.

A workflow can use all three interfaces. Intent interpretation may generate a bounded contract, a scheduler may compute joint feasibility, and a selector may choose a certified action. The classification follows each step's output responsibility, independent of the vendor and of whether the step uses a language model internally. Comparing implementations requires the same usable output, available information and downstream checks.

\section{Review Method}
\label{sec:review-method}

\subsection{Scope and Source Identification}

We reviewed semantic and intent-based decision making for network configuration, orchestration and diagnosis. Eligible studies identify a network decision output and range from explicitly described conceptual designs to deployed systems, with or without latency measurements and positive results. For mixed-domain studies, we extract the identifiable network tasks. We exclude pure resource prediction without a configuration, orchestration or diagnostic decision.

Source identification began with 92 records cited in the manuscript and added three fixed arXiv title-and-abstract queries, publisher-page discovery and one backward-citation round. Publisher discovery was limited to individual pages from the Institute of Electrical and Electronics Engineers (IEEE) and Association for Computing Machinery (ACM). The queries cover 1 January 2016 through 28 September 2026. The citation seeds are INTA~\cite{wei2025inta} for configuration, Chat-Driven Optimal Management~\cite{miyaoka2025chatdriven} for orchestration and Network Arena~\cite{wang2025network} for diagnosis. The three complete arXiv result sets contain 159 hits and 151 unique identifiers. Combining these with the starting bibliography and targeted additions produces 250 records. Figure~\ref{fig:e0-screening} records duplicates, cross-source overlaps, exclusions and report grouping.

Assistant-led retrieval, deduplication and screening identified 145 records for review of the relevant full-text sections. These checks retained 139 source/version records in 133 reading groups and moved six records to background material. Appendix~\ref{app:e0-protocol} gives the search routes, complete queries, retrieval date, eligibility criteria and coding definitions.

Two researchers independently coded the retained groups. Their review excluded one study that predicted central processing unit (CPU) consumption without an identifiable in-scope network decision. This screen yields 132 paper/report families and 138 source/version records, mapped to 137 bibliographic entries. Figure~\ref{fig:e0-screening} distinguishes initial screening from independent human coding. A second screening of all exclusions adds seven families ($\S$\ref{sec:review-verification}), giving 139 families and 145 source/version records.

\subsection{Screening Verification and Search Coverage}\label{sec:review-verification}

Two independent coders re-screened all 111 records that the initial screen had set aside. These comprise 79 background and 26 out-of-scope records from title-and-abstract screening and the six records moved to background after full-text review. The coders were two language models from different families, one GPT and one Claude model. Each worked from the published eligibility criterion, blind to the original decision and to the other coder. Title-and-abstract agreement over include, background and exclude was Cohen's $\kappa=0.661$~\cite{cohen1960agreement}, with Gwet's agreement coefficient (AC1)~\cite{gwet2008computing} at 0.738. After one exchange of rationales, 24 records went to full-text review, where both coders judged seven eligible. Because the original reasons for these seven applied a narrower operational scope than the published criterion, the seven enter the corpus as new families. Two further records remained split after reconciliation and stay outside the corpus.

The same two model coders applied the unchanged 18-field codebook to the seven families. Before reconciliation they agreed on 114 of 126 family-level field judgments (90.5\%), including all four timing-claim fields. A field whose remaining disagreement after reconciliation changed family-level positivity is coded unclear. Seven field judgments are coded this way, none of them in the four timing-claim fields. This agreement is kept separate from the researchers' agreement in Table~\ref{tab:e0-agreement}.

A recall check ran the three query concepts over OpenAlex, a multi-publisher index, with the same date window. The 966 unique works include 119 records already in the corpus. The two coders screened the other 847 by title and abstract with three-class $\kappa=0.741$ (AC1 0.812) and $\kappa=0.828$ for include versus not. Of these, 594 passed.

Before checking access to any full text, we drew a stratified random sample of 100 records, 50 from language-model work and 50 from earlier intent networking. Full text was retrieved for 98 draws, and the two coders applied the full-text criterion and the four timing fields of the codebook. Before reconciliation they agreed on eligibility for 90.8\% of the draws and on the timing fields for 60.2--89.8\%. Of the 98 draws, 77 are eligible and 2 remain unresolved. The frame-weighted rates are 32.4\% for explicit loop claims (95\% interval 22.2--42.9\%), 6.5\% for matched support (1.3--12.4\%), 5.3\% for p95 reporting (1.2--10.8\%) and 2.6\% for deadline attainment (0--6.9\%). Each interval contains the corresponding corpus rate. Bounds that count the two unretrieved draws and every unresolved code as negative or as positive give 30.7--45.4\% for loop claims and 6.1--20.9\% for matched support. Appendix~\ref{app:coverage-audit} gives the queries, strata, per-stratum counts and estimator. Records outside the primary search routes therefore report loop claims and matched support at rates consistent with the corpus. The 139 families remain the analyzed set, and their denominators count families from the primary routes.

\subsection{Coding Units and Adjudication}

Each coding unit identifies a work, task, decision step and evidence scope. The final inventory contains 429 scope records, made up of 393 step definitions and 36 linked scopes. The two researchers coded 405 of them, and the two model coders coded the 24 scopes of the added families. Linked scopes retain distinct implementations, versions or aggregate measurements without automatically creating another step. Shared measurements are linked instead of being copied to each component of a pipeline. Studies that share a benchmark remain separate families.

The codebook contains 18 fields in four groups (Table~\ref{tab:e0-codebook}). Eight record timing, workload and comparison evidence such as tail latency, deadline attainment and a non-learning comparator. Four record the control-loop claim, its measured support and the deployment context. Four record the three correctness gates and whether a scope distinguishes them, and two record public code and data pointers. Missing, unclear and not-applicable evidence retain distinct codes.

Of the 405 scope pairs, 231 compare the initial answers and 174 compare independent rechecks made after the scope boundaries were aligned. Table~\ref{tab:e0-agreement} reports these strata separately and in combination. The combined estimate summarizes both stages before adjudication. Coverage estimates use consensus values, whereas agreement estimates use the independent answers.

For single-choice fields, we report unweighted $\kappa$ and AC1. AC1 uses the codebook's category universe. The two multi-select fields use exact-set agreement and option-wise binary $\kappa$/AC1. Table~\ref{tab:e0-agreement-options} gives the option-wise coefficients with each researcher's selection count, so that common non-selection stays visible.

After comparing the independent answers, the researchers clarified the load and stability definitions during adjudication (Appendix~\ref{app:e0-c1}). Operational arrival or traffic load is distinct from single-request complexity, and stability concerns behavior over operational time or load. This clarification changed 65 of 810 load/stability values.

\subsection{Aggregation and Uncertainty}

The primary estimand is reporting coverage across all 139 families. A family is positive for an indicator when at least one scope has the target code. Missing, unclear and not-applicable records remain in this denominator. Table~\ref{tab:e0-audit} also reports coverage over 429 scopes and over the 393 step definitions, and Table~\ref{tab:e0-categories} gives the scope-level category counts. Aggregate measurements count at the scope where they are recorded.

Descriptive 95\% confidence intervals (CIs) use 10,000 bootstrap resamples of whole paper/report families with seed~42. Resampling whole families preserves the dependence among scopes of the same family. These intervals summarize variation among families in the assembled corpus under the adjudicated codes. Zero or complete observed coverage produces a degenerate empirical interval.

The task-group figure uses overlapping screening labels. Configuration covers 66 families, orchestration 52 and diagnosis 38. These labels were assigned while preparing the reading groups and were not independently coded. Because multi-task families appear in more than one group, comparisons between groups are descriptive.

\section{Systematization}\label{sec:systematization}
\subsection{Three Axes}
We locate each network task on three axes. The decision interface distinguishes selection, generation and deterministic computation. The control or execution path distinguishes offline preparation from request-driven operation and identifies the functional role consuming the output. Check ownership records who performs three checks. The observation check supplies or reconciles the observed state a decision uses. The feasibility check evaluates joint feasibility, and the coverage check detects absent candidate coverage.

A single paper can contain several interfaces and measurement scopes. Table~\ref{tab:e0-systematization} maps all 139 families on the three axes. Where the reviewed evidence specifies them, check entries give the owning component and the type of evidence at the reported task boundary, from a proposed design or model self-check to deterministic validation, a finite execution test or formal verification. A check whose responsible component remains unclear in the reviewed evidence has unresolved ownership.

Two model coders, one from the GPT family and one from the Claude family, recoded the three axes blind from full text under a frozen codebook. A third blind pass mapped the published cells into the same fields (Table~\ref{tab:tx-reliability}). Over the 129 families outside a 10-family pilot, the coders agree on the execution path with $\kappa=0.69$ and on the generation and computation interfaces with $\kappa=0.72$ and $0.67$. Selection is less reproducible ($\kappa=0.50$). Whether a family places an observation, feasibility or coverage check is reproduced with $\kappa$ from 0.62 to 0.73, and the owner category of a shared check matches exactly in 65 to 68\% of cases.

Agreement with the published entries is lower for check presence ($\kappa$ from 0.10 to 0.42). Both coders agree that 40 of the 129 families place an observation check and 43 a coverage check, where Table~\ref{tab:e0-systematization} records 9 and 11. Of those 9 observation checks, both coders find 6, and of the 11 coverage checks they find 7. Neither coder finds one of the observation checks and two of the coverage checks. The codebook excludes plain reading of current state, which several published observation entries include. The check columns therefore support no prevalence estimate, and the paper draws none from them.

\subsection{Patterns Across the 139 Families}\label{sec:taxonomy-patterns}
Generation is the most common decision interface. It appears in 94 of the 139 families, selection in 76 and deterministic computation in 67, and 83 families combine two or more interfaces (Table~\ref{tab:taxonomy-patterns}).

In the published entries, check ownership concentrates in families with a deterministic computation step. Families with deterministic computation name a feasibility owner in 45 of 67 cases and a coverage owner in 11. Families without it name a feasibility owner in 23 of 72 and a coverage owner in 2. Operational load is reported by 35 of the 67 and by 13 of the 72.

In the evidence audit ($\S$\ref{sec:evidence-audit}), families without a computation step make 22 of the 50 loop claims, and none of the 22 is supported by matched measurement. All four matched claims come from families that combine computation with selection or generation. These counts are descriptive and overlap with task groups. The lenses in $\S$\ref{sec:cross-study} test what the missing responsibilities change.

\begin{table}[t]
\centering\footnotesize
\setlength{\tabcolsep}{2.6pt}
\renewcommand{\arraystretch}{1.05}
\caption{Reporting evidence and check ownership by decision interface across the 139 families. Rows partition the families by the interface letters in Table~\ref{tab:e0-systematization}: selection (S), generation (G), deterministic computation (C) and unspecified (U). Cells count families. Reporting columns use the evidence audit, and check columns count families whose published entry names an owner.}\label{tab:taxonomy-patterns}
\begin{tabular}{@{}lrrrrrrrr@{}}
\toprule
\rowcolor{ebBlue} & & \multicolumn{2}{c}{Loop claim} & & & & \multicolumn{2}{c}{Check owned} \\
\rowcolor{ebBlue} Interface & $n$ & Any & Matched & p95+ & Deadline & Load & Feasib. & Cover. \\
\midrule
S only & 11 & 4 & 0 & 0 & 1 & 4 & 1 & 0 \\
G only & 24 & 5 & 0 & 1 & 1 & 3 & 9 & 0 \\
U only & 2 & 0 & 0 & 0 & 0 & 0 & 1 & 0 \\
Mixed, no C & 35 & 13 & 0 & 1 & 0 & 6 & 12 & 2 \\
\midrule
C only & 19 & 3 & 0 & 2 & 0 & 11 & 14 & 1 \\
Mixed with C & 48 & 25 & 4 & 5 & 2 & 24 & 31 & 10 \\
\midrule
All & 139 & 50 & 4 & 9 & 4 & 48 & 68 & 13 \\
\bottomrule
\end{tabular}
\end{table}

\subsection{Functional Roles and Check Ownership}
Table~\ref{tab:responsibilities} maps tasks to model outputs, controller responsibilities and comparison baselines. In an O-RAN workflow, an xApp or rApp can consume an interpreted objective, but measurements, policy communication and device control remain separate functions. In management, the intent handler decomposes objectives and an orchestrator binds them to resources. A software-defined networking (SDN) controller or policy function installs decisions. These roles can place the observation, feasibility and coverage checks in different components. The implemented path fixes where each check runs.

\begin{table*}[!t]\centering\footnotesize
\caption{Model roles and other workflow responsibilities for representative network tasks. Baselines reflect the output responsibility of each comparison.}
\label{tab:responsibilities}\setlength{\tabcolsep}{4pt}\renewcommand{\arraystretch}{1.12}
\def\familytab{\begin{tabular}{llll}\toprule
\rowcolor{ebBlue} Task & Model role & Other workflow steps & Baselines \\\midrule
Intent~\cite{manias2024intentbased,li2026replacing} & Classify attributes & Values; binding; joint checks & Classifier; structured LLM \\
Documents~\cite{nam2026llmbased,wei2025inta} & Rank or reject & Retrieval; extraction & Retrieval score; re-ranker \\
Configuration~\cite{wang2024netconfeval,el-hassany2018netcomplete} & Select template or tool & Generation; validation & Constrained generator \\
Properties~\cite{khurshid2013veriflow,hossain2025netintent} & Interpret requirement & Formalization; checking & Symbolic verifier \\
Diagnosis~\cite{natu2006active,clark2026jev} & Rank tests or actions & Hypotheses; probes; execution & Probing; procedures; tool agents \\
Reuse~\cite{bang2023gptcache,li2026fast} & Trigger reinterpretation & Refresh; recomputation & Cache with controller \\\bottomrule
\end{tabular}}
\setlength{\tabcolsep}{0pt}\sbox0{\familytab}
\dimen0=\linewidth\advance\dimen0 by -1pt\advance\dimen0 by -\wd0\divide\dimen0 by 8
\ifdim\dimen0<1.5pt \dimen0=1.5pt\fi
\setlength{\tabcolsep}{\dimen0}\familytab
\end{table*}

Check ownership determines when validation can affect an action. NetComplete computes a completion of an operator's sketch under modeled requirements~\cite{el-hassany2018netcomplete}, and VeriFlow evaluates forwarding invariants~\cite{khurshid2013veriflow}. INTA generates configurations after retrieval~\cite{wei2025inta}. In configuration-repair evaluation, a final scoring oracle evaluates the completed configuration, whereas an agent-feedback checker guides intermediate repairs~\cite{asadli2026evaluating}. A deployment guard checks actions before admission. Table~\ref{tab:e0-systematization} records the reported check responsibilities.

\subsection{Candidate Coverage and Conditional Selection}
For decision instance $i$, let $C_i$ be the permitted output set, $A_i$ the acceptable output set under the task predicate and $\hat a_i\in C_i$ the selection. The coverage indicator $K_i=\mathbf{1}\{C_i\cap A_i\ne\varnothing\}$ is one exactly when an acceptable output is available. If $\Pr(K_i=1)>0$, then
\begin{equation}
\Pr(\hat a_i\in A_i)=\Pr(K_i=1)\Pr(\hat a_i\in A_i\mid K_i=1).
\label{eq:coverage}
\end{equation}
Valid rejection or clarification belongs in both sets when the task allows it, and multi-field answers require an acceptable complete output. Invalid interface returns are counted separately. Equation~\eqref{eq:coverage} separates availability of an acceptable response from selection quality conditional on availability.

When the permitted output is an escalation request, a correct selection initiates catalogue construction or service recovery. Candidate preparation, construction of missing content, selection, execution and verification each have their own outcome. The tests in $\S$\ref{sec:cross-study} examine how escalation success changes across tasks.

\section{Literature Reporting Coverage}
\label{sec:evidence-audit}

\begin{figure*}[tp]
\centering
\includegraphics[width=7in]{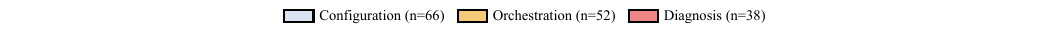}\\[-2pt]
\subfloat[Timing evidence]{\includegraphics[width=1.70in]{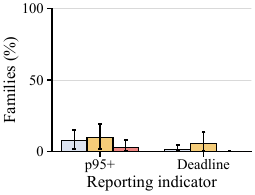}}\hfill
\subfloat[Load and waiting]{\includegraphics[width=1.70in]{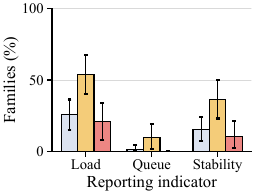}}\hfill
\subfloat[Comparators and checks]{\includegraphics[width=1.70in]{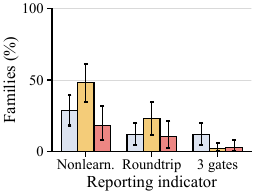}}\hfill
\subfloat[Code and data]{\includegraphics[width=1.70in]{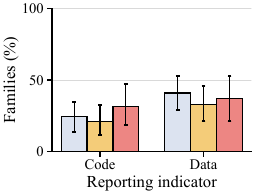}}
\caption{Reporting coverage by overlapping task group: (a) tail and deadline timing, (b) load, waiting and stability, (c) non-learning comparators, round trips and three-gate separation, and (d) code and data pointers. Bars show positive-family percentages with descriptive 95\% whole-family bootstrap intervals (10,000 draws). Task labels come from screening. Evidence fields were independently coded and adjudicated. Group sizes are in the legend.}
\label{fig:e0-audit}
\end{figure*}

Table~\ref{tab:e0-audit} reports all 18 indicators over the 139 families with their denominators and family-level intervals, and Figure~\ref{fig:e0-audit} compares reporting across overlapping task groups.

\subsection{Timing Coverage and Operational Conditions}

Among 139 families, 92 (66.2\%) identify a timing endpoint and 49 (35.3\%) specify a metric denominator under the codebook. Tail and deadline evidence is much less common. Nine families report latency at the 95th percentile (p95) or higher (6.5\%, 95\% interval 2.9--10.8\%), and four report deadline attainment (2.9\%, 0.7--5.8\%).

Operational input load is reported in 48 families (34.5\%, 26.6--42.4\%), queue or waiting evidence in six (4.3\%, 1.4--7.9\%) and stability evidence in 31 (22.3\%, 15.8--29.5\%). The stability indicator records analytical, simulated or testbed examination of behavior over operational time or load. Single-request scaling is coded separately.

An explicit non-learning comparator appears in 48 families (34.5\%, 26.6--42.4\%). Twenty-three families explicitly include network round trips in a relevant timing measurement (16.5\%, 10.8--23.0\%). The denominator is all 139 families, including those for which network round trips are inapplicable. Table~\ref{tab:e0-categories} distinguishes explicit exclusion, unclear boundaries and inapplicability.

\begingroup\footnotesize
\setlength{\tabcolsep}{4pt}
\setlength{\aboverulesep}{0pt}\setlength{\belowrulesep}{0pt}
\renewcommand{\arraystretch}{1.10}
\begin{table*}[tp]
\centering
\caption{Literature reporting coverage. Cells give count/denominator, percentage and 95\% paper-family bootstrap interval (10,000 draws). A positive family has at least one scope with the code. Unclear may overlap positive coverage, and all-NA means every scope is inapplicable. All remain in the stated denominators. Step definitions exclude the 36 linked scopes.}
\label{tab:e0-audit}
\def\familytab{\begin{tabular}{lrrrrr}
\toprule
\rowcolor{ebBlue} Field & Families (139) & All scopes (429) & Step definitions (393) & \shortstack{Unclear\\families} & \shortstack{All-NA\\families} \\
\midrule
\rowcolor{ebAmber!45}\multicolumn{6}{l}{\textbf{\itshape Timing, workload, and comparison}} \\*
Named timing endpoint & 92/139 (66.2\%) \textcolor{black!60}{[58.3, 74.1]} & 164/429 (38.2\%) \textcolor{black!60}{[31.9, 44.9]} & 143/393 (36.4\%) \textcolor{black!60}{[30.1, 43.1]} & 2 & 0 \\
\addlinespace[1.5pt]
Specified metric denominator & 49/139 (35.3\%) \textcolor{black!60}{[27.3, 43.2]} & 86/429 (20.0\%) \textcolor{black!60}{[14.6, 26.1]} & 69/393 (17.6\%) \textcolor{black!60}{[12.5, 23.4]} & 55 & 46 \\
\addlinespace[1.5pt]
p95 or higher & 9/139 (6.5\%) \textcolor{black!60}{[2.9, 10.8]} & 14/429 (3.3\%) \textcolor{black!60}{[1.2, 5.8]} & 11/393 (2.8\%) \textcolor{black!60}{[1.1, 4.9]} & 1 & 0 \\
\addlinespace[1.5pt]
Deadline attainment & 4/139 (2.9\%) \textcolor{black!60}{[0.7, 5.8]} & 7/429 (1.6\%) \textcolor{black!60}{[0.2, 3.6]} & 5/393 (1.3\%) \textcolor{black!60}{[0.2, 2.8]} & 2 & 0 \\
\addlinespace[1.5pt]
Input load & 48/139 (34.5\%) \textcolor{black!60}{[26.6, 42.4]} & 87/429 (20.3\%) \textcolor{black!60}{[15.0, 26.0]} & 74/393 (18.8\%) \textcolor{black!60}{[13.5, 24.5]} & 1 & 0 \\
\addlinespace[1.5pt]
Queue / waiting & 6/139 (4.3\%) \textcolor{black!60}{[1.4, 7.9]} & 11/429 (2.6\%) \textcolor{black!60}{[0.7, 4.9]} & 8/393 (2.0\%) \textcolor{black!60}{[0.5, 3.9]} & 2 & 0 \\
\addlinespace[1.5pt]
Stability evidence & 31/139 (22.3\%) \textcolor{black!60}{[15.8, 29.5]} & 49/429 (11.4\%) \textcolor{black!60}{[7.6, 15.5]} & 36/393 (9.2\%) \textcolor{black!60}{[6.0, 12.6]} & 0 & 0 \\
\addlinespace[1.5pt]
Non-learning comparison & 48/139 (34.5\%) \textcolor{black!60}{[26.6, 42.4]} & 68/429 (15.9\%) \textcolor{black!60}{[11.7, 20.3]} & 57/393 (14.5\%) \textcolor{black!60}{[10.5, 18.8]} & 3 & 0 \\
\addlinespace[1.5pt]
\rowcolor{ebAmber!45}\multicolumn{6}{l}{\textbf{\itshape Control-loop claim and deployment}} \\*
Explicit loop claim & 50/139 (36.0\%) \textcolor{black!60}{[28.1, 43.9]} & 127/429 (29.6\%) \textcolor{black!60}{[22.0, 37.5]} & 112/393 (28.5\%) \textcolor{black!60}{[21.1, 36.3]} & 0 & 0 \\
\addlinespace[1.5pt]
Supported loop claim & 4/139 (2.9\%) \textcolor{black!60}{[0.7, 5.8]} & 7/429 (1.6\%) \textcolor{black!60}{[0.2, 3.6]} & 5/393 (1.3\%) \textcolor{black!60}{[0.2, 2.8]} & 0 & 2 \\
\addlinespace[1.5pt]
Named execution location & 86/139 (61.9\%) \textcolor{black!60}{[54.0, 69.8]} & 214/429 (49.9\%) \textcolor{black!60}{[41.8, 57.5]} & 191/393 (48.6\%) \textcolor{black!60}{[40.5, 56.6]} & 51 & 16 \\
\addlinespace[1.5pt]
Network round trip included & 23/139 (16.5\%) \textcolor{black!60}{[10.8, 23.0]} & 39/429 (9.1\%) \textcolor{black!60}{[5.1, 13.9]} & 31/393 (7.9\%) \textcolor{black!60}{[4.1, 12.8]} & 58 & 61 \\
\addlinespace[1.5pt]
\rowcolor{ebAmber!45}\multicolumn{6}{l}{\textbf{\itshape Validation responsibilities}} \\*
Format check reported & 71/139 (51.1\%) \textcolor{black!60}{[42.4, 59.0]} & 138/429 (32.2\%) \textcolor{black!60}{[25.8, 38.7]} & 123/393 (31.3\%) \textcolor{black!60}{[25.1, 37.7]} & 0 & 0 \\
\addlinespace[1.5pt]
Semantic check reported & 133/139 (95.7\%) \textcolor{black!60}{[92.1, 98.6]} & 355/429 (82.8\%) \textcolor{black!60}{[78.1, 87.1]} & 323/393 (82.2\%) \textcolor{black!60}{[77.6, 86.6]} & 4 & 0 \\
\addlinespace[1.5pt]
Final-state check reported & 54/139 (38.8\%) \textcolor{black!60}{[30.9, 46.8]} & 83/429 (19.3\%) \textcolor{black!60}{[14.8, 24.2]} & 72/393 (18.3\%) \textcolor{black!60}{[14.1, 22.9]} & 9 & 19 \\
\addlinespace[1.5pt]
Three gates distinguished & 9/139 (6.5\%) \textcolor{black!60}{[2.9, 10.8]} & 12/429 (2.8\%) \textcolor{black!60}{[1.1, 5.0]} & 11/393 (2.8\%) \textcolor{black!60}{[1.1, 4.8]} & 2 & 0 \\
\addlinespace[1.5pt]
\rowcolor{ebAmber!45}\multicolumn{6}{l}{\textbf{\itshape Artifact pointers}} \\*
Public code pointer & 34/139 (24.5\%) \textcolor{black!60}{[17.3, 31.7]} & 96/429 (22.4\%) \textcolor{black!60}{[15.4, 30.1]} & 92/393 (23.4\%) \textcolor{black!60}{[16.1, 31.4]} & 8 & 0 \\
\addlinespace[1.5pt]
Public data pointer & 51/139 (36.7\%) \textcolor{black!60}{[28.8, 44.6]} & 140/429 (32.6\%) \textcolor{black!60}{[24.4, 40.9]} & 129/393 (32.8\%) \textcolor{black!60}{[24.3, 41.5]} & 5 & 0 \\
\addlinespace[1.5pt]
\bottomrule
\end{tabular}}
\setlength{\tabcolsep}{0pt}\sbox0{\familytab}
\dimen0=\linewidth\advance\dimen0 by -1pt\advance\dimen0 by -\wd0\divide\dimen0 by 12
\ifdim\dimen0<1.5pt \dimen0=1.5pt\fi
\setlength{\tabcolsep}{\dimen0}\familytab
\end{table*}
\endgroup

\subsection{Correctness Gates and Reusable Artifacts}

Semantic checks are reported in 133 families (95.7\%), format checks in 71 (51.1\%) and final-state checks in 54 (38.8\%). These counts can refer to different scopes within a family. Nine families distinguish all three gates in some scope (6.5\%, 2.9--10.8\%).

Public code pointers appear in 34 families (24.5\%) and public data pointers in 51 (36.7\%). The indicators record links supplied in the paper or supplement.

\subsection{Coding Reliability}

Combined pre-adjudication exact agreement is 99.8\% for tail reporting and 98.3\% for deadline reporting, but 27.9\% for load, 39.3\% for queue evidence and 30.6\% for stability. For these three fields, the $\kappa$/AC1 pairs are 0.083/0.088, 0.041/0.281 and 0.075/0.143. Agreement is also low for network round-trip inclusion (42.5\%) and final-state checks (42.2\%).

Coverage of these five fields therefore carries greater coding uncertainty than tail and deadline coverage. Appendix~\ref{app:e0-c1} gives the load and stability definitions clarified during adjudication. Table~\ref{tab:e0-agreement} reports agreement before adjudication, and Appendix~\ref{app:agreement} gives the category counts and multi-select agreement.

\subsection{A Minimum Reporting Record}\label{sec:reporting-record}

Each network decision step needs a reporting record that connects its input to the claimed outcome. Identify the input, decision output, candidates and checks assigned to the model and controller. Define the measured events from arrival and queue entry through decision, execution acknowledgement and service verification, and mark unavailable events explicitly.

Report each correctness gate with its own denominator and retain failure, timeout, rejection, escalation and fallback counts. Give the latency distribution and deadline attainment for the claimed control loop. Report them with the operational arrival or traffic load, decision concurrency, queueing or waiting evidence and the observation window.

Include a non-learning comparator that uses the same information and constraints and undergoes the same execution checks. Identify deployment location and whether communication, retries and validation enter the timing boundary. Provide code and data pointers, and state which measurements are inapplicable.

\section{Unified Evaluation Framework}
\label{sec:unified-framework}

A decision engine must be evaluated at the point where its output changes a network workflow. Our framework defines common events, correctness gates and perturbations for service contracts, RAN policies and application-layer control tasks. Fixed-input comparisons examine semantic decisions. Load traces measure queueing and deadline attainment. Fixed-action interventions isolate decision delay in transport and edge execution stacks.

\subsection{Timing and Decision Capacity}
\label{sec:framework-events}

An intent arrives at $t_0$, leaves the decision queue at $t_q$ and produces a decision accepted by the declared schema or controller checks at $t_d$. The execution interface acknowledges the action at $t_a$, and a domain-specific probe verifies the requested result at $t_b$. We record queueing, decision time and the two downstream elapsed times separately:
\begin{align}
 Q &= t_q-t_0, & D &= t_d-t_q, \label{eq:framework-decision}\\
 T_a &= t_a-t_0, & T_b &= t_b-t_0. \label{eq:framework-endpoints}
\end{align}
Semantic correctness is evaluated separately against the task predicate (Gate~2, $\S$\ref{sec:framework-gates}). The meaning of $t_a$ follows the interface contract. In the edge and transport intervention, $t_a$ is an API or configuration acknowledgement and $t_b$ verifies the new workload or route. Table~\ref{tab:framework-endpoints} maps these events and the milestones recorded by each execution interface.

\begin{table}[t]
\centering
\caption{Timing events and measurement boundaries in the common event model.}
\label{tab:framework-endpoints}
\footnotesize
\setlength{\tabcolsep}{3pt}
\renewcommand{\arraystretch}{1.12}
\newcommand{\eventname}[1]{\parbox[c][28pt][c]{\linewidth}{\raggedright\bfseries\strut #1}}
\newcommand{\eventcell}[3]{\parbox[c][28pt][c]{\linewidth}{\raggedright\strut #1\\\strut #2\\\strut #3}}
\begin{tabular}{>{\raggedright\arraybackslash}m{.27\columnwidth}>{\raggedright\arraybackslash}m{\dimexpr.73\columnwidth-4\tabcolsep-1pt\relax}}
\toprule
\rowcolor{ebBlue} Boundary & What the record establishes \\
\midrule
\eventname{Client return} & \eventcell{$L_{\mathrm{ret}}$ is observed call duration.}{Format validity reported separately.}{Service completion needs its own probe.} \\
\eventname{Decision / slot} & \eventcell{$t_d$ is the decision after stated checks.}{$S$ spans slot acquisition to actual release.}{Includes waits, retries and timeouts.} \\
\eventname{Interface ACK} & \eventcell{$t_a$ is Kubernetes API acknowledgement}{or FRRouting configuration submission.}{RAN uses base-station acknowledgement.} \\
\eventname{Verified result} & \eventcell{$t_b$ is a new edge replica Ready plus}{a correct image response or required}{route within its round-trip time bound.} \\
\eventname{RAN observation} & \eventcell{$t_K$ is the first recorded KPM change.}{Records a measurement update.}{Service verification uses a separate $t_b$.} \\
\bottomrule
\end{tabular}
\end{table}

For all arrivals $\mathcal{A}$, let $T_{e,i}=t_{e,i}-t_{0,i}$ be the elapsed time to endpoint $e\in\{a,b\}$ for arrival $i$. Deadline attainment at budget $B$ is
\begin{equation}
 \widehat{p}_e(B)=\frac{1}{|\mathcal{A}|}
 \sum_{i\in\mathcal{A}}\mathbf{1}\{t_{e,i}\text{ observed},\ T_{e,i}\leq B\}.
 \label{eq:framework-budget}
\end{equation}
Equation~\eqref{eq:framework-budget} counts an unreached endpoint as a miss. The record keeps its failure reason and the elapsed time to timeout or termination. Endpoint quantiles use the requests that reach that endpoint and are reported with the observed count and all-arrival attainment. Ratios $D/T_a$ and $D/T_b$ use the corresponding observed pairs. We report medians of these per-request ratios. Segment medians and total-duration medians are computed separately. We denote the 50th, 95th and 99th percentiles by p50, p95 and p99.

The slot can remain occupied after $t_d$. The quantity $S=t_{\mathrm{rel}}-t_q$ uses actual release time $t_{\mathrm{rel}}$. The offered arrival rate $\lambda$, mean occupied time $\mathbb{E}[S]$ and number of decision slots $c$ define the load indicator
\begin{equation}
 \rho=\lambda\mathbb{E}[S]/c.
 \label{eq:framework-rho}
\end{equation}
For an open workload admitted without shedding, $\rho<1$ is a necessary capacity condition. Deadline attainment also depends on burstiness, service-time variation and buffering. Admission-controlled traces distinguish offered and accepted rates. In service admission, the offered-load estimate uses service times of dispatched requests. Service times for rejected requests are unobserved. The indicator $\rho$ is unavailable for fixed-input and serial runs without open arrivals.

For client-return records, we report $\Pr(L_{\mathrm{ret}}\leq B)$ and the joint probability of format-valid return within $B$. RAN records supply an acknowledgement and a KPM observation at $t_K$. In some of these records the first KPM change precedes the acknowledgement, and the signed KPM-minus-acknowledgement difference is then negative.

\subsection{Three Gates with Explicit Denominators}
\label{sec:framework-gates}

Gate~1 is schema or interface validity over all arrivals. Gate~2 is satisfaction of the task predicate over applicable arrivals, including minimum-cost selection when required by that task. Deliberately unsupported requests are shown separately. Gate~3 is a verified correct terminal outcome over all arrivals. It can be a successful requested service or a completed rejection or escalation specified by the protocol. Failures, timeouts and fallbacks stay in their denominators.

A rejection meets Gate~2 when the task predicate requires it. Online evaluation reports first-decision correctness, direct execution success and final success after fallback. Each task's native predicate specifies what must be correct. Service-contract tasks report service top-1 and full-contract exact match. RAN-policy tasks report target-cluster correctness on telemetry-dependent inputs and full-policy exact match.

\subsection{Perturbations and Diagnostic Lenses}
\label{sec:framework-perturbations}

Table~\ref{tab:framework-perturbations} maps perturbations to their experimental conditions, diagnostic lenses and measured outcomes. Structural expansion varies output fields, searched telemetry rows or interacting requirements along each task's axis. Catalogue changes separate covered selection from rejection or escalation when coverage disappears. Unsupported service requests provide an analogous condition. The check-ownership intervention compares model self-check with a public-gate workflow for coupled placement. It changes instructions and refresh behavior together. The execution-path intervention estimates delay effects within domains and compares paths descriptively across domains.

\begin{table*}[!t]
\centering
\caption{Perturbation families, diagnostic lenses, experimental conditions and measured outcomes. Outcomes follow the task predicates and timing endpoints. N/A denotes no corresponding arm. ACK denotes acknowledgement.}
\label{tab:framework-perturbations}
\footnotesize
\setlength{\tabcolsep}{3.5pt}
\renewcommand{\arraystretch}{1.14}
\newdimen\frameworkwidth
\frameworkwidth=\dimexpr\textwidth-10\tabcolsep-1pt\relax
\newcommand{\mapcell}[2]{\parbox[c][20pt][c]{\linewidth}{\raggedright\strut #1\\\strut #2}}
\begin{tabular}{>{\raggedright\arraybackslash}m{.16\frameworkwidth}>{\raggedright\arraybackslash}m{.19\frameworkwidth}>{\raggedright\arraybackslash}m{.20\frameworkwidth}>{\raggedright\arraybackslash}m{.25\frameworkwidth}>{\raggedright\arraybackslash}m{.20\frameworkwidth}}
\toprule
\rowcolor{ebBlue} Family / lens & Service contracts~\cite{li2026replacing} & RAN policies~\cite{li2026intent} & Application tasks / intervention & Measured outcomes \\
\midrule
\mapcell{\textbf{Structure}}{L4} & 4, 6 or 8 output fields & \mapcell{Fresh with 3, 7, 21 or 57}{telemetry cells} & \mapcell{4, 16 or 64 interacting}{requirements} & \mapcell{Task correctness}{and return time} \\
\mapcell{\textbf{Irrelevant text}}{L4} & \mapcell{Padding up to}{16,384 tokens} & N/A & \mapcell{Irrelevant context}{and repeated requirements} & \mapcell{Task correctness}{and return time} \\
\mapcell{\textbf{Observations}}{L3/L4} & N/A & \mapcell{Fresh / stale / noisy /}{contradictory telemetry} & \mapcell{Complete / prose / missing / stale}{conflict / resolved conflict} & \mapcell{Correct handling of}{observations and return time} \\
\mapcell{\textbf{Covered catalogue}}{L3/L4} & \mapcell{Scale / unseen services}{partial replacement} & N/A & \mapcell{Stable IDs / rekeying}{replacement with two candidates} & \mapcell{Selection correctness}{and return time} \\
\mapcell{\textbf{Absent coverage}}{L3} & \mapcell{Unsupported requests}{as an analogy} & N/A & \mapcell{Missing contract candidate}{or infeasible path} & \mapcell{Correct rejection or escalation}{and return time} \\
\mapcell{\textbf{Arrivals}}{L2} & Open admission traces & Open intent-arrival traces & N/A & \mapcell{Queue time and slot load $\rho$}{with native deadline attainment} \\
\mapcell{\textbf{Check ownership}}{L3} & \mapcell{Cached interpretation}{and numerical scheduling} & N/A & \mapcell{Model self-check versus}{public-gate workflow} & \mapcell{Task correctness}{and return time} \\
\mapcell{\textbf{Execution path}}{L1/L2} & \mapcell{Real optical character}{recognition (OCR)} & \mapcell{A1/E2 ACKs}{and KPM observations} & \mapcell{Transport/edge physical paths}{and real queues with replay} & \mapcell{Stage and outcome times}{with queue amplification} \\
\bottomrule
\end{tabular}
\end{table*}

\subsection{Questions and Comparison Families}
\label{sec:framework-hypotheses}

The questions were formulated after inspecting the measurement records and are exploratory.

\begin{table*}[t]
\centering
\caption{Diagnostic lenses, exploratory questions, estimands, declared comparison families with their contrast counts, and data sources. Each family retains all eligible contrasts defined before analysis. Edge and RAN denote the source studies~\cite{li2026replacing,li2026intent}.}
\label{tab:framework-lenses}
\footnotesize
\setlength{\tabcolsep}{3pt}
\renewcommand{\arraystretch}{1.12}
\newdimen\lenswidth
\lenswidth=\dimexpr\textwidth-10\tabcolsep-1pt\relax
\begin{tabular}{>{\raggedright\arraybackslash}p{.15\lenswidth}>{\raggedright\arraybackslash}p{.33\lenswidth}>{\raggedright\arraybackslash}p{.22\lenswidth}>{\raggedright\arraybackslash}p{.18\lenswidth}>{\raggedright\arraybackslash}p{.12\lenswidth}}
\toprule
\rowcolor{ebBlue} Lens & Question & Estimands & Comparison families & Data \\
\midrule
L1 Measurement boundary & How does the endpoint change timing effects and admission verdicts? & Stage times, endpoint slopes, decision fractions & Decision delay (4 slope tests) & SoK, RAN \\
L2 Workload summary & How do median, tail and deadline summaries change under queued work? & Return and queue quantiles, attainment, $\rho$, modeled completion & Deadline and capacity (83 load cells) & SoK, Edge, RAN \\
L3 Check placement & What changes when observation, coverage or feasibility checks move within the workflow? & Correctness, validity, return p50/p95, online outcomes & Observation (66), coverage (39), workflow (45) & SoK, Edge, RAN \\
L4 Robustness scope & How do structure, length, representation and catalogue change bound a result? & Correctness, validity, return p50/p95, paired changes & Structure (89) & SoK, Edge, RAN \\
\bottomrule
\end{tabular}
\end{table*}

Each family sets the varied condition against a declared reference. For the timing lenses, the decision-delay family tests endpoint slopes against the practical band $[0.9,1.1]$, and the deadline and capacity family spans all 83 load cells with pointwise time-block intervals. For check placement, the observation family compares placement prose and the missing, stale, conflicting and resolved-conflict states with complete state. It also compares stale, noisy and contradictory RAN telemetry with fresh telemetry at 57 cells. The coverage family sets rekeyed, replaced, two-candidate and absent contracts against the base catalogue and absent paths against present ones. The workflow family holds three public-gate minus self-check contrasts and shared minus independent quota under both workflows for every method. For robustness, the structure family compares access-policy conditions and service-contract padding with the base condition, and fresh RAN conditions with 7, 21 and 57 cells with three cells on the telemetry-dependent subset. Contract field-count contrasts stay descriptive because requests differ.

Each finding receives a three-part verdict. The effect is \emph{repeated}, \emph{task-dependent}, \emph{unresolved} or \emph{descriptive}. Breadth is \emph{multi-task}, \emph{single-task} or \emph{single-study}. Ownership is \emph{SoK-owned}, \emph{borrowed-and-reanalysed} or \emph{borrowed context}.

We call an effect repeated when multiple tasks manipulate the same construct and use comparable outcomes. Their within-task intervals must also agree in direction beyond a declared negligible-effect reference. Different constructs or outcomes yield a task-dependent verdict. We use reference ranges of $\pm3$ percentage points for correctness and attainment-rate differences. The latency range is $\pm50$\,ms. Each task retains its own effect estimate.

\subsection{Estimation and Dependence}
\label{sec:framework-statistics}

Fixed-input tables report native proportions with 95\% Wilson intervals~\cite{wilson1927probable}, alongside return-time p50/p95 and the count with observed timing. In result tables, $n$ is the number of evaluated cases or arrivals in the stated condition. Within-task paired contrasts match input identities and pairing strata. Service-contract field-count and catalogue-size conditions with different requests use descriptive condition-level summaries. Valid pairs are resampled within their original task or wording strata using 10,000 bootstrap draws with seed~42. Latency contrasts use the median of per-input differences for pairs with observed timing.

Paired correctness contrasts use the two-sided exact McNemar test~\cite{mcnemar1947note}, implemented as a binomial test on discordant pairs with null probability $1/2$. Holm adjustment~\cite{holm1979simple} is applied separately to each declared correctness-comparison family. The notation $p_{\mathrm{Holm}}$ denotes the adjusted p-value. Bootstrap latency intervals remain pointwise. The six-case and 24-case online subsets are descriptive.

The timing analysis stratifies results by task, deployment, hardware and collection period. Its intervals resample occupied 30\,s time blocks, with 60 and 120\,s sensitivity views. Cells with fewer than eight occupied blocks are marked as having weak temporal coverage. Intervals are unavailable when timestamps are missing. The 10\,ms, 100\,ms, 1\,s and 10\,s budgets span control-loop timescales. Per-cell p99 is descriptive. For RAN Qwen, the primary record is the attempt returning Hypertext Transfer Protocol (HTTP) status~200 per case. Timing measures that attempt's duration.

\subsection{Execution-Path Intervention}
\label{sec:framework-intervention}

The measurement-boundary lens combines a sparse physical arm with a load arm using execution-time replay. In each domain, 30 fixed-action scenes receive delays of 0, 0.1, 0.5 and 2\,s. Ten physical repeats are nested within each scene and delay. The sparse arm has zero decision queueing time ($t_q=t_0$). It measures each endpoint's slope against actual decision delay and the associated decision fraction. Whole-scene bootstrap resampling retains nested repeats and delay conditions. Four mean-slope tests, two domains by two endpoints, form one Holm family.

The sparse arm completes 1,200 verified executions in each domain. Both stacks use the same guest environment. The scene, action, input and baseline recovery criteria remain fixed across delay arms. Seed~42 randomizes their order within each repeat.

Transport uses containerlab 0.79 and FRRouting (FRR) 10.5.1 on an eight-router ring. The ring runs Open Shortest Path First (OSPF) with full-mesh internal Border Gateway Protocol (BGP) sessions. Each scene fixes a source, service prefix and link-metric change selecting the required lower-delay path. Configuration submission defines $t_a$. The route, traceroute and round-trip time (RTT) criterion defines $t_b$.

Baseline recovery checks all eight link-state databases and waits six seconds of protocol quiet. It then verifies the forwarding information base (FIB) and ping. Native protocol timers are retained.

Edge uses k3d 5.9 and k3s 1.35.5. Its 30 migration or scaling scenes span two worker nodes and five image workloads. A real Pillow service returns a checked grayscale image. New Pod identities, readiness and the output pixel hash define verification. Kubernetes API acknowledgement defines $t_a$. Link emulation, API proxy and probe overhead are included in the measured path.

The load arm uses a real first-in, first-out (FIFO) decision queue with one slot per condition. The delay elapses inside that slot, and the slot is released afterwards. An isolated 100\,ms baseline is measured 200 times. The baseline sets three arrival rates with baseline load $\rho_0$ at 0.5, 0.8 and 0.95. Those rates remain fixed across delay conditions. Each Poisson arrival trace uses seed~42, a 60\,s warm-up and twenty subsequent 10\,s measurement blocks. The same arrivals and physical-residual draws are shared across delay conditions. Four controlled delays and 24 empirical timing strata from the workload-summary lens define 84 queues. The empirical conditions replay recorded durations by task, implementation, deployment and collection period.

After the queue, the load arm samples measured zero-delay physical residuals, including verification failures, to estimate outcome times under those residual distributions. Time-block intervals describe the finite observation window, with adjacent-block sensitivity at 20\,s. A two-hour drain limit bounds observation of unfinished queues. Endpoint attainment retains every measured arrival. The edge 2\,s and transport 10\,s targets are exploratory task budgets. A common sensitivity grid covers 0.5/1/2/5/10\,s. We estimate delay effects within each domain and compare the resulting paths across topologies, actions, stacks and probes.

The load arm completes all 84 real single-slot FIFO queues. It drains all 166,432 arrivals, including warm-up. The analysis uses the 126,588 post-warm-up arrivals in the twenty measurement blocks of each queue. The measured 100\,ms baseline sets rates of approximately 4.992, 7.987 and 9.484 arrivals/s.

Execution after the decision queue uses paired replay from each domain's 300 zero-delay physical observations. Twenty-four empirical timing strata supply recorded decision durations for further replay conditions~\cite{li2026replacing,li2026intent}.

\section{Consequences of the Systematized Gaps}\label{sec:cross-study}

We apply the event model in $\S$\ref{sec:framework-events} to service contracts, RAN policies and constructed application-layer tasks. Edge and RAN measurements are reused from their source studies. Each lens tests one gap exposed by the systematization and the audit. The measurement-boundary lens (L1) and the workload-summary lens (L2) test the missing timing and load evidence ($\S$\ref{sec:evidence-audit}). The check-placement lens (L3) tests the consequence of check ownership that the published entries leave unnamed, most often in families without a computation step ($\S$\ref{sec:taxonomy-patterns}). The robustness-scope lens (L4) bounds the scope of the other verdicts. The measurements show consequences and do not estimate their prevalence.

Labels identify the measured model versions. Service-contract and RAN-policy tasks use \jev, \deepseek, \glm{} and \qwenflash. Their task-specific comparators are \semif, \qwenjson, Laya and \anyjev. Application-control tasks use \sokjev, \deepseek, \gemini, \sokglm{} and local \sokqwen. Rules, solvers and retrieval baselines remain task-specific.

Bold tinted cells mark the best observed value in the direction specified by each caption, including ties. Grey brackets show intervals. N/A marks an unavailable or inapplicable measurement. Appendix~\ref{app:native-cases} reports native condition matrices. Appendix~\ref{app:supporting-lenses} gives supporting results by lens.

\subsection{L1 Measurement Boundary}\label{sec:lens-boundary}\label{sec:rq-f}

Where a decision is timed changes the benefit of a faster engine. The audit finds explicit loop claims in 50/139 families, and 4/139 support the claim with measurement matched to its stated budget and boundary ($\S$\ref{sec:evidence-audit}). A claim checked at the decision return can differ from the same claim checked at the acknowledgement or verified outcome.

With the action held fixed, verified outcome time follows decision delay approximately one for one, and both slopes lie within the practical band $[0.9,1.1]$. The mean scene slope of $T_b$ against $D$ is 1.00025 in transport, with a 95\% scene interval of 0.99970--1.00080. The edge slope is 0.99096, with an interval of 0.95906--1.02186. Removing the two-second delay saves 2.00070\,s in transport (1.99954--2.00181) and 2.00924\,s in edge (1.92801--2.08679).

The share that the decision occupies depends on the endpoint. A 0.5\,s decision takes 91.9\% of the transport acknowledgement time $T_a$ and 47.2\% of the verified outcome time $T_b$. In edge, the same delay takes 98.9\% of $T_a$ and 22.8\% of $T_b$ (Table~\ref{tab:e2-results}). The average per-scene relative saving from the two-second arm to zero delay is 78.1\% in transport and 55.8\% in edge.

\begin{figure*}[tp]\centering
\includegraphics[width=7in]{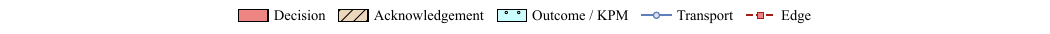}\\[-2pt]
\subfloat[RAN observed KPM]{\includegraphics[width=1.70in]{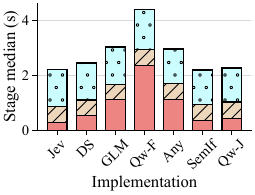}}\hfill
\subfloat[Transport physical path]{\includegraphics[width=1.70in]{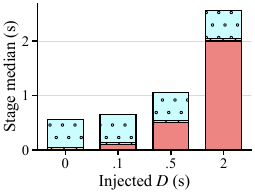}}\hfill
\subfloat[Edge physical path]{\includegraphics[width=1.70in]{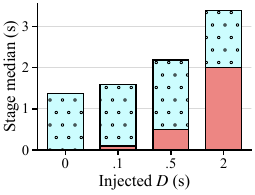}}\hfill
\subfloat[Sparse verified outcome]{\includegraphics[width=1.70in]{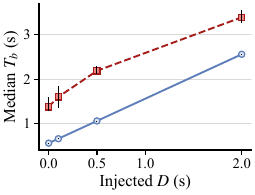}}
\caption{Stage medians and sparse verified-outcome medians. Panel (a) retains 30/30/28/29/27/29/30 endpoint pairs from 30 RAN arrivals per interpreter~\cite{li2026intent}. Transport and edge use 30 scenes with ten repeats per delay. Stacks sum stage medians. Outcome intervals resample scenes. KPM is an observation bound.}
\label{fig:e2-paths}\end{figure*}

The RAN study~\cite{li2026intent} shows the same separation at the radio boundary, and the RAN values in this paragraph come from it. Median interpretation takes 0.286--2.352\,s, against 15.5--19.1\,ms for A1 and 2.4--3.1\,ms for E2, so the decision dominates the acknowledged control path. At the radio outcome, ideal enforcement moves the affected-class SLA violation by 3.96 percentage points, and no hosted model shows a resolved violation increase over Jev at the base point. In that base-point experiment, interpreters that miss the near-RT budget at the acknowledgement show no resolved penalty at the service outcome. Figure~\ref{fig:e2-paths} separates decision, acknowledgement and observed or verified outcomes. Its RAN panel contains all 210 arrivals of the RAN study, and its stacks use the 203 records with both acknowledgement and KPM observations. Six signed KPM-minus-ACK gaps are negative.

\noindent\textit{Verdict.} Repeated effect, multi-task evidence and SoK-owned results with borrowed-and-reanalysed context.

\subsection{L2 Workload Summary}\label{sec:lens-workload}\label{sec:rq-d}

A latency summary taken without load can admit an engine that the loaded loop rejects. Only 9/139 families report p95 or higher, and 4/139 report deadline attainment ($\S$\ref{sec:evidence-audit}). Because the queue field has low coding agreement, the audit gives no reliable queue prevalence.

The load arm, which queues real decisions ahead of measured execution residuals, shows the verdict change directly. A 0.5\,s decision meets the 10\,s transport budget for all 300 isolated requests. At 4.992 arrivals/s, where the 100\,ms baseline fills half of the decision slot, the same delay raises $\rho$ to 2.500. Attainment then falls to 0.0\% and the median verified outcome reaches 233.749\,s (Table~\ref{tab:e2-results}). In edge at $\rho_0=0.95$, a 100\,ms decision cuts 2\,s attainment from 76.0\% in isolation to 18.6\% (9.3--29.0).

Table~\ref{tab:e1-timing} contrasts medians with tails and deadlines. It keeps task-specific endpoints and marks each task family's ownership. At 16 service arrivals/s, Jev has $\rho=1.084$, queue p95 1.492\,s and native completion 90.7\%~\cite{li2026replacing}.

\begin{table*}[tp]
\centering\scriptsize
\caption{Format-valid return by budget with task-family ownership. In the access-policy base condition, model rows summarize the same 81 base cases from the second collection batch, spread over all four request families, while rule rows summarize all 120 base cases. Return quantiles are in seconds. The p99 is descriptive. Probability intervals use 30\,s blocks. An asterisk marks fewer than eight occupied blocks. High-load $\rho$ uses 16 service arrivals/s and 2 RAN intents/s. Edge and RAN values are borrowed and reanalysed~\cite{li2026replacing,li2026intent}.}
\label{tab:e1-timing}
\setlength{\tabcolsep}{2.5pt}
\renewcommand{\arraystretch}{0.94}
\setlength{\aboverulesep}{0pt}\setlength{\belowrulesep}{0pt}\setlength{\extrarowheight}{0pt}
\def\familytab{\begin{tabular}{llrrrrr}
\toprule
\rowcolor{ebBlue}
Model & Ownership & $n$ & \shortstack{Return (s)\\p50 / p95 / p99} & \shortstack{$\leq100$ ms\\(\%) [CI]} & \shortstack{$\leq1$ s\\(\%) [CI]} & \shortstack{High-load\\$\rho$} \\
\midrule
\rowcolor{ebAmber!45}
\multicolumn{7}{l}{\textbf{\itshape Access policy, SoK-owned base condition}} \\
\rowcolor{ebCoral!10}\sokjev & SoK-owned & 81 & 0.350 / 0.483 / 0.620 & 0.0 \textcolor{black!60}{[0.0,0.0]} & \cellcolor{ebCoral!35}\textbf{100.0} \textcolor{black!60}{[100.0,100.0]} & N/A \\
\deepseek & SoK-owned & 81 & 0.469 / 1.623 / 6.123 & 0.0 \textcolor{black!60}{[0.0,0.0]} & 88.9 \textcolor{black!60}{[81.2,95.4]} & N/A \\
\gemini & SoK-owned & 81 & 0.923 / 1.144 / 1.371 & 0.0 \textcolor{black!60}{[0.0,0.0]} & 74.1 \textcolor{black!60}{[64.9,82.9]} & N/A \\
\sokglm & SoK-owned & 81 & 0.812 / 1.504 / 7.866 & 0.0 \textcolor{black!60}{[0.0,0.0]} & 72.8 \textcolor{black!60}{[62.3,82.7]} & N/A \\
\sokqwen & SoK-owned & 81 & 0.329 / 0.388 / 0.435 & 0.0 \textcolor{black!60}{[0.0,0.0]} & \cellcolor{ebCoral!35}\textbf{100.0} \textcolor{black!60}{[100.0,100.0]} & N/A \\
Full rule & SoK-owned & 120 & 0.000 / 0.000 / 0.000 & \cellcolor{ebCoral!35}\textbf{100.0} \textcolor{black!60}{[N/A]} & \cellcolor{ebCoral!35}\textbf{100.0} \textcolor{black!60}{[N/A]} & N/A \\
Cheapest & SoK-owned & 120 & 0.000 / 0.000 / 0.000 & \cellcolor{ebCoral!35}\textbf{100.0} \textcolor{black!60}{[N/A]} & \cellcolor{ebCoral!35}\textbf{100.0} \textcolor{black!60}{[N/A]} & N/A \\
First & SoK-owned & 120 & \cellcolor{ebCoral!35}\textbf{0.000} / \cellcolor{ebCoral!35}\textbf{0.000} / \cellcolor{ebCoral!35}\textbf{0.000} & \cellcolor{ebCoral!35}\textbf{100.0} \textcolor{black!60}{[N/A]} & \cellcolor{ebCoral!35}\textbf{100.0} \textcolor{black!60}{[N/A]} & N/A \\
\midrule
\rowcolor{ebAmber!45}
\multicolumn{7}{l}{\textbf{\itshape Service contract, reanalysed from the Edge study~\cite{li2026replacing}}} \\
\rowcolor{ebCoral!10}\jev & From A & 300 & 0.274 / 0.381 / 0.536 & 0.0 \textcolor{black!60}{[0.0,0.0]} & \cellcolor{ebCoral!35}\textbf{100.0} \textcolor{black!60}{[100.0,100.0]} & 1.084 \\
\semif & From A & 300 & 0.124 / 0.296 / 0.323 & 0.3 \textcolor{black!60}{[0.0,0.5]}* & \cellcolor{ebCoral!35}\textbf{100.0} \textcolor{black!60}{[100.0,100.0]}* & 10.170 \\
\laya & From A & 300 & 0.055 / 0.222 / 0.242 & 83.3 \textcolor{black!60}{[N/A]} & \cellcolor{ebCoral!35}\textbf{100.0} \textcolor{black!60}{[N/A]} & 1.037 \\
\deepseek & From A & 300 & 0.378 / 1.054 / 4.933 & 0.0 \textcolor{black!60}{[0.0,0.0]} & 94.7 \textcolor{black!60}{[88.8,98.7]} & 1.842 \\
\glm & From A & 300 & 0.646 / 2.685 / 3.363 & 0.0 \textcolor{black!60}{[0.0,0.0]} & 81.0 \textcolor{black!60}{[71.3,89.4]} & 2.972 \\
\qwenflash & From A & 300 & 1.774 / 3.668 / 5.774 & 0.0 \textcolor{black!60}{[0.0,0.0]} & 0.0 \textcolor{black!60}{[0.0,0.0]} & 5.876 \\
\qwenjson & From A & 300 & 0.774 / 0.868 / 1.036 & 0.0 \textcolor{black!60}{[0.0,0.0]} & 98.7 \textcolor{black!60}{[97.2,99.7]} & 15.823 \\
Rule & From A & 300 & 0.000 / 0.000 / 0.000 & \cellcolor{ebCoral!35}\textbf{100.0} \textcolor{black!60}{[N/A]} & \cellcolor{ebCoral!35}\textbf{100.0} \textcolor{black!60}{[N/A]} & N/A \\
Oracle & From A & 300 & \cellcolor{ebCoral!35}\textbf{0.000} / \cellcolor{ebCoral!35}\textbf{0.000} / \cellcolor{ebCoral!35}\textbf{0.000} & \cellcolor{ebCoral!35}\textbf{100.0} \textcolor{black!60}{[N/A]} & \cellcolor{ebCoral!35}\textbf{100.0} \textcolor{black!60}{[N/A]} & 0.002 \\
\midrule
\rowcolor{ebAmber!45}
\multicolumn{7}{l}{\textbf{\itshape RAN policy at 57 fresh cells, reanalysed from the RAN study~\cite{li2026intent}}} \\
\rowcolor{ebCoral!10}\jev & From B & 300 & \cellcolor{ebCoral!35}\textbf{0.275} / \cellcolor{ebCoral!35}\textbf{0.376} / 0.429 & \cellcolor{ebCoral!35}\textbf{0.0} \textcolor{black!60}{[0.0,0.0]}* & 99.7 \textcolor{black!60}{[99.0,100.0]}* & 0.149 \\
\semif & From B & 300 & 0.431 / 0.524 / 0.556 & \cellcolor{ebCoral!35}\textbf{0.0} \textcolor{black!60}{[0.0,0.0]}* & \cellcolor{ebCoral!35}\textbf{100.0} \textcolor{black!60}{[100.0,100.0]}* & 0.261 \\
\anyjev & From B & 300 & 0.403 / 0.422 / \cellcolor{ebCoral!35}\textbf{0.428} & \cellcolor{ebCoral!35}\textbf{0.0} \textcolor{black!60}{[0.0,0.0]}* & \cellcolor{ebCoral!35}\textbf{100.0} \textcolor{black!60}{[100.0,100.0]}* & 1.169 \\
\deepseek & From B & 300 & 0.520 / 0.812 / 0.984 & \cellcolor{ebCoral!35}\textbf{0.0} \textcolor{black!60}{[0.0,0.0]}* & 99.0 \textcolor{black!60}{[97.7,100.0]}* & 0.403 \\
\glm & From B & 300 & 1.128 / 2.111 / 4.195 & \cellcolor{ebCoral!35}\textbf{0.0} \textcolor{black!60}{[0.0,0.0]}* & 24.0 \textcolor{black!60}{[18.5,28.8]}* & 0.350 \\
\qwenflash & From B & 300 & 2.220 / 2.945 / 4.124 & \cellcolor{ebCoral!35}\textbf{0.0} \textcolor{black!60}{[0.0,0.0]} & 0.0 \textcolor{black!60}{[0.0,0.0]} & 1.085 \\
\qwenjson & From B & 300 & 0.429 / 0.439 / 0.443 & \cellcolor{ebCoral!35}\textbf{0.0} \textcolor{black!60}{[0.0,0.0]}* & \cellcolor{ebCoral!35}\textbf{100.0} \textcolor{black!60}{[100.0,100.0]}* & 0.217 \\
\bottomrule
\end{tabular}}
\setlength{\tabcolsep}{0pt}\sbox0{\familytab}
\dimen0=\linewidth\advance\dimen0 by -1pt\advance\dimen0 by -\wd0\divide\dimen0 by 14
\ifdim\dimen0<1.5pt \dimen0=1.5pt\fi
\setlength{\tabcolsep}{\dimen0}\familytab
\end{table*}

At 2 RAN intents/s, Jev has $\rho=0.149$ and queue p95 2.13\,ms. Qwen 3.8 reaches $\rho=1.085$ and queue p95 20.30\,s. AnyJev reaches $\rho=1.169$ and queue p95 34.53\,s. DeepSeek has $\rho=0.403$ and queue p95 6.37\,s~\cite{li2026intent}.

At $\rho_0=0.95$, the transport and edge replays share one decision queue, and its p95 is 2.568\,s for the 100\,ms arm. The 500\,ms arm reaches $\rho=4.749$ and queue p95 970.063\,s. The 2\,s arm reaches $\rho=18.977$ and queue p95 4,623.773\,s.

\begin{figure}[!t]\centering
\includegraphics[width=3.36in]{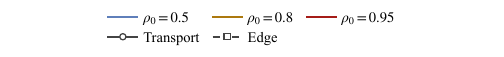}\\[-2pt]
\subfloat[Outcome tail]{\includegraphics[width=1.68in]{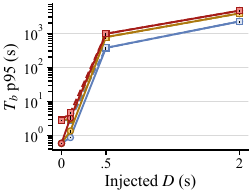}}\hfill
\subfloat[Budget attainment]{\includegraphics[width=1.68in]{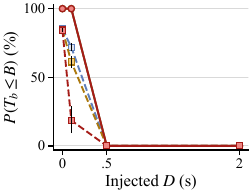}}
\caption{Finite-window queue amplification at fixed arrival rates. Colour identifies baseline load $\rho_0$. Line and marker identify domain. Outcome tails use measured execution residuals after real decision queues. Budget attainment uses all arrivals with $B=10$\,s for transport and $2$\,s for edge.}
\label{fig:e2-load}\end{figure}

These overloaded queues accumulate work during the arrival window and drain after arrivals stop. Their intervals describe this finite nonstationary window. Figure~\ref{fig:e2-load} shows how the same arrivals yield different outcome tails and deadline attainment when slot occupancy changes.

\noindent\textit{Verdict.} Task-dependent effect, multi-task evidence and SoK-owned plus borrowed-and-reanalysed data.

\subsection{L3 Check Placement}\label{sec:lens-checks}\label{sec:rq-b}\label{sec:rq-c}\label{sec:rq-e}

An engine that owns a check can pass it under one condition and fail it under another. The check columns of the systematization support no prevalence estimate for observation reconciliation, coverage escalation or joint feasibility in the controller ($\S$\ref{sec:systematization}).

Engine-owned observation reconciliation depends on the conflict type. Jev gives the correct placement action in 50/120 current-conflict cases and 11/120 resolved-conflict cases. The latter condition requires precedence between newer evidence and an older contradiction. Figure~\ref{fig:rq-b} and Table~\ref{tab:rq-b-native} in Appendix~\ref{app:supporting-lenses} give the details.

\begin{figure*}[tp]\centering
\includegraphics[width=7in]{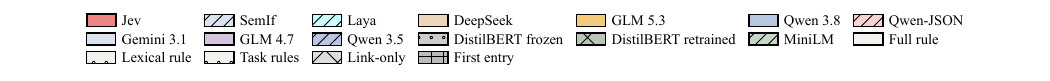}\\[-2pt]
\subfloat[Service churn 50\%]{\includegraphics[width=1.70in]{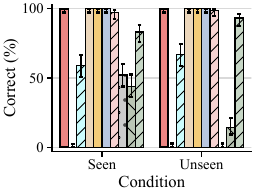}}\hfill
\subfloat[Contract catalogue]{\includegraphics[width=1.70in]{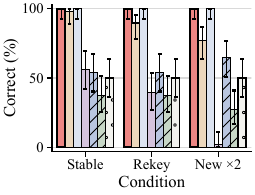}}\hfill
\subfloat[Route coverage]{\includegraphics[width=1.70in]{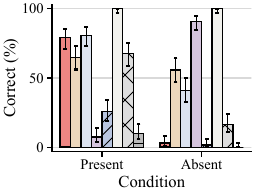}}\hfill
\subfloat[Escalation]{\includegraphics[width=1.70in]{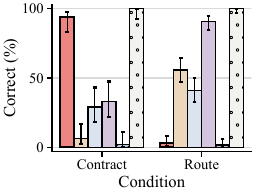}}
\caption{Covered selection and absent-coverage escalation. Panel (a) reports seen and unseen service top-1 with $n=136/134$ from the Edge study~\cite{li2026replacing}. Contract candidates use $n=48$. Paths use $n=120$. Bars retain native predicates with 95\% Wilson intervals.}
\label{fig:rq-c}\end{figure*}

Coverage checks are also task-dependent. Jev requests new contract candidates in 45/48 absent-candidate cases. It escalates only 4/120 infeasible forwarding instances. Figure~\ref{fig:rq-c} compares the two coverage checks.

The public-gate condition is a bundled workflow redesign. Instructions, refresh actions and check ownership change together. Jev's independent-quota correctness rises from 29/120 to 84/120. The paired gain is 45.8\,pp with a 95\% interval of 37.5--54.2.

\begin{figure*}[tp]\centering
\includegraphics[width=7in]{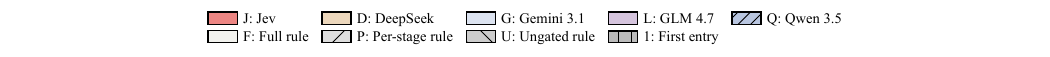}\\[-2pt]
\subfloat[Independent quota]{\includegraphics[width=1.70in]{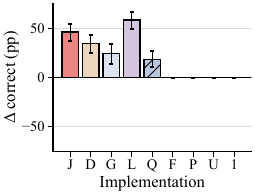}}\hfill
\subfloat[Shared quota]{\includegraphics[width=1.70in]{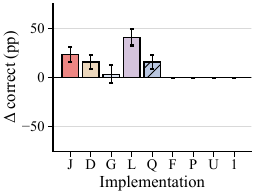}}\hfill
\subfloat[Quota and separation]{\includegraphics[width=1.70in]{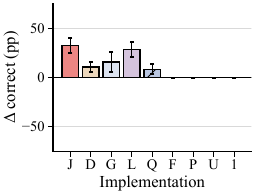}}\hfill
\subfloat[Shared minus independent]{\includegraphics[width=1.70in]{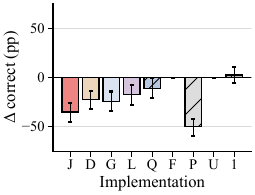}}
\caption{Public-gate minus self-check correctness in panels (a)--(c). Panel (d) shows shared minus independent quota under public gating. Intervals resample 120 paired cases within worker-count and demand strata. The intervention jointly changes instructions, refresh actions and check ownership.}
\label{fig:rq-e}\end{figure*}

With shared quotas, the gain is 23.3\,pp with an interval of 15.8--30.8. Shared quotas plus fault-domain separation give 32.5\,pp with an interval of 25.0--40.0. The per-stage rule fails when shared quotas require joint accounting. Figure~\ref{fig:rq-e} reports the bundled contrasts.

\noindent\textit{Verdict.} Task-dependent effect, multi-task evidence and SoK-owned results with borrowed-and-reanalysed context.

\subsection{L4 Robustness Scope}\label{sec:lens-robustness}\label{sec:rq-a}

Robustness scope bounds where the verdicts of the other three lenses hold. The audit did not code robustness to structure, length, representation or catalogue change and therefore provides no prevalence estimate for this lens.

Each task grows along its own structural axis. Jev's low-density contract exact matches are 277/300, 236/300 and 173/300 for 4, 6 and 8 fields~\cite{li2026replacing}. Its access-policy counts are 107/120, 97/120 and 47/120 for 4, 16 and 64 requirements.

In records from the RAN study~\cite{li2026intent}, Jev selects the correct RAN target in 150 and 145 of 150 cases with 3 and 7 fresh cells, and in 135 and 136 of 150 with 21 and 57 cells. Our reanalysis gives a contrast between 57 and 3 cells of $-9.33$\,pp, with a paired 95\% interval of $[-14.00,-5.32]$ and structure-family Holm-adjusted $p=0.00757$.

\begin{figure*}[tp]\centering
\includegraphics[width=7in]{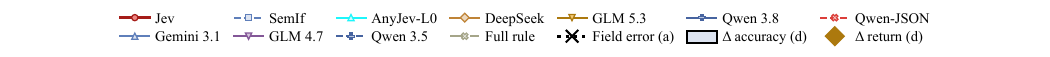}\\[-2pt]
\subfloat[Contract fields]{\includegraphics[width=1.70in]{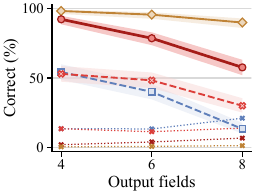}}\hfill
\subfloat[Telemetry cells]{\includegraphics[width=1.70in]{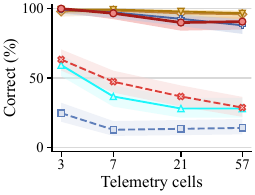}}\hfill
\subfloat[Policy requirements]{\includegraphics[width=1.70in]{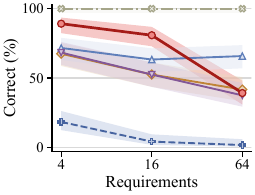}}\hfill
\subfloat[Padding effects]{\includegraphics[width=1.70in]{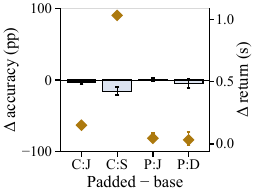}}
\caption{Structural growth and paired padding effects. Panels (a) and the contract component of (d) are borrowed and reanalysed from the Edge study~\cite{li2026replacing}. Panel (b) is borrowed and reanalysed from the RAN study~\cite{li2026intent}. Panel (c) and the policy component of (d) are SoK-owned. Bands in panels (a)--(c) show 95\% Wilson intervals. Panel (d) uses paired stratified-bootstrap intervals.}
\label{fig:rq-a}\end{figure*}

Padding and repetition keep the requirement set but change its presentation. Figure~\ref{fig:rq-a} shows the four robustness scopes. Its contract and padding components come from the Edge study~\cite{li2026replacing} and its telemetry component from the RAN study~\cite{li2026intent}, while the SoK owns the access-policy arm.

\noindent\textit{Verdict.} Task-dependent effect, multi-task evidence and SoK-owned plus borrowed-and-reanalysed data.

\section{Design Rules}\label{sec:design-rules}
Each rule starts from a gap in the systematization or the evidence audit, adds the consequence measured under its lens, and names a controller action to close it. The boundary and workload rules rest on audit prevalence. The check-placement rules start from the ownership the published entries name and rest on measured consequences, because the check columns support no prevalence estimate ($\S$\ref{sec:systematization}). The robustness rule starts from a field the audit does not code.

\smallskip\noindent\textbf{Budget the Decision at the Required Outcome.} Fifty of the 139 families claim that their engine fits a control loop or time budget, and four support the claim with measurement at the stated boundary. The endpoint determines the share of time the decision occupies. A 0.5\,s decision takes 91.9\% of the transport acknowledgement time and 47.2\% of the verified outcome time ($\S$\ref{sec:lens-boundary}). In the RAN study, interpretation dominates the acknowledged control path, yet the base-point experiment shows no resolved penalty at the service outcome~\cite{li2026intent}. State which endpoint the loop budget governs. Measure $D/T_a$, $D/T_b$ and the offered load before replacing an engine with a faster one, since a sparse verified outcome shortens by about the decision waiting removed.

\smallskip\noindent\textbf{Admit an Engine by Attainment at Offered Load.} Nine families report latency at p95 or higher, and four report deadline attainment. An isolated-request summary can pass an engine that the loaded loop rejects. A 0.5\,s transport decision meets the 10\,s budget for all 300 isolated requests and for none at 4.992 arrivals/s when decisions queue ahead of measured execution times ($\S$\ref{sec:lens-workload}). In edge at $\rho_0=0.95$, a 100\,ms decision lowers attainment of the 2\,s budget from 76.0\% to 18.6\%. Specify the budget at the required endpoint, a decision concurrency limit and an overload policy. Report attainment at the offered load with the occupied decision slot and observation window.

\smallskip\noindent\textbf{Reconcile Observation Versions Before the Decision.} The published entries name an observation owner for 11 of the 139 families ($\S$\ref{sec:taxonomy-patterns}). Missing, stale, contradictory and superseded observations require different actions. Jev gives the correct placement action in 50/120 current-conflict cases and in 11/120 resolved-conflict cases, where newer evidence supersedes an older contradiction ($\S$\ref{sec:lens-checks}). Keep version, precedence and unresolved-conflict state in the controller. Pass the decision a reconciled or explicitly unresolved state.

\smallskip\noindent\textbf{Make Absent Coverage a Controller State.} The published entries name a coverage owner for 13 families, and for 2 of the 72 without a computation step. Jev requests a new candidate for 45 of 48 contracts without one and escalates 4 of 120 infeasible forwarding instances ($\S$\ref{sec:lens-checks}). The same engine passes one coverage check and fails the other. Implement an explicit no-feasible-candidate outcome in the catalogue service or the path-computation workflow. Test covered selection and absent-coverage escalation separately, including how the controller obtains or certifies alternatives.

\smallskip\noindent\textbf{Compute Joint Feasibility in the Controller.} The published entries name a feasibility owner for 45 of the 67 families with a computation step and for 23 of the 72 without one. The public-gate redesign changes instructions, refresh actions and check ownership together. It raises Jev's independent-quota correctness from 29/120 to 84/120, a paired gain of 45.8\,pp (37.5--54.2). The per-stage rule fails when shared quotas require joint accounting. Where the access-policy, placement and forwarding tasks define complete feasibility and cost predicates, the full rule satisfies every measured condition with sub-millisecond computation (Tables~\ref{tab:retained-policy-scale},~\ref{tab:retained-quota} and~\ref{tab:retained-route-catalogue}). Assign formalized constraints to a scheduler or solver after the engine interprets the intent. For open-vocabulary or underspecified objectives, construct the contract that defines this computation before the decision.

\smallskip\noindent\textbf{Report the Robustness Scope With Each Verdict.} The evidence audit does not code the structural scope a verdict covers and gives no count for it. Each task grows along its own structural axis. Jev's contract exact matches fall from 277/300 to 173/300 between four and eight fields~\cite{li2026replacing}. Its access-policy counts fall from 107/120 to 47/120 between 4 and 64 requirements. Its RAN target selection loses 9.33\,pp between 3 and 57 cells~\cite{li2026intent}. Padding and repetition keep the requirement set but change its presentation ($\S$\ref{sec:lens-robustness}). Prune or preaggregate inputs that cannot affect the decision. Evaluate output width, observation retrieval and constraint interaction separately, and state each verdict's tested scope.

\section{Limitations}\label{sec:limitations}
The evidence audit measures what the literature reports. A family coded without matched timing evidence may still operate within its claimed loop. The audit therefore establishes missing support for admission claims. The measured consequences in $\S$\ref{sec:cross-study} show why that missing support matters. They come from the Edge and RAN source studies and the SoK-owned tasks. The surveyed network systems of other authors remain untested.

Each experiment uses one random stream. Its intervals come from resampling within the run, by input, scene or time block, and do not describe variation across independent replications. Because the synthesis questions and comparison families were formulated after the measurement records were inspected, their tests are exploratory. Holm adjustment controls error within each declared family only.

The access-policy, placement and forwarding tasks are constructed application-layer tasks with complete predicates. They isolate selector behavior under controlled structure, observation and coverage. Their graph costs do not represent measured network delay. Because the public-gate contrast changes instructions, refresh actions and check ownership together, it estimates the effect of a workflow redesign rather than the effect of moving one check.

Because the load arm samples measured physical residuals after a real FIFO decision queue, its post-queue actions do not run concurrently. Near saturation, its intervals describe the finite observation window rather than a steady state. Transport and edge differ in topology, action, stack and verification probe. Comparisons between the two domains are therefore descriptive. Radio closed-loop results are taken from the published summaries of the RAN study~\cite{li2026intent}.

\section{Open Problems}\label{sec:open-problems}
\smallskip\noindent\textbf{Guard Complementarity.} Parsers, configuration verifiers and execution probes check different properties. Evaluating their complementarity requires measuring the errors each guard detects, its false alarms and the violations that survive their combination. Configuration translation and routing-stack semantics are part of this comparison. The three gates supply its events and denominators.

\smallskip\noindent\textbf{Candidate Construction and Certification.} Typed selection needs a candidate set that contains an acceptable output. How to certify this coverage under changing constraints, and when to synthesize a new candidate instead, remain open. Measuring the full cost of candidate construction and certification is also open. A correct escalation initiates catalogue construction or service recovery, and coverage restoration and verified recovery require separate evaluation.

\smallskip\noindent\textbf{Dependency-Driven Refresh and Cross-Domain Conflict.} Reusable interpretation needs explicit dependencies on catalogue, contract and policy state. A cached interpretation can skip a decision call while the request, contract version and relevant catalogue semantics stay unchanged, but dependent network state requires recomputed feasibility. Refreshing everything on each telemetry change repeats unaffected computations. Refreshing only the model can retain invalid downstream assumptions. Dependency tracking must identify what a state change invalidates. For cross-domain intents, arbitration must resolve competing resources, priorities and authority and verify the shared outcome.

\smallskip\noindent\textbf{Replication Across Deployments.} The next step tests how decision-slot occupancy and execution overlap interact across sites and implementations. Experiments should vary arrival processes, provider concurrency and concurrent physical actions while matching task responsibilities and recording the complete event path. Independent traffic realizations would separate workload variation from the physical-repeat variation measured in $\S$\ref{sec:lens-boundary}. These comparisons can determine how deployment changes the relation between faster decisions, queue tails and verified outcomes.

\section{Conclusion}\label{sec:conclusion}
We systematized 139 paper families on semantic decision engines in network control and coded each by decision interface, execution path and check ownership. Of these, 50 claimed that their engine fit a control loop or time budget, and 4 supported the claim at the stated boundary. Families with a deterministic computation step made 28 of the claims and all four supported ones. The 72 families without such a step made 22, none supported, and only 2 of them named a coverage owner. Bounded tests showed that the endpoint set the decision's share of elapsed time and that load turned a decision that met every isolated deadline into one that met none. The same engine passed some checks and failed others. We concluded that a loop claim needed to state its endpoint, load and check ownership, and derived a reporting record, design rules and a research agenda for doing so.

\FloatBarrier
\bibliographystyle{IEEEtran}
\bibliography{references}
\clearpage\onecolumn\appendices
\raggedbottom
\makeatletter
\setlength{\@fptop}{0pt}
\setlength{\@fpsep}{16pt}
\setlength{\@fpbot}{0pt plus 1fil}
\makeatother
\section{Complete Native Conditions}\label{app:native-cases}
The following matrices report each task's native conditions, baselines and denominators. They supplement the lens-based analysis in $\S$\ref{sec:cross-study}. Bold tinted cells mark the largest displayed correct fraction in each condition and column, including ties.
\begingroup\begin{table}[!htbp]\centering\scriptsize
\caption{Policy input structure. Changed/Preserved each contain 60 cases, distinguished by whether all 64 requirements change the minimum-cost feasible choice. Conditions retain a feasible candidate and pair the same cases and candidates. Long irrelevant context and repetition retain four distinct requirements. Tokens gives observed ranges at matched character budgets. Best correctness per condition is in bold.}
\label{tab:retained-policy-scale}
\setlength{\tabcolsep}{3pt}
\setlength{\aboverulesep}{0pt}\setlength{\belowrulesep}{0pt}
\renewcommand{\arraystretch}{0.94}
\def\retainedtab{\begin{tabular}{lrrrrrrr}
\toprule
\rowcolor{ebBlue} \multicolumn{1}{c}{} & \multicolumn{3}{c}{Correct decisions $\uparrow$} & \multicolumn{1}{c}{Valid $\uparrow$} & \multicolumn{2}{c}{Response (s) $\downarrow$} & \multicolumn{1}{c}{Input} \\
\rowcolor{ebBlue} Method & Changed & Preserved & All & All & Median & p95 & Tokens \\
\midrule
\rowcolor{ebAmber!45}\multicolumn{8}{l}{\textbf{\textit{Short input: 4 requirements, 6k characters}}} \\
\rowcolor{ebCoral!10}\sokjev & 52/60 & 55/60 & 107/120 & 120/120 & 0.382 & 0.646 & 1699--2640 \\
\deepseek & 38/60 & 43/60 & 81/120 & 119/120 & 0.558 & 1.441 & 1262--1630 \\
\gemini & 39/60 & 47/60 & 86/120 & 120/120 & 0.987 & 1.252 & 1396--2321 \\
\sokglm & 43/60 & 39/60 & 82/120 & 120/120 & 0.883 & 4.645 & 1174--1559 \\
\sokqwen & 13/60 & 9/60 & 22/120 & 120/120 & 0.328 & 0.389 & 1322--2261 \\
Full rule & \cellcolor{ebCoral!35}\textbf{60/60} & \cellcolor{ebCoral!35}\textbf{60/60} & \cellcolor{ebCoral!35}\textbf{120/120} & 120/120 & $<.001$ & $<.001$ & N/A \\
Cheapest & 48/60 & 45/60 & 93/120 & 120/120 & $<.001$ & $<.001$ & N/A \\
First entry & 20/60 & 17/60 & 37/120 & 120/120 & $<.001$ & $<.001$ & N/A \\
\rowcolor{ebAmber!45}\multicolumn{8}{l}{\textbf{\textit{Long irrelevant context: 4 requirements, 24k characters}}} \\
\rowcolor{ebCoral!10}\sokjev & 52/60 & 56/60 & 108/120 & 120/120 & 0.468 & 0.988 & 4469--5410 \\
\deepseek & 38/60 & 37/60 & 75/120 & 119/120 & 0.608 & 1.046 & 4264--4633 \\
\gemini & 38/60 & 50/60 & 88/120 & 120/120 & 1.167 & 1.470 & 4397--5323 \\
\sokglm & 41/60 & 37/60 & 78/120 & 120/120 & 1.034 & 2.964 & 3944--4331 \\
\sokqwen & 9/60 & 10/60 & 19/120 & 120/120 & 0.568 & 0.648 & 4092--5032 \\
Full rule & \cellcolor{ebCoral!35}\textbf{60/60} & \cellcolor{ebCoral!35}\textbf{60/60} & \cellcolor{ebCoral!35}\textbf{120/120} & 120/120 & $<.001$ & $<.001$ & N/A \\
Cheapest & 48/60 & 45/60 & 93/120 & 120/120 & $<.001$ & $<.001$ & N/A \\
First entry & 20/60 & 17/60 & 37/120 & 120/120 & $<.001$ & $<.001$ & N/A \\
\rowcolor{ebAmber!45}\multicolumn{8}{l}{\textbf{\textit{Repeated requirements: 4 requirements, 24k characters}}} \\
\rowcolor{ebCoral!10}\sokjev & 55/60 & 56/60 & 111/120 & 120/120 & 0.431 & 0.687 & 9186--9816 \\
\deepseek & 38/60 & 39/60 & 77/120 & 120/120 & 0.630 & 1.446 & 8628--9440 \\
\gemini & 40/60 & 48/60 & 88/120 & 120/120 & 1.161 & 1.462 & 8815--9444 \\
\sokglm & 43/60 & 33/60 & 76/120 & 120/120 & 1.171 & 2.055 & 6928--7161 \\
\sokqwen & 16/60 & 16/60 & 32/120 & 120/120 & 0.984 & 1.048 & 8801--9436 \\
Full rule & \cellcolor{ebCoral!35}\textbf{60/60} & \cellcolor{ebCoral!35}\textbf{60/60} & \cellcolor{ebCoral!35}\textbf{120/120} & 120/120 & $<.001$ & $<.001$ & N/A \\
Cheapest & 48/60 & 45/60 & 93/120 & 120/120 & $<.001$ & $<.001$ & N/A \\
First entry & 20/60 & 17/60 & 37/120 & 120/120 & $<.001$ & $<.001$ & N/A \\
\rowcolor{ebAmber!45}\multicolumn{8}{l}{\textbf{\textit{16 distinct requirements: 16 requirements, 24k characters}}} \\
\rowcolor{ebCoral!10}\sokjev & 43/60 & 54/60 & 97/120 & 120/120 & 0.415 & 0.673 & 4553--5486 \\
\deepseek & 24/60 & 39/60 & 63/120 & 120/120 & 0.602 & 1.205 & 4329--4685 \\
\gemini & 29/60 & 47/60 & 76/120 & 120/120 & 1.125 & 1.393 & 4476--5393 \\
\sokglm & 26/60 & 37/60 & 63/120 & 120/120 & 1.013 & 1.929 & 3992--4372 \\
\sokqwen & 2/60 & 3/60 & 5/120 & 120/120 & 0.576 & 0.644 & 4176--5105 \\
Full rule & \cellcolor{ebCoral!35}\textbf{60/60} & \cellcolor{ebCoral!35}\textbf{60/60} & \cellcolor{ebCoral!35}\textbf{120/120} & 120/120 & $<.001$ & $<.001$ & N/A \\
Cheapest & 28/60 & 45/60 & 73/120 & 120/120 & $<.001$ & $<.001$ & N/A \\
First entry & 18/60 & 17/60 & 35/120 & 120/120 & $<.001$ & $<.001$ & N/A \\
\rowcolor{ebAmber!45}\multicolumn{8}{l}{\textbf{\textit{64 distinct requirements: 64 requirements, 24k characters}}} \\
\rowcolor{ebCoral!10}\sokjev & 11/60 & 36/60 & 47/120 & 120/120 & 0.406 & 0.758 & 4880--5818 \\
\deepseek & 11/60 & 39/60 & 50/120 & 120/120 & 0.634 & 1.601 & 4580--4902 \\
\gemini & 32/60 & 47/60 & 79/120 & 120/120 & 1.140 & 1.393 & 4782--5705 \\
\sokglm & 3/60 & 42/60 & 45/120 & 120/120 & 1.033 & 2.793 & 4173--4558 \\
\sokqwen & 0/60 & 2/60 & 2/120 & 120/120 & 0.601 & 0.688 & 4501--5440 \\
Full rule & \cellcolor{ebCoral!35}\textbf{60/60} & \cellcolor{ebCoral!35}\textbf{60/60} & \cellcolor{ebCoral!35}\textbf{120/120} & 120/120 & $<.001$ & $<.001$ & N/A \\
Cheapest & 0/60 & 45/60 & 45/120 & 120/120 & $<.001$ & $<.001$ & N/A \\
First entry & 13/60 & 17/60 & 30/120 & 120/120 & $<.001$ & $<.001$ & N/A \\
\bottomrule\end{tabular}}
\sbox0{\retainedtab}
\setlength{\tabcolsep}{\dimexpr\tabcolsep+(\linewidth-\wd0-1pt)/16\relax}
\ifdim\tabcolsep<1.5pt\setlength{\tabcolsep}{1.5pt}\fi
\sbox0{\retainedtab}
\typeout{RETAINED policy-scale width=\the\wd0\space limit=\the\linewidth}
\retainedtab\end{table}\endgroup

\begingroup\begin{table}[!htbp]\centering\scriptsize
\caption{Observation conditions. Each node-count group has 40 cases across four demand families, balanced by shared-quota binding. Inputs contain four candidates and 9,000 characters. Correct requires minimum-cost feasible selection or justified refresh. The complete/shared condition is the same measurement reused in Table~\ref{tab:retained-quota}. Response time ends at decision return. Best correctness per condition is in bold.}
\label{tab:retained-observation}
\setlength{\tabcolsep}{3pt}
\setlength{\aboverulesep}{0pt}\setlength{\belowrulesep}{0pt}
\renewcommand{\arraystretch}{0.94}
\def\retainedtab{\begin{tabular}{lrrrrrrr}
\toprule
\rowcolor{ebBlue} \multicolumn{1}{c}{} & \multicolumn{4}{c}{Correct decisions $\uparrow$} & \multicolumn{1}{c}{Valid $\uparrow$} & \multicolumn{2}{c}{Response (s) $\downarrow$} \\
\rowcolor{ebBlue} Method & 3 nodes & 5 nodes & 8 nodes & All & All & Median & p95 \\
\midrule
\rowcolor{ebAmber!45}\multicolumn{8}{l}{\textbf{\textit{Complete / shared quota}}} \\
\rowcolor{ebCoral!10}\sokjev & 10/40 & 2/40 & 2/40 & 14/120 & 120/120 & 0.364 & 0.619 \\
\deepseek & 10/40 & 4/40 & 0/40 & 14/120 & 120/120 & 0.499 & 0.754 \\
\gemini & 17/40 & 10/40 & 9/40 & 36/120 & 120/120 & 1.027 & 1.584 \\
\sokglm & 0/40 & 1/40 & 0/40 & 1/120 & 120/120 & 0.884 & 8.373 \\
\sokqwen & 14/40 & 5/40 & 1/40 & 20/120 & 120/120 & 0.247 & 0.352 \\
Full rule & \cellcolor{ebCoral!35}\textbf{40/40} & \cellcolor{ebCoral!35}\textbf{40/40} & \cellcolor{ebCoral!35}\textbf{40/40} & \cellcolor{ebCoral!35}\textbf{120/120} & 120/120 & $<.001$ & $<.001$ \\
Ungated rule & \cellcolor{ebCoral!35}\textbf{40/40} & \cellcolor{ebCoral!35}\textbf{40/40} & \cellcolor{ebCoral!35}\textbf{40/40} & \cellcolor{ebCoral!35}\textbf{120/120} & 120/120 & $<.001$ & $<.001$ \\
\rowcolor{ebAmber!45}\multicolumn{8}{l}{\textbf{\textit{Equivalent prose}}} \\
\rowcolor{ebCoral!10}\sokjev & 10/40 & 2/40 & 1/40 & 13/120 & 120/120 & 0.355 & 0.585 \\
\deepseek & 8/40 & 3/40 & 0/40 & 11/120 & 120/120 & 0.477 & 0.859 \\
\gemini & 14/40 & 12/40 & 11/40 & 37/120 & 120/120 & 0.986 & 1.236 \\
\sokglm & 0/40 & 0/40 & 1/40 & 1/120 & 120/120 & 0.783 & 6.672 \\
\sokqwen & 15/40 & 8/40 & 3/40 & 26/120 & 120/120 & 0.247 & 0.340 \\
Full rule & \cellcolor{ebCoral!35}\textbf{40/40} & \cellcolor{ebCoral!35}\textbf{40/40} & \cellcolor{ebCoral!35}\textbf{40/40} & \cellcolor{ebCoral!35}\textbf{120/120} & 120/120 & $<.001$ & $<.001$ \\
Ungated rule & \cellcolor{ebCoral!35}\textbf{40/40} & \cellcolor{ebCoral!35}\textbf{40/40} & \cellcolor{ebCoral!35}\textbf{40/40} & \cellcolor{ebCoral!35}\textbf{120/120} & 120/120 & $<.001$ & $<.001$ \\
\rowcolor{ebAmber!45}\multicolumn{8}{l}{\textbf{\textit{Missing required report}}} \\
\rowcolor{ebCoral!10}\sokjev & \cellcolor{ebCoral!35}\textbf{40/40} & \cellcolor{ebCoral!35}\textbf{40/40} & \cellcolor{ebCoral!35}\textbf{40/40} & \cellcolor{ebCoral!35}\textbf{120/120} & 120/120 & 0.353 & 0.583 \\
\deepseek & \cellcolor{ebCoral!35}\textbf{40/40} & \cellcolor{ebCoral!35}\textbf{40/40} & \cellcolor{ebCoral!35}\textbf{40/40} & \cellcolor{ebCoral!35}\textbf{120/120} & 120/120 & 0.477 & 0.738 \\
\gemini & \cellcolor{ebCoral!35}\textbf{40/40} & \cellcolor{ebCoral!35}\textbf{40/40} & \cellcolor{ebCoral!35}\textbf{40/40} & \cellcolor{ebCoral!35}\textbf{120/120} & 120/120 & 0.962 & 1.257 \\
\sokglm & 39/40 & \cellcolor{ebCoral!35}\textbf{40/40} & \cellcolor{ebCoral!35}\textbf{40/40} & 119/120 & 120/120 & 0.776 & 3.059 \\
\sokqwen & 0/40 & 0/40 & 0/40 & 0/120 & 120/120 & 0.250 & 0.331 \\
Full rule & \cellcolor{ebCoral!35}\textbf{40/40} & \cellcolor{ebCoral!35}\textbf{40/40} & \cellcolor{ebCoral!35}\textbf{40/40} & \cellcolor{ebCoral!35}\textbf{120/120} & 120/120 & $<.001$ & $<.001$ \\
Ungated rule & 0/40 & 0/40 & 0/40 & 0/120 & 120/120 & $<.001$ & $<.001$ \\
\rowcolor{ebAmber!45}\multicolumn{8}{l}{\textbf{\textit{Stale required report}}} \\
\rowcolor{ebCoral!10}\sokjev & 39/40 & \cellcolor{ebCoral!35}\textbf{40/40} & \cellcolor{ebCoral!35}\textbf{40/40} & 119/120 & 119/120 & 0.353 & 0.594 \\
\deepseek & \cellcolor{ebCoral!35}\textbf{40/40} & \cellcolor{ebCoral!35}\textbf{40/40} & \cellcolor{ebCoral!35}\textbf{40/40} & \cellcolor{ebCoral!35}\textbf{120/120} & 120/120 & 0.494 & 0.811 \\
\gemini & \cellcolor{ebCoral!35}\textbf{40/40} & \cellcolor{ebCoral!35}\textbf{40/40} & \cellcolor{ebCoral!35}\textbf{40/40} & \cellcolor{ebCoral!35}\textbf{120/120} & 120/120 & 0.963 & 1.249 \\
\sokglm & \cellcolor{ebCoral!35}\textbf{40/40} & \cellcolor{ebCoral!35}\textbf{40/40} & \cellcolor{ebCoral!35}\textbf{40/40} & \cellcolor{ebCoral!35}\textbf{120/120} & 120/120 & 0.855 & 18.322 \\
\sokqwen & 1/40 & 2/40 & 6/40 & 9/120 & 120/120 & 0.250 & 0.329 \\
Full rule & \cellcolor{ebCoral!35}\textbf{40/40} & \cellcolor{ebCoral!35}\textbf{40/40} & \cellcolor{ebCoral!35}\textbf{40/40} & \cellcolor{ebCoral!35}\textbf{120/120} & 120/120 & $<.001$ & $<.001$ \\
Ungated rule & 0/40 & 0/40 & 0/40 & 0/120 & 120/120 & $<.001$ & $<.001$ \\
\rowcolor{ebAmber!45}\multicolumn{8}{l}{\textbf{\textit{Conflicting current reports}}} \\
\rowcolor{ebCoral!10}\sokjev & 10/40 & 20/40 & 20/40 & 50/120 & 120/120 & 0.355 & 0.606 \\
\deepseek & \cellcolor{ebCoral!35}\textbf{40/40} & \cellcolor{ebCoral!35}\textbf{40/40} & \cellcolor{ebCoral!35}\textbf{40/40} & \cellcolor{ebCoral!35}\textbf{120/120} & 120/120 & 0.499 & 0.897 \\
\gemini & 5/40 & 9/40 & 17/40 & 31/120 & 120/120 & 0.986 & 1.397 \\
\sokglm & \cellcolor{ebCoral!35}\textbf{40/40} & \cellcolor{ebCoral!35}\textbf{40/40} & \cellcolor{ebCoral!35}\textbf{40/40} & \cellcolor{ebCoral!35}\textbf{120/120} & 120/120 & 0.812 & 12.777 \\
\sokqwen & 0/40 & 0/40 & 2/40 & 2/120 & 120/120 & 0.248 & 0.330 \\
Full rule & \cellcolor{ebCoral!35}\textbf{40/40} & \cellcolor{ebCoral!35}\textbf{40/40} & \cellcolor{ebCoral!35}\textbf{40/40} & \cellcolor{ebCoral!35}\textbf{120/120} & 120/120 & $<.001$ & $<.001$ \\
Ungated rule & 0/40 & 0/40 & 0/40 & 0/120 & 120/120 & $<.001$ & $<.001$ \\
\rowcolor{ebAmber!45}\multicolumn{8}{l}{\textbf{\textit{Older conflict resolved}}} \\
\rowcolor{ebCoral!10}\sokjev & 9/40 & 1/40 & 1/40 & 11/120 & 120/120 & 0.374 & 0.590 \\
\deepseek & 1/40 & 0/40 & 1/40 & 2/120 & 120/120 & 0.507 & 0.813 \\
\gemini & 14/40 & 10/40 & 7/40 & 31/120 & 120/120 & 1.015 & 1.276 \\
\sokglm & 0/40 & 1/40 & 0/40 & 1/120 & 120/120 & 0.861 & 6.439 \\
\sokqwen & 2/40 & 0/40 & 0/40 & 2/120 & 120/120 & 0.249 & 0.334 \\
Full rule & \cellcolor{ebCoral!35}\textbf{40/40} & \cellcolor{ebCoral!35}\textbf{40/40} & \cellcolor{ebCoral!35}\textbf{40/40} & \cellcolor{ebCoral!35}\textbf{120/120} & 120/120 & $<.001$ & $<.001$ \\
Ungated rule & \cellcolor{ebCoral!35}\textbf{40/40} & \cellcolor{ebCoral!35}\textbf{40/40} & \cellcolor{ebCoral!35}\textbf{40/40} & \cellcolor{ebCoral!35}\textbf{120/120} & 120/120 & $<.001$ & $<.001$ \\
\bottomrule\end{tabular}}
\sbox0{\retainedtab}
\setlength{\tabcolsep}{\dimexpr\tabcolsep+(\linewidth-\wd0-1pt)/16\relax}
\ifdim\tabcolsep<1.5pt\setlength{\tabcolsep}{1.5pt}\fi
\sbox0{\retainedtab}
\typeout{RETAINED observation width=\the\wd0\space limit=\the\linewidth}
\retainedtab\end{table}\endgroup

\begingroup\begin{table}[!htbp]\centering\scriptsize
\caption{Contract interpretation on 48 configurations per condition. Replacement introduces two incoming candidates. Absent requires new candidates. Valid and median response refer only to Stable. Rule is a lexical contract matcher. MiniLM re-ranks descriptions. Best correctness per condition is in bold.}
\label{tab:retained-contract-catalogue}
\setlength{\tabcolsep}{3pt}
\setlength{\aboverulesep}{0pt}\setlength{\belowrulesep}{0pt}
\renewcommand{\arraystretch}{0.94}
\def\retainedtab{\begin{tabular}{lrrrrrr}
\toprule
\rowcolor{ebBlue} \multicolumn{1}{c}{} & \multicolumn{4}{c}{Correct decisions $\uparrow$} & \multicolumn{2}{c}{Stable only} \\
\rowcolor{ebBlue} Method & Stable & Rekeyed & Replacement & Absent & Valid & Median (s) \\
\midrule
\rowcolor{ebAmber!45}\multicolumn{7}{l}{\textbf{\textit{A. Contract interpretation: 48 configurations per condition}}} \\
\rowcolor{ebCoral!10}\sokjev & \cellcolor{ebCoral!35}\textbf{48/48} & \cellcolor{ebCoral!35}\textbf{48/48} & \cellcolor{ebCoral!35}\textbf{48/48} & 45/48 & 48/48 & 0.323 \\
\deepseek & 47/48 & 43/48 & 37/48 & 3/48 & 48/48 & 0.447 \\
\gemini & \cellcolor{ebCoral!35}\textbf{48/48} & \cellcolor{ebCoral!35}\textbf{48/48} & \cellcolor{ebCoral!35}\textbf{48/48} & 14/48 & 48/48 & 0.893 \\
\sokglm & 27/48 & 19/48 & 1/48 & 16/48 & 48/48 & 0.775 \\
\sokqwen & 26/48 & 26/48 & 31/48 & 1/48 & 48/48 & 0.243 \\
Rule & 24/48 & 24/48 & 24/48 & \cellcolor{ebCoral!35}\textbf{48/48} & 48/48 & $<0.001$ \\
MiniLM & 18/48 & 18/48 & 13/48 & 24/48 & 48/48 & 0.041 \\
\bottomrule\end{tabular}}
\sbox0{\retainedtab}
\setlength{\tabcolsep}{\dimexpr\tabcolsep+(\linewidth-\wd0-1pt)/14\relax}
\ifdim\tabcolsep<1.5pt\setlength{\tabcolsep}{1.5pt}\fi
\sbox0{\retainedtab}
\typeout{RETAINED contract-catalogue width=\the\wd0\space limit=\the\linewidth}
\retainedtab\end{table}\endgroup

\begingroup\begin{table}[!htbp]\centering\scriptsize
\caption{Held-out route catalogues, 20 new graphs per contract and 120 per condition, with eight candidates. Link-only chooses the cheapest connected listed path without the other contract checks. Correct requires minimum-cost feasible selection or justified catalogue rebuilding, which loads a preconstructed certified catalogue. All fixed responses are valid. Response time ends at decision return. Best correctness per condition is in bold.}
\label{tab:retained-route-catalogue}
\setlength{\tabcolsep}{3pt}
\setlength{\aboverulesep}{0pt}\setlength{\belowrulesep}{0pt}
\renewcommand{\arraystretch}{0.94}
\def\retainedtab{\begin{tabular}{lrrrrrrrrr}
\toprule
\rowcolor{ebBlue} \multicolumn{1}{c}{} & \multicolumn{7}{c}{Correct decisions $\uparrow$} & \multicolumn{2}{c}{Response (s) $\downarrow$} \\
\rowcolor{ebBlue} Method & Connectivity & Required & Avoid & Hops & Budget & Ordered & Overall & Median & p95 \\
\midrule
\rowcolor{ebAmber!45}\multicolumn{10}{l}{\textbf{\textit{A. Feasible candidate present}}} \\
\rowcolor{ebCoral!10}\sokjev & 17/20 & 15/20 & 16/20 & 19/20 & 18/20 & 10/20 & 95/120 & 0.351 & 0.484 \\
\deepseek & 14/20 & 16/20 & 13/20 & 17/20 & 6/20 & 12/20 & 78/120 & 0.525 & 0.722 \\
\gemini & 15/20 & 16/20 & 19/20 & 16/20 & 18/20 & 13/20 & 97/120 & 0.954 & 1.208 \\
\sokglm & 1/20 & 1/20 & 1/20 & 5/20 & 0/20 & 1/20 & 9/120 & 0.872 & 6.666 \\
\sokqwen & 5/20 & 4/20 & 6/20 & 7/20 & 3/20 & 6/20 & 31/120 & 0.357 & 0.447 \\
Full rule & \cellcolor{ebCoral!35}\textbf{20/20} & \cellcolor{ebCoral!35}\textbf{20/20} & \cellcolor{ebCoral!35}\textbf{20/20} & \cellcolor{ebCoral!35}\textbf{20/20} & \cellcolor{ebCoral!35}\textbf{20/20} & \cellcolor{ebCoral!35}\textbf{20/20} & \cellcolor{ebCoral!35}\textbf{120/120} & $<.001$ & $<.001$ \\
Link-only & \cellcolor{ebCoral!35}\textbf{20/20} & 6/20 & 16/20 & 18/20 & \cellcolor{ebCoral!35}\textbf{20/20} & 1/20 & 81/120 & $<.001$ & $<.001$ \\
First entry & 3/20 & 1/20 & 1/20 & 3/20 & 1/20 & 3/20 & 12/120 & $<.001$ & $<.001$ \\
\rowcolor{ebAmber!45}\multicolumn{10}{l}{\textbf{\textit{B. No feasible candidate listed}}} \\
\rowcolor{ebCoral!10}\sokjev & 0/20 & 1/20 & 0/20 & 0/20 & 1/20 & 2/20 & 4/120 & 0.350 & 0.507 \\
\deepseek & 3/20 & 4/20 & 12/20 & 13/20 & 18/20 & 17/20 & 67/120 & 0.512 & 0.744 \\
\gemini & 1/20 & 3/20 & 16/20 & 7/20 & 9/20 & 13/20 & 49/120 & 0.950 & 1.322 \\
\sokglm & 18/20 & 16/20 & 19/20 & 19/20 & \cellcolor{ebCoral!35}\textbf{20/20} & 17/20 & 109/120 & 0.853 & 5.507 \\
\sokqwen & 0/20 & 0/20 & 2/20 & 0/20 & 0/20 & 0/20 & 2/120 & 0.363 & 0.451 \\
Full rule & \cellcolor{ebCoral!35}\textbf{20/20} & \cellcolor{ebCoral!35}\textbf{20/20} & \cellcolor{ebCoral!35}\textbf{20/20} & \cellcolor{ebCoral!35}\textbf{20/20} & \cellcolor{ebCoral!35}\textbf{20/20} & \cellcolor{ebCoral!35}\textbf{20/20} & \cellcolor{ebCoral!35}\textbf{120/120} & $<.001$ & $<.001$ \\
Link-only & \cellcolor{ebCoral!35}\textbf{20/20} & 0/20 & 0/20 & 0/20 & 0/20 & 0/20 & 20/120 & $<.001$ & $<.001$ \\
First entry & 0/20 & 0/20 & 0/20 & 0/20 & 0/20 & 0/20 & 0/120 & $<.001$ & $<.001$ \\
\bottomrule\end{tabular}}
\sbox0{\retainedtab}
\setlength{\tabcolsep}{\dimexpr\tabcolsep+(\linewidth-\wd0-1pt)/20\relax}
\ifdim\tabcolsep<1.5pt\setlength{\tabcolsep}{1.5pt}\fi
\sbox0{\retainedtab}
\typeout{RETAINED route-catalogue width=\the\wd0\space limit=\the\linewidth}
\retainedtab\end{table}\endgroup

\begingroup\begin{table}[!htbp]\centering\scriptsize
\caption{Observation gates and coupled quotas. Each node-count group has 40 cases across four demand families, balanced by shared-quota binding. Correct requires minimum-cost feasible selection or justified refresh. Per-stage rule checks capacities separately, omitting combined demand on a shared worker. Complete/shared reuses Table~\ref{tab:retained-observation}. Public gating preserves state bytes and candidates while changing instructions and removing refresh. Quota conditions are paired within case.}
\label{tab:retained-quota}
\setlength{\tabcolsep}{3pt}
\setlength{\aboverulesep}{0pt}\setlength{\belowrulesep}{0pt}
\renewcommand{\arraystretch}{0.94}
\def\retainedtab{\begin{tabular}{lrrrrrrr}
\toprule
\rowcolor{ebBlue} \multicolumn{1}{c}{} & \multicolumn{4}{c}{Correct decisions $\uparrow$} & \multicolumn{1}{c}{Valid $\uparrow$} & \multicolumn{2}{c}{Response (s) $\downarrow$} \\
\rowcolor{ebBlue} Method & 3 nodes & 5 nodes & 8 nodes & All & All & Median & p95 \\
\midrule
\rowcolor{ebAmber!45}\multicolumn{8}{l}{\textbf{\textit{Independent quota}}} \\
\rowcolor{ebCoral!10}\sokjev & 17/40 & 7/40 & 5/40 & 29/120 & 120/120 & 0.358 & 0.642 \\
\deepseek & 12/40 & 5/40 & 2/40 & 19/120 & 120/120 & 0.502 & 0.861 \\
\gemini & 18/40 & 18/40 & 4/40 & 40/120 & 120/120 & 0.994 & 1.326 \\
\sokglm & 0/40 & 1/40 & 0/40 & 1/120 & 120/120 & 0.813 & 9.581 \\
\sokqwen & 17/40 & 10/40 & 3/40 & 30/120 & 120/120 & 0.246 & 0.329 \\
Full rule & \cellcolor{ebCoral!35}\textbf{40/40} & \cellcolor{ebCoral!35}\textbf{40/40} & \cellcolor{ebCoral!35}\textbf{40/40} & \cellcolor{ebCoral!35}\textbf{120/120} & 120/120 & $<.001$ & $<.001$ \\
Per-stage rule & \cellcolor{ebCoral!35}\textbf{40/40} & \cellcolor{ebCoral!35}\textbf{40/40} & \cellcolor{ebCoral!35}\textbf{40/40} & \cellcolor{ebCoral!35}\textbf{120/120} & 120/120 & $<.001$ & $<.001$ \\
\rowcolor{ebAmber!45}\multicolumn{8}{l}{\textbf{\textit{Complete / shared quota}}} \\
\rowcolor{ebCoral!10}\sokjev & 10/40 & 2/40 & 2/40 & 14/120 & 120/120 & 0.364 & 0.619 \\
\deepseek & 10/40 & 4/40 & 0/40 & 14/120 & 120/120 & 0.499 & 0.754 \\
\gemini & 17/40 & 10/40 & 9/40 & 36/120 & 120/120 & 1.027 & 1.584 \\
\sokglm & 0/40 & 1/40 & 0/40 & 1/120 & 120/120 & 0.884 & 8.373 \\
\sokqwen & 14/40 & 5/40 & 1/40 & 20/120 & 120/120 & 0.247 & 0.352 \\
Full rule & \cellcolor{ebCoral!35}\textbf{40/40} & \cellcolor{ebCoral!35}\textbf{40/40} & \cellcolor{ebCoral!35}\textbf{40/40} & \cellcolor{ebCoral!35}\textbf{120/120} & 120/120 & $<.001$ & $<.001$ \\
Per-stage rule & 20/40 & 20/40 & 20/40 & 60/120 & 120/120 & $<.001$ & $<.001$ \\
\rowcolor{ebAmber!45}\multicolumn{8}{l}{\textbf{\textit{Shared quota + domain separation}}} \\
\rowcolor{ebCoral!10}\sokjev & 6/40 & 0/40 & 0/40 & 6/120 & 120/120 & 0.354 & 0.627 \\
\deepseek & 3/40 & 1/40 & 0/40 & 4/120 & 120/120 & 0.499 & 0.790 \\
\gemini & 14/40 & 14/40 & 14/40 & 42/120 & 120/120 & 0.989 & 1.220 \\
\sokglm & 0/40 & 0/40 & 0/40 & 0/120 & 120/120 & 0.823 & 2.839 \\
\sokqwen & 2/40 & 1/40 & 1/40 & 4/120 & 120/120 & 0.248 & 0.340 \\
Full rule & \cellcolor{ebCoral!35}\textbf{40/40} & \cellcolor{ebCoral!35}\textbf{40/40} & \cellcolor{ebCoral!35}\textbf{40/40} & \cellcolor{ebCoral!35}\textbf{120/120} & 120/120 & $<.001$ & $<.001$ \\
Per-stage rule & \cellcolor{ebCoral!35}\textbf{40/40} & \cellcolor{ebCoral!35}\textbf{40/40} & \cellcolor{ebCoral!35}\textbf{40/40} & \cellcolor{ebCoral!35}\textbf{120/120} & 120/120 & $<.001$ & $<.001$ \\
\rowcolor{ebAmber!45}\multicolumn{8}{l}{\textbf{\textit{Public gate / independent quota}}} \\
\rowcolor{ebCoral!10}\sokjev & 29/40 & 25/40 & 30/40 & 84/120 & 120/120 & 0.358 & 0.604 \\
\deepseek & 22/40 & 21/40 & 17/40 & 60/120 & 120/120 & 0.509 & 1.175 \\
\gemini & 27/40 & 22/40 & 20/40 & 69/120 & 120/120 & 1.009 & 1.241 \\
\sokglm & 26/40 & 21/40 & 24/40 & 71/120 & 119/120 & 0.843 & 3.553 \\
\sokqwen & 21/40 & 18/40 & 13/40 & 52/120 & 120/120 & 0.311 & 0.411 \\
Full rule & \cellcolor{ebCoral!35}\textbf{40/40} & \cellcolor{ebCoral!35}\textbf{40/40} & \cellcolor{ebCoral!35}\textbf{40/40} & \cellcolor{ebCoral!35}\textbf{120/120} & 120/120 & $<.001$ & $<.001$ \\
Per-stage rule & \cellcolor{ebCoral!35}\textbf{40/40} & \cellcolor{ebCoral!35}\textbf{40/40} & \cellcolor{ebCoral!35}\textbf{40/40} & \cellcolor{ebCoral!35}\textbf{120/120} & 120/120 & $<.001$ & $<.001$ \\
\rowcolor{ebAmber!45}\multicolumn{8}{l}{\textbf{\textit{Public gate / shared quota}}} \\
\rowcolor{ebCoral!10}\sokjev & 16/40 & 13/40 & 13/40 & 42/120 & 120/120 & 0.353 & 0.715 \\
\deepseek & 15/40 & 10/40 & 8/40 & 33/120 & 120/120 & 0.501 & 1.087 \\
\gemini & 17/40 & 13/40 & 10/40 & 40/120 & 120/120 & 1.009 & 1.234 \\
\sokglm & 17/40 & 16/40 & 17/40 & 50/120 & 120/120 & 0.840 & 8.439 \\
\sokqwen & 15/40 & 13/40 & 11/40 & 39/120 & 120/120 & 0.319 & 0.389 \\
Full rule & \cellcolor{ebCoral!35}\textbf{40/40} & \cellcolor{ebCoral!35}\textbf{40/40} & \cellcolor{ebCoral!35}\textbf{40/40} & \cellcolor{ebCoral!35}\textbf{120/120} & 120/120 & $<.001$ & $<.001$ \\
Per-stage rule & 20/40 & 20/40 & 20/40 & 60/120 & 120/120 & $<.001$ & $<.001$ \\
\rowcolor{ebAmber!45}\multicolumn{8}{l}{\textbf{\textit{Public gate / domain separation}}} \\
\rowcolor{ebCoral!10}\sokjev & 17/40 & 11/40 & 17/40 & 45/120 & 120/120 & 0.375 & 0.585 \\
\deepseek & 9/40 & 6/40 & 2/40 & 17/120 & 120/120 & 0.506 & 0.999 \\
\gemini & 16/40 & 18/40 & 27/40 & 61/120 & 120/120 & 1.021 & 1.276 \\
\sokglm & 9/40 & 12/40 & 13/40 & 34/120 & 120/120 & 0.808 & 3.075 \\
\sokqwen & 3/40 & 6/40 & 5/40 & 14/120 & 120/120 & 0.322 & 0.401 \\
Full rule & \cellcolor{ebCoral!35}\textbf{40/40} & \cellcolor{ebCoral!35}\textbf{40/40} & \cellcolor{ebCoral!35}\textbf{40/40} & \cellcolor{ebCoral!35}\textbf{120/120} & 120/120 & $<.001$ & $<.001$ \\
Per-stage rule & \cellcolor{ebCoral!35}\textbf{40/40} & \cellcolor{ebCoral!35}\textbf{40/40} & \cellcolor{ebCoral!35}\textbf{40/40} & \cellcolor{ebCoral!35}\textbf{120/120} & 120/120 & $<.001$ & $<.001$ \\
\bottomrule\end{tabular}}
\sbox0{\retainedtab}
\setlength{\tabcolsep}{\dimexpr\tabcolsep+(\linewidth-\wd0-1pt)/16\relax}
\ifdim\tabcolsep<1.5pt\setlength{\tabcolsep}{1.5pt}\fi
\sbox0{\retainedtab}
\typeout{RETAINED quota width=\the\wd0\space limit=\the\linewidth}
\retainedtab\end{table}\endgroup

\section{Supporting Results by Lens}\label{app:supporting-lenses}

The supporting material follows the lens order. For L1, the tables give the borrowed radio context and the complete SoK-owned endpoint results. The full deadline curves and native load traces behind the L2 synthesis in $\S$\ref{sec:lens-workload} follow. Observation-state detail and the declared coverage and workflow comparison families support L3, and the L4 tables give the native structural conditions and paired length effects.
\begin{table*}[!htbp]
\centering\scriptsize
\caption{Radio-loop simulation borrowed from the RAN study~\cite{li2026intent}. Latency-only contrasts use 0.3 intents/s, 30\,km/h and five UEs per cell. Values are challenger minus anchor SLA violation with paired time-block intervals and Holm adjustment across eight tests.}
\label{tab:sixg-published-radio}
\setlength{\tabcolsep}{3pt}
\renewcommand{\arraystretch}{0.94}
\setlength{\aboverulesep}{0pt}\setlength{\belowrulesep}{0pt}\setlength{\extrarowheight}{0pt}
\def\familytab{\begin{tabular}{lrrrr}
\toprule
\rowcolor{ebBlue}
Challenger - anchor & Endpoint & $\Delta$ violation pp [95\% CI] & $p_{\mathrm{Holm}}$ & Contrast direction \\
\midrule
\rowcolor{ebAmber!45}
\multicolumn{5}{l}{\textbf{\itshape Radio base point: latency-only arms, event-triggered mode}} \\
DeepSeek - Jev & Affected class & 0.03 \textcolor{black!60}{[-1.45,1.55]} & 1 & Unresolved \\
DeepSeek - Jev & Network-wide & -0.66 \textcolor{black!60}{[-2.14,0.46]} & 1 & Unresolved \\
GLM 5.3 - Jev & Affected class & -2.45 \textcolor{black!60}{[-5.13,-0.63]} & 0.0072 & Lower violation \\
GLM 5.3 - Jev & Network-wide & -2.66 \textcolor{black!60}{[-4.94,-1.03]} & 0.0007999 & Lower violation \\
Qwen 3.8 - Jev & Affected class & -0.98 \textcolor{black!60}{[-2.21,0.23]} & 0.581 & Unresolved \\
Qwen 3.8 - Jev & Network-wide & -0.66 \textcolor{black!60}{[-2.44,0.50]} & 1 & Unresolved \\
Qwen JSON - SemIf & Affected class & 1.46 \textcolor{black!60}{[-0.31,3.97]} & 0.6232 & Unresolved \\
Qwen JSON - SemIf & Network-wide & 1.12 \textcolor{black!60}{[0.46,1.97]} & 0.0007999 & Higher violation \\
\bottomrule
\end{tabular}}
\setlength{\tabcolsep}{0pt}\sbox0{\familytab}
\dimen0=\linewidth\advance\dimen0 by -1pt\advance\dimen0 by -\wd0\divide\dimen0 by 10
\ifdim\dimen0<1.5pt \dimen0=1.5pt\fi
\setlength{\tabcolsep}{\dimen0}\familytab
\end{table*}

\begin{table*}[!htbp]
\centering
\caption{Fixed-action timing and load. Conditional p50/p95 are in seconds. The term $n_v/n$ gives verified outcomes over all arrivals. Attainment uses all arrivals with $B=10$\,s for transport and $2$\,s for edge. Load outcomes use measured residuals after real FIFO queues.}
\label{tab:e2-results}
\scriptsize
\setlength{\tabcolsep}{4pt}
\renewcommand{\arraystretch}{0.94}
\setlength{\aboverulesep}{0pt}\setlength{\belowrulesep}{0pt}\setlength{\extrarowheight}{0pt}
\def\familytab{\begin{tabular}{rrrrrrrr}
\toprule
\rowcolor{ebBlue}
$D$ (s) & $n_v/n$ & $T_a$ p50 / p95 & $T_b$ p50 / p95 & $D/T_a$ & $D/T_b$ & $P(T_b\leq B)$ & $\rho$ \\
\midrule
\rowcolor{ebAmber!45}
\multicolumn{8}{l}{\textbf{\itshape Transport--sparse physical execution}} \\
0 & 300/300 & \cellcolor{ebCoral!35}\textbf{0.044 / 0.051} & \cellcolor{ebCoral!35}\textbf{0.561 / 0.586} & 0.1 & 0.0 & \cellcolor{ebCoral!35}\textbf{100.0 \textcolor{black!60}{[100.0, 100.0]}} & N/A \\
0.1 & 300/300 & 0.144 / 0.151 & 0.659 / 0.687 & 69.3 & 15.2 & \cellcolor{ebCoral!35}\textbf{100.0 \textcolor{black!60}{[100.0, 100.0]}} & N/A \\
0.5 & 300/300 & 0.544 / 0.551 & 1.060 / 1.086 & 91.9 & 47.2 & \cellcolor{ebCoral!35}\textbf{100.0 \textcolor{black!60}{[100.0, 100.0]}} & N/A \\
2 & 300/300 & 2.045 / 2.051 & 2.561 / 2.587 & 97.8 & 78.1 & \cellcolor{ebCoral!35}\textbf{100.0 \textcolor{black!60}{[100.0, 100.0]}} & N/A \\
\midrule
\rowcolor{ebAmber!45}
\multicolumn{8}{l}{\textbf{\itshape Transport--queue + execution replay, $\rho_0=0.5$, $\lambda=4.992$\,/s}} \\
0 & 1016/1016 & \cellcolor{ebCoral!35}\textbf{0.044 / 0.051} & \cellcolor{ebCoral!35}\textbf{0.561 / 0.588} & 0.1 & 0.0 & \cellcolor{ebCoral!35}\textbf{100.0 \textcolor{black!60}{[100.0, 100.0]}} & 0.000 \\
0.1 & 1016/1016 & 0.154 / 0.348 & 0.683 / 0.869 & 65.3 & 14.7 & \cellcolor{ebCoral!35}\textbf{100.0 \textcolor{black!60}{[100.0, 100.0]}} & 0.502 \\
0.5 & 1016/1016 & 233.246 / 372.255 & 233.749 / 372.767 & 0.2 & 0.2 & 0.0 \textcolor{black!60}{[0.0, 0.0]} & 2.500 \\
2 & 1016/1016 & 1405.524 / 2229.573 & 1406.028 / 2230.085 & 0.1 & 0.1 & 0.0 \textcolor{black!60}{[0.0, 0.0]} & 9.987 \\
\midrule
\rowcolor{ebAmber!45}
\multicolumn{8}{l}{\textbf{\itshape Transport--queue + execution replay, $\rho_0=0.8$, $\lambda=7.987$\,/s}} \\
0 & 1576/1576 & \cellcolor{ebCoral!35}\textbf{0.044 / 0.051} & \cellcolor{ebCoral!35}\textbf{0.561 / 0.586} & 0.1 & 0.0 & \cellcolor{ebCoral!35}\textbf{100.0 \textcolor{black!60}{[100.0, 100.0]}} & 0.000 \\
0.1 & 1576/1576 & 0.287 / 0.838 & 0.802 / 1.355 & 35.0 & 12.6 & \cellcolor{ebCoral!35}\textbf{100.0 \textcolor{black!60}{[100.0, 100.0]}} & 0.804 \\
0.5 & 1576/1576 & 508.040 / 773.329 & 508.566 / 773.853 & 0.1 & 0.1 & 0.0 \textcolor{black!60}{[0.0, 0.0]} & 3.999 \\
2 & 1576/1576 & 2509.895 / 3838.564 & 2510.421 / 3839.087 & 0.1 & 0.1 & 0.0 \textcolor{black!60}{[0.0, 0.0]} & 15.981 \\
\midrule
\rowcolor{ebAmber!45}
\multicolumn{8}{l}{\textbf{\itshape Transport--queue + execution replay, $\rho_0=0.95$, $\lambda=9.484$\,/s}} \\
0 & 1929/1929 & \cellcolor{ebCoral!35}\textbf{0.045 / 0.051} & \cellcolor{ebCoral!35}\textbf{0.561 / 0.587} & 0.1 & 0.0 & \cellcolor{ebCoral!35}\textbf{100.0 \textcolor{black!60}{[100.0, 100.0]}} & 0.000 \\
0.1 & 1929/1929 & 1.480 / 2.711 & 2.002 / 3.231 & 6.8 & 5.0 & \cellcolor{ebCoral!35}\textbf{100.0 \textcolor{black!60}{[100.0, 100.0]}} & 0.955 \\
0.5 & 1929/1929 & 627.838 / 970.613 & 628.367 / 971.128 & 0.1 & 0.1 & 0.0 \textcolor{black!60}{[0.0, 0.0]} & 4.749 \\
2 & 1929/1929 & 2981.335 / 4625.823 & 2981.864 / 4626.337 & 0.1 & 0.1 & 0.0 \textcolor{black!60}{[0.0, 0.0]} & 18.977 \\
\midrule
\rowcolor{ebAmber!45}
\multicolumn{8}{l}{\textbf{\itshape Edge--sparse physical execution}} \\
0 & 300/300 & \cellcolor{ebCoral!35}\textbf{0.005 / 0.006} & \cellcolor{ebCoral!35}\textbf{1.379 / 2.777} & 1.2 & 0.0 & \cellcolor{ebCoral!35}\textbf{85.0 \textcolor{black!60}{[78.3, 90.0]}} & N/A \\
0.1 & 300/300 & 0.105 / 0.106 & 1.602 / 2.739 & 94.9 & 6.2 & 76.0 \textcolor{black!60}{[66.3, 84.3]} & N/A \\
0.5 & 300/300 & 0.506 / 0.506 & 2.189 / 3.001 & 98.9 & 22.8 & 29.7 \textcolor{black!60}{[17.3, 43.0]} & N/A \\
2 & 300/300 & 2.006 / 2.007 & 3.394 / 4.758 & 99.7 & 58.9 & 0.0 \textcolor{black!60}{[0.0, 0.0]} & N/A \\
\midrule
\rowcolor{ebAmber!45}
\multicolumn{8}{l}{\textbf{\itshape Edge--queue + execution replay, $\rho_0=0.5$, $\lambda=4.992$\,/s}} \\
0 & 1016/1016 & \cellcolor{ebCoral!35}\textbf{0.005 / 0.006} & \cellcolor{ebCoral!35}\textbf{1.382 / 2.734} & 0.6 & 0.0 & \cellcolor{ebCoral!35}\textbf{85.2 \textcolor{black!60}{[83.6, 87.0]}} & 0.000 \\
0.1 & 1016/1016 & 0.115 / 0.307 & 1.555 / 2.867 & 87.9 & 6.5 & 71.9 \textcolor{black!60}{[69.0, 75.0]} & 0.502 \\
0.5 & 1016/1016 & 233.210 / 372.216 & 235.246 / 374.051 & 0.2 & 0.2 & 0.0 \textcolor{black!60}{[0.0, 0.0]} & 2.500 \\
2 & 1016/1016 & 1405.488 / 2229.533 & 1407.557 / 2230.965 & 0.1 & 0.1 & 0.0 \textcolor{black!60}{[0.0, 0.0]} & 9.987 \\
\midrule
\rowcolor{ebAmber!45}
\multicolumn{8}{l}{\textbf{\itshape Edge--queue + execution replay, $\rho_0=0.8$, $\lambda=7.987$\,/s}} \\
0 & 1576/1576 & \cellcolor{ebCoral!35}\textbf{0.005 / 0.006} & \cellcolor{ebCoral!35}\textbf{1.382 / 2.798} & 0.6 & 0.0 & \cellcolor{ebCoral!35}\textbf{84.6 \textcolor{black!60}{[82.6, 86.6]}} & 0.000 \\
0.1 & 1576/1576 & 0.248 / 0.795 & 1.774 / 3.063 & 40.6 & 5.7 & 61.2 \textcolor{black!60}{[56.8, 65.6]} & 0.804 \\
0.5 & 1576/1576 & 507.999 / 773.286 & 509.794 / 774.647 & 0.1 & 0.1 & 0.0 \textcolor{black!60}{[0.0, 0.0]} & 3.999 \\
2 & 1576/1576 & 2509.854 / 3838.521 & 2512.032 / 3839.745 & 0.1 & 0.1 & 0.0 \textcolor{black!60}{[0.0, 0.0]} & 15.981 \\
\midrule
\rowcolor{ebAmber!45}
\multicolumn{8}{l}{\textbf{\itshape Edge--queue + execution replay, $\rho_0=0.95$, $\lambda=9.484$\,/s}} \\
0 & 1929/1929 & \cellcolor{ebCoral!35}\textbf{0.005 / 0.006} & \cellcolor{ebCoral!35}\textbf{1.382 / 2.812} & 0.5 & 0.0 & \cellcolor{ebCoral!35}\textbf{84.3 \textcolor{black!60}{[82.8, 85.9]}} & 0.000 \\
0.1 & 1929/1929 & 1.441 / 2.674 & 2.954 / 4.655 & 7.0 & 3.4 & 18.6 \textcolor{black!60}{[9.3, 29.0]} & 0.955 \\
0.5 & 1929/1929 & 627.802 / 970.569 & 628.943 / 972.304 & 0.1 & 0.1 & 0.0 \textcolor{black!60}{[0.0, 0.0]} & 4.749 \\
2 & 1929/1929 & 2981.299 / 4625.779 & 2982.439 / 4627.480 & 0.1 & 0.1 & 0.0 \textcolor{black!60}{[0.0, 0.0]} & 18.977 \\
\bottomrule
\end{tabular}}
\setlength{\tabcolsep}{0pt}\sbox0{\familytab}
\dimen0=\linewidth\advance\dimen0 by -1pt\advance\dimen0 by -\wd0\divide\dimen0 by 16
\ifdim\dimen0<1.5pt \dimen0=1.5pt\fi
\setlength{\tabcolsep}{\dimen0}\familytab
\end{table*}

\begin{figure*}[!htbp]\centering
\includegraphics[width=7in]{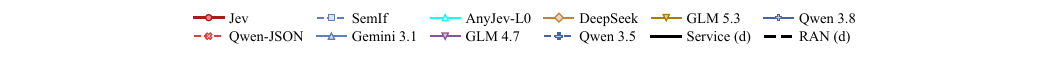}\\[-2pt]
\subfloat[Access policy]{\includegraphics[width=1.70in]{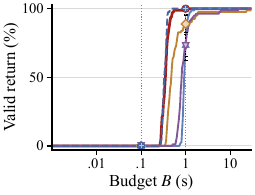}}\hfill
\subfloat[Service contracts]{\includegraphics[width=1.70in]{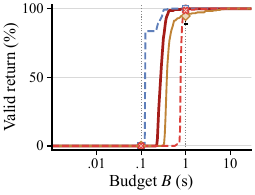}}\hfill
\subfloat[RAN telemetry]{\includegraphics[width=1.70in]{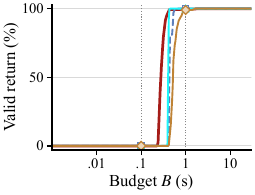}}\hfill
\subfloat[Decision-slot demand]{\includegraphics[width=1.70in]{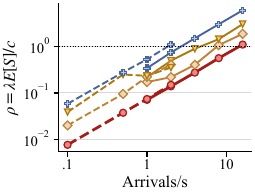}}
\caption{Task-specific format-valid return within budget and shared decision-slot demand. Panels (b)--(d) reuse measurements from the Edge and RAN studies~\cite{li2026replacing,li2026intent}. Whiskers at 100\,ms and 1\,s are 30\,s time-block intervals.}
\label{fig:e1}\end{figure*}
\begin{table*}[!htbp]
\centering\scriptsize
\caption{Service admission at 1 and 2 arrivals/s, borrowed and reanalysed from the Edge study~\cite{li2026replacing}. Offered $\rho$ estimates dispatched service demand. Return and native modeled completion use all arrivals. Queue quantiles condition on observed waits.}
\label{tab:rq-d-capacity}
\setlength{\tabcolsep}{2.5pt}
\renewcommand{\arraystretch}{0.94}
\setlength{\aboverulesep}{0pt}\setlength{\belowrulesep}{0pt}\setlength{\extrarowheight}{0pt}
\def\familytab{\begin{tabular}{lrrrrr}
\toprule
\rowcolor{ebBlue}
Implementation & \shortstack{Offered $\rho$\\{}[CI]} & \shortstack{Accepted\\$\rho$} & \shortstack{Queue (s)\\p50 / p95} & \shortstack{Return $\leq B$\\(\%) [CI]} & \shortstack{Native completion\\(\%) [CI]} \\
\midrule
\rowcolor{ebAmber!45}
\multicolumn{6}{l}{\textbf{\itshape Service admission: 1 arrivals/s, B = 2 s. 300 arrivals per row}} \\
\rowcolor{ebCoral!10}\jev & 0.071 \textcolor{black!60}{[0.070,0.073]} & 0.071 & 0.002 / 0.002 & 100.0 \textcolor{black!60}{[100.0,100.0]} & 94.0 \textcolor{black!60}{[90.7,96.8]} \\
\semif & 0.170 \textcolor{black!60}{[0.156,0.184]} & 0.170 & 0.002 / 0.002 & 100.0 \textcolor{black!60}{[100.0,100.0]} & 63.0 \textcolor{black!60}{[56.6,69.9]} \\
\laya & 0.028 \textcolor{black!60}{[0.027,0.029]} & 0.028 & 0.002 / 0.002 & 100.0 \textcolor{black!60}{[100.0,100.0]} & 11.7 \textcolor{black!60}{[7.3,15.7]} \\
\deepseek & 0.171 \textcolor{black!60}{[0.094,0.273]} & 0.165 & 0.002 / 0.002 & 92.3 \textcolor{black!60}{[83.3,100.0]} & 90.7 \textcolor{black!60}{[81.7,97.9]} \\
\glm & 0.217 \textcolor{black!60}{[0.202,0.231]} & 0.217 & 0.002 / 0.002 & 93.3 \textcolor{black!60}{[90.7,96.1]} & 91.7 \textcolor{black!60}{[88.7,94.8]} \\
\qwenflash & 0.338 \textcolor{black!60}{[0.326,0.350]} & 0.338 & 0.002 / 0.002 & 95.0 \textcolor{black!60}{[92.6,97.6]} & 90.0 \textcolor{black!60}{[87.5,92.7]} \\
\qwenjson & 0.739 \textcolor{black!60}{[0.506,0.984]} & 0.643 & 0.002 / 1.807 & 38.0 \textcolor{black!60}{[15.1,63.7]} & 19.3 \textcolor{black!60}{[6.8,33.4]} \\
Fixed 0.3 s & 0.077 \textcolor{black!60}{[0.077,0.077]} & 0.077 & 0.002 / 0.002 & 100.0 \textcolor{black!60}{[100.0,100.0]} & \cellcolor{ebCoral!35}\textbf{100.0} \textcolor{black!60}{[100.0,100.0]} \\
Fixed 0.6 s & 0.152 \textcolor{black!60}{[0.152,0.152]} & 0.152 & 0.002 / 0.002 & 100.0 \textcolor{black!60}{[100.0,100.0]} & \cellcolor{ebCoral!35}\textbf{100.0} \textcolor{black!60}{[100.0,100.0]} \\
Fixed 1.2 s & 0.302 \textcolor{black!60}{[0.302,0.302]} & 0.302 & 0.002 / 0.002 & 100.0 \textcolor{black!60}{[100.0,100.0]} & \cellcolor{ebCoral!35}\textbf{100.0} \textcolor{black!60}{[100.0,100.0]} \\
Oracle & 0.000 \textcolor{black!60}{[0.000,0.000]} & 0.000 & 0.002 / 0.002 & 100.0 \textcolor{black!60}{[100.0,100.0]} & \cellcolor{ebCoral!35}\textbf{100.0} \textcolor{black!60}{[100.0,100.0]} \\
\midrule
\rowcolor{ebAmber!45}
\multicolumn{6}{l}{\textbf{\itshape Service admission: 2 arrivals/s, B = 2 s. 300 arrivals per row}} \\
\rowcolor{ebCoral!10}\jev & 0.139 \textcolor{black!60}{[0.133,0.144]} & 0.138 & 0.002 / 0.002 & 100.0 \textcolor{black!60}{[100.0,100.0]} & 95.3 \textcolor{black!60}{[92.4,98.2]} \\
\semif & 1.164 \textcolor{black!60}{[1.027,1.268]} & 0.966 & 1.074 / 1.910 & 11.0 \textcolor{black!60}{[0.4,29.1]} & 8.7 \textcolor{black!60}{[0.3,24.5]} \\
\laya & 0.049 \textcolor{black!60}{[0.048,0.050]} & 0.049 & 0.002 / 0.002 & 100.0 \textcolor{black!60}{[100.0,100.0]} & 11.7 \textcolor{black!60}{[7.9,16.8]} \\
\deepseek & 0.217 \textcolor{black!60}{[0.199,0.236]} & 0.217 & 0.002 / 0.002 & 99.3 \textcolor{black!60}{[98.1,100.0]} & 97.7 \textcolor{black!60}{[95.8,99.4]} \\
\glm & 0.490 \textcolor{black!60}{[0.384,0.607]} & 0.489 & 0.002 / 0.574 & 85.3 \textcolor{black!60}{[74.2,95.4]} & 82.3 \textcolor{black!60}{[70.0,93.4]} \\
\qwenflash & 0.733 \textcolor{black!60}{[0.719,0.749]} & 0.727 & 0.002 / 1.439 & 73.0 \textcolor{black!60}{[60.9,86.3]} & 66.0 \textcolor{black!60}{[54.0,79.3]} \\
\qwenjson & 2.154 \textcolor{black!60}{[2.114,2.196]} & 1.019 & 1.658 / 1.943 & 0.7 \textcolor{black!60}{[0.0,2.0]} & 0.3 \textcolor{black!60}{[0.0,1.0]} \\
Fixed 0.3 s & 0.154 \textcolor{black!60}{[0.154,0.154]} & 0.154 & 0.002 / 0.002 & 100.0 \textcolor{black!60}{[100.0,100.0]} & \cellcolor{ebCoral!35}\textbf{100.0} \textcolor{black!60}{[100.0,100.0]} \\
Fixed 0.6 s & 0.304 \textcolor{black!60}{[0.304,0.304]} & 0.304 & 0.002 / 0.002 & 100.0 \textcolor{black!60}{[100.0,100.0]} & \cellcolor{ebCoral!35}\textbf{100.0} \textcolor{black!60}{[100.0,100.0]} \\
Fixed 1.2 s & 0.604 \textcolor{black!60}{[0.604,0.604]} & 0.603 & 0.002 / 0.449 & 99.0 \textcolor{black!60}{[97.8,100.0]} & 98.3 \textcolor{black!60}{[96.5,100.0]} \\
Oracle & 0.000 \textcolor{black!60}{[0.000,0.000]} & 0.000 & 0.002 / 0.002 & 100.0 \textcolor{black!60}{[100.0,100.0]} & \cellcolor{ebCoral!35}\textbf{100.0} \textcolor{black!60}{[100.0,100.0]} \\
\bottomrule
\end{tabular}}
\setlength{\tabcolsep}{0pt}\sbox0{\familytab}
\dimen0=\linewidth\advance\dimen0 by -1pt\advance\dimen0 by -\wd0\divide\dimen0 by 12
\ifdim\dimen0<1.5pt \dimen0=1.5pt\fi
\setlength{\tabcolsep}{\dimen0}\familytab
\end{table*}

\begin{table*}[!htbp]
\centering\scriptsize
\caption{Service admission at 4, 8 and 16 arrivals/s, borrowed and reanalysed from the Edge study~\cite{li2026replacing}. Columns and uncertainty follow Table~\ref{tab:rq-d-capacity}. Return and native modeled completion use all arrivals. Offered $\rho$ estimates dispatched service demand.}
\label{tab:rq-d-capacity-edge-high}
\setlength{\tabcolsep}{2.5pt}
\renewcommand{\arraystretch}{0.94}
\setlength{\aboverulesep}{0pt}\setlength{\belowrulesep}{0pt}\setlength{\extrarowheight}{0pt}
\def\familytab{\begin{tabular}{lrrrrr}
\toprule
\rowcolor{ebBlue}
Implementation & \shortstack{Offered $\rho$\\{}[CI]} & \shortstack{Accepted\\$\rho$} & \shortstack{Queue (s)\\p50 / p95} & \shortstack{Return $\leq B$\\(\%) [CI]} & \shortstack{Native completion\\(\%) [CI]} \\
\midrule
\rowcolor{ebAmber!45}
\multicolumn{6}{l}{\textbf{\itshape Service admission: 4 arrivals/s, B = 2 s. 300 arrivals per row}} \\
\rowcolor{ebCoral!10}\jev & 0.272 \textcolor{black!60}{[0.268,0.277]} & 0.272 & 0.002 / 0.002 & 99.7 \textcolor{black!60}{[99.2,100.0]} & 93.7 \textcolor{black!60}{[89.8,95.3]} \\
\semif & 2.637 \textcolor{black!60}{[2.602,2.667]} & 1.037 & 1.812 / 1.986 & 1.0 \textcolor{black!60}{[0.0,2.7]} & 1.0 \textcolor{black!60}{[0.0,2.7]} \\
\laya & 0.102 \textcolor{black!60}{[0.097,0.106]} & 0.102 & 0.002 / 0.002 & 100.0 \textcolor{black!60}{[100.0,100.0]} & 11.7 \textcolor{black!60}{[8.0,15.6]} \\
\deepseek & 0.401 \textcolor{black!60}{[0.399,0.406]} & 0.401 & 0.002 / 0.005 & 100.0 \textcolor{black!60}{[100.0,100.0]} & 98.3 \textcolor{black!60}{[96.6,99.2]} \\
\glm & 0.883 \textcolor{black!60}{[0.806,1.018]} & 0.871 & 0.349 / 1.508 & 84.3 \textcolor{black!60}{[62.7,90.3]} & 80.7 \textcolor{black!60}{[57.6,86.7]} \\
\qwenflash & 1.478 \textcolor{black!60}{[1.396,1.738]} & 1.019 & 1.649 / 1.942 & 3.7 \textcolor{black!60}{[2.7,4.7]} & 2.7 \textcolor{black!60}{[2.3,3.4]} \\
\qwenjson & 4.203 \textcolor{black!60}{[4.052,4.359]} & 1.050 & 1.776 / 1.960 & 0.7 \textcolor{black!60}{[0.0,1.8]} & 0.3 \textcolor{black!60}{[0.0,0.9]} \\
Fixed 0.3 s & 0.307 \textcolor{black!60}{[0.307,0.307]} & 0.307 & 0.002 / 0.002 & 100.0 \textcolor{black!60}{[100.0,100.0]} & \cellcolor{ebCoral!35}\textbf{100.0} \textcolor{black!60}{[100.0,100.0]} \\
Fixed 0.6 s & 0.607 \textcolor{black!60}{[0.607,0.607]} & 0.607 & 0.002 / 0.234 & 100.0 \textcolor{black!60}{[100.0,100.0]} & \cellcolor{ebCoral!35}\textbf{100.0} \textcolor{black!60}{[100.0,100.0]} \\
Fixed 1.2 s & 1.207 \textcolor{black!60}{[1.207,1.208]} & 1.018 & 1.480 / 1.964 & 15.3 \textcolor{black!60}{[11.9,18.6]} & 11.7 \textcolor{black!60}{[8.6,15.9]} \\
Oracle & 0.001 \textcolor{black!60}{[0.001,0.001]} & 0.001 & 0.002 / 0.002 & 100.0 \textcolor{black!60}{[100.0,100.0]} & \cellcolor{ebCoral!35}\textbf{100.0} \textcolor{black!60}{[100.0,100.0]} \\
\midrule
\rowcolor{ebAmber!45}
\multicolumn{6}{l}{\textbf{\itshape Service admission: 8 arrivals/s, B = 2 s. 300 arrivals per row}} \\
\rowcolor{ebCoral!10}\jev & 0.556 \textcolor{black!60}{[0.554,0.560]} & 0.555 & 0.002 / 0.078 & 100.0 \textcolor{black!60}{[100.0,100.0]} & 94.3 \textcolor{black!60}{[89.8,95.4]} \\
\semif & 5.179 \textcolor{black!60}{[5.150,5.303]} & 1.087 & 1.914 / 1.981 & 1.0 \textcolor{black!60}{[0.0,1.2]} & 1.0 \textcolor{black!60}{[0.0,1.2]} \\
\laya & 0.232 \textcolor{black!60}{[0.217,0.235]} & 0.231 & 0.002 / 0.002 & 100.0 \textcolor{black!60}{[100.0,100.0]} & 11.7 \textcolor{black!60}{[10.2,12.0]} \\
\deepseek & 1.053 \textcolor{black!60}{[0.938,1.084]} & 0.954 & 0.717 / 1.858 & 65.3 \textcolor{black!60}{[56.8,100.0]} & 60.3 \textcolor{black!60}{[51.5,96.6]} \\
\glm & 1.416 \textcolor{black!60}{[1.349,1.432]} & 1.047 & 1.782 / 1.977 & 9.3 \textcolor{black!60}{[1.7,11.2]} & 7.7 \textcolor{black!60}{[0.0,9.5]} \\
\qwenflash & 2.954 \textcolor{black!60}{[2.919,2.963]} & 1.083 & 1.862 / 1.984 & 1.3 \textcolor{black!60}{[0.0,1.7]} & 1.3 \textcolor{black!60}{[0.0,1.7]} \\
\qwenjson & 8.424 \textcolor{black!60}{[8.352,8.819]} & 1.094 & 1.901 / 1.990 & 0.7 \textcolor{black!60}{[0.0,0.8]} & 0.3 \textcolor{black!60}{[0.0,0.4]} \\
Fixed 0.3 s & 0.615 \textcolor{black!60}{[0.615,0.615]} & 0.615 & 0.002 / 0.125 & 100.0 \textcolor{black!60}{[100.0,100.0]} & \cellcolor{ebCoral!35}\textbf{100.0} \textcolor{black!60}{[100.0,100.0]} \\
Fixed 0.6 s & 1.215 \textcolor{black!60}{[1.213,1.215]} & 1.044 & 1.683 / 1.968 & 21.3 \textcolor{black!60}{[6.8,24.9]} & 19.0 \textcolor{black!60}{[5.1,22.4]} \\
Fixed 1.2 s & 2.414 \textcolor{black!60}{[2.413,2.415]} & 1.054 & 1.855 / 1.987 & 1.7 \textcolor{black!60}{[0.0,2.1]} & 1.7 \textcolor{black!60}{[0.0,2.1]} \\
Oracle & 0.001 \textcolor{black!60}{[0.001,0.001]} & 0.001 & 0.002 / 0.002 & 100.0 \textcolor{black!60}{[100.0,100.0]} & \cellcolor{ebCoral!35}\textbf{100.0} \textcolor{black!60}{[100.0,100.0]} \\
\midrule
\rowcolor{ebAmber!45}
\multicolumn{6}{l}{\textbf{\itshape Service admission: 16 arrivals/s, B = 2 s. 300 arrivals per row}} \\
\rowcolor{ebCoral!10}\jev & 1.084 \textcolor{black!60}{[N/A]} & 1.084 & 0.603 / 1.492 & 98.3 \textcolor{black!60}{[N/A]} & 90.7 \textcolor{black!60}{[N/A]} \\
\semif & 10.170 \textcolor{black!60}{[N/A]} & 1.186 & 1.953 / 1.996 & 1.0 \textcolor{black!60}{[N/A]} & 1.0 \textcolor{black!60}{[N/A]} \\
\laya & 1.037 \textcolor{black!60}{[N/A]} & 1.002 & 0.434 / 1.966 & 75.3 \textcolor{black!60}{[N/A]} & 9.3 \textcolor{black!60}{[N/A]} \\
\deepseek & 1.842 \textcolor{black!60}{[N/A]} & 1.110 & 1.839 / 1.982 & 10.3 \textcolor{black!60}{[N/A]} & 9.3 \textcolor{black!60}{[N/A]} \\
\glm & 2.972 \textcolor{black!60}{[N/A]} & 1.119 & 1.924 / 1.996 & 2.3 \textcolor{black!60}{[N/A]} & 2.3 \textcolor{black!60}{[N/A]} \\
\qwenflash & 5.876 \textcolor{black!60}{[N/A]} & 1.155 & 1.933 / 1.987 & 0.3 \textcolor{black!60}{[N/A]} & 0.3 \textcolor{black!60}{[N/A]} \\
\qwenjson & 15.823 \textcolor{black!60}{[N/A]} & 1.212 & 1.936 / 1.994 & 0.3 \textcolor{black!60}{[N/A]} & 0.3 \textcolor{black!60}{[N/A]} \\
Fixed 0.3 s & 1.229 \textcolor{black!60}{[N/A]} & 1.097 & 1.445 / 1.954 & 57.3 \textcolor{black!60}{[N/A]} & 50.7 \textcolor{black!60}{[N/A]} \\
Fixed 0.6 s & 2.428 \textcolor{black!60}{[N/A]} & 1.116 & 1.880 / 1.988 & 5.3 \textcolor{black!60}{[N/A]} & 5.0 \textcolor{black!60}{[N/A]} \\
Fixed 1.2 s & 4.829 \textcolor{black!60}{[N/A]} & 1.142 & 1.905 / 1.983 & 1.3 \textcolor{black!60}{[N/A]} & 1.3 \textcolor{black!60}{[N/A]} \\
Oracle & 0.002 \textcolor{black!60}{[N/A]} & 0.002 & 0.002 / 0.002 & 100.0 \textcolor{black!60}{[N/A]} & \cellcolor{ebCoral!35}\textbf{100.0} \textcolor{black!60}{[N/A]} \\
\bottomrule
\end{tabular}}
\setlength{\tabcolsep}{0pt}\sbox0{\familytab}
\dimen0=\linewidth\advance\dimen0 by -1pt\advance\dimen0 by -\wd0\divide\dimen0 by 12
\ifdim\dimen0<1.5pt \dimen0=1.5pt\fi
\setlength{\tabcolsep}{\dimen0}\familytab
\end{table*}

\begin{table*}[!htbp]
\centering\scriptsize
\caption{RAN-policy arrival load borrowed and reanalysed from the RAN study~\cite{li2026intent}. Completion is an installable policy within modeled A1/E2 timing. Return and completion use all arrivals. Queue quantiles condition on observed waits.}
\label{tab:rq-d-capacity-sixg}
\setlength{\tabcolsep}{2.5pt}
\renewcommand{\arraystretch}{0.94}
\setlength{\aboverulesep}{0pt}\setlength{\belowrulesep}{0pt}\setlength{\extrarowheight}{0pt}
\def\familytab{\begin{tabular}{lrrrrr}
\toprule
\rowcolor{ebBlue}
Implementation & \shortstack{Offered $\rho$\\{}[CI]} & \shortstack{Accepted\\$\rho$} & \shortstack{Queue (s)\\p50 / p95} & \shortstack{Return $\leq B$\\(\%) [CI]} & \shortstack{Native completion\\(\%) [CI]} \\
\midrule
\rowcolor{ebAmber!45}
\multicolumn{6}{l}{\textbf{\itshape RAN policy: 0.1 arrivals/s, B = 1 s. 300 arrivals per row}} \\
\rowcolor{ebCoral!10}\jev & 0.008 \textcolor{black!60}{[0.008,0.008]} & 0.008 & 0.001 / 0.002 & 100.0 \textcolor{black!60}{[100.0,100.0]} & \cellcolor{ebCoral!35}\textbf{100.0} \textcolor{black!60}{[100.0,100.0]} \\
\semif & 0.011 \textcolor{black!60}{[0.007,0.019]} & 0.012 & 0.006 / 0.027 & 98.7 \textcolor{black!60}{[95.7,100.0]} & 98.7 \textcolor{black!60}{[95.7,100.0]} \\
\anyjev & 0.017 \textcolor{black!60}{[0.017,0.018]} & 0.019 & 0.005 / 0.026 & 92.7 \textcolor{black!60}{[88.6,96.4]} & 74.3 \textcolor{black!60}{[69.0,79.7]} \\
\deepseek & 0.020 \textcolor{black!60}{[0.018,0.023]} & 0.022 & 0.001 / 0.002 & 89.0 \textcolor{black!60}{[85.1,92.8]} & 89.0 \textcolor{black!60}{[85.1,92.8]} \\
\glm & 0.039 \textcolor{black!60}{[0.035,0.044]} & 0.042 & 0.001 / 0.002 & 23.3 \textcolor{black!60}{[17.7,29.5]} & 22.0 \textcolor{black!60}{[16.4,28.0]} \\
\qwenflash & 0.059 \textcolor{black!60}{[0.055,0.063]} & 0.063 & 0.002 / 0.002 & 0.0 \textcolor{black!60}{[0.0,0.0]} & 0.0 \textcolor{black!60}{[0.0,0.0]} \\
\qwenjson & 0.010 \textcolor{black!60}{[0.010,0.011]} & 0.011 & 0.006 / 0.028 & 99.7 \textcolor{black!60}{[98.9,100.0]} & 99.7 \textcolor{black!60}{[98.9,100.0]} \\
\midrule
\rowcolor{ebAmber!45}
\multicolumn{6}{l}{\textbf{\itshape RAN policy: 0.5 arrivals/s, B = 1 s. 300 arrivals per row}} \\
\rowcolor{ebCoral!10}\jev & 0.038 \textcolor{black!60}{[0.037,0.039]} & 0.040 & 0.002 / 0.002 & 99.7 \textcolor{black!60}{[98.9,100.0]} & 99.7 \textcolor{black!60}{[98.9,100.0]} \\
\semif & 0.040 \textcolor{black!60}{[0.037,0.042]} & 0.042 & 0.002 / 0.006 & 100.0 \textcolor{black!60}{[100.0,100.0]} & \cellcolor{ebCoral!35}\textbf{100.0} \textcolor{black!60}{[100.0,100.0]} \\
\anyjev & 0.119 \textcolor{black!60}{[0.105,0.135]} & 0.128 & 0.001 / 0.006 & 63.7 \textcolor{black!60}{[54.0,74.2]} & 52.3 \textcolor{black!60}{[43.5,61.6]} \\
\deepseek & 0.095 \textcolor{black!60}{[0.072,0.134]} & 0.102 & 0.002 / 0.002 & 93.0 \textcolor{black!60}{[87.3,97.2]} & 92.7 \textcolor{black!60}{[86.9,96.9]} \\
\glm & 0.248 \textcolor{black!60}{[0.187,0.310]} & 0.265 & 0.002 / 1.437 & 25.3 \textcolor{black!60}{[14.6,37.2]} & 24.7 \textcolor{black!60}{[14.1,36.4]} \\
\qwenflash & 0.277 \textcolor{black!60}{[0.267,0.288]} & 0.296 & 0.002 / 0.002 & 0.0 \textcolor{black!60}{[0.0,0.0]} & 0.0 \textcolor{black!60}{[0.0,0.0]} \\
\qwenjson & 0.052 \textcolor{black!60}{[0.052,0.052]} & 0.056 & 0.002 / 0.006 & 100.0 \textcolor{black!60}{[100.0,100.0]} & \cellcolor{ebCoral!35}\textbf{100.0} \textcolor{black!60}{[100.0,100.0]} \\
\midrule
\rowcolor{ebAmber!45}
\multicolumn{6}{l}{\textbf{\itshape RAN policy: 1 arrivals/s, B = 1 s. 300 arrivals per row}} \\
\rowcolor{ebCoral!10}\jev & 0.074 \textcolor{black!60}{[0.073,0.076]} & 0.079 & 0.002 / 0.002 & 100.0 \textcolor{black!60}{[100.0,100.0]} & \cellcolor{ebCoral!35}\textbf{100.0} \textcolor{black!60}{[100.0,100.0]} \\
\semif & 0.095 \textcolor{black!60}{[0.080,0.113]} & 0.102 & 0.001 / 0.003 & 97.3 \textcolor{black!60}{[93.2,100.0]} & 97.3 \textcolor{black!60}{[93.2,100.0]} \\
\anyjev & 0.347 \textcolor{black!60}{[0.291,0.404]} & 0.371 & 0.001 / 1.474 & 36.7 \textcolor{black!60}{[27.4,46.3]} & 29.0 \textcolor{black!60}{[21.1,38.0]} \\
\deepseek & 0.221 \textcolor{black!60}{[0.161,0.298]} & 0.236 & 0.002 / 0.914 & 84.0 \textcolor{black!60}{[70.2,94.5]} & 84.0 \textcolor{black!60}{[70.2,94.5]} \\
\glm & 0.232 \textcolor{black!60}{[0.206,0.256]} & 0.248 & 0.002 / 0.002 & 73.0 \textcolor{black!60}{[61.4,85.8]} & 70.7 \textcolor{black!60}{[59.5,83.6]} \\
\qwenflash & 0.524 \textcolor{black!60}{[0.513,0.538]} & 0.561 & 0.002 / 1.260 & 0.0 \textcolor{black!60}{[0.0,0.0]} & 0.0 \textcolor{black!60}{[0.0,0.0]} \\
\qwenjson & 0.106 \textcolor{black!60}{[0.104,0.107]} & 0.113 & 0.001 / 0.003 & 100.0 \textcolor{black!60}{[100.0,100.0]} & \cellcolor{ebCoral!35}\textbf{100.0} \textcolor{black!60}{[100.0,100.0]} \\
\midrule
\rowcolor{ebAmber!45}
\multicolumn{6}{l}{\textbf{\itshape RAN policy: 2 arrivals/s, B = 1 s. 300 arrivals per row}} \\
\rowcolor{ebCoral!10}\jev & 0.149 \textcolor{black!60}{[0.146,0.152]} & 0.159 & 0.002 / 0.002 & 99.7 \textcolor{black!60}{[98.8,100.0]} & 99.7 \textcolor{black!60}{[98.8,100.0]} \\
\semif & 0.261 \textcolor{black!60}{[0.193,0.318]} & 0.279 & 0.001 / 0.259 & 90.0 \textcolor{black!60}{[81.3,100.0]} & 88.7 \textcolor{black!60}{[79.5,100.0]} \\
\anyjev & 1.169 \textcolor{black!60}{[1.134,1.210]} & 1.250 & 25.171 / 34.533 & 0.7 \textcolor{black!60}{[0.0,1.8]} & 0.7 \textcolor{black!60}{[0.0,1.8]} \\
\deepseek & 0.403 \textcolor{black!60}{[0.290,0.599]} & 0.431 & 0.002 / 6.366 & 84.7 \textcolor{black!60}{[63.2,98.7]} & 84.7 \textcolor{black!60}{[63.2,98.7]} \\
\glm & 0.350 \textcolor{black!60}{[0.334,0.361]} & 0.374 & 0.002 / 0.231 & 92.7 \textcolor{black!60}{[90.1,95.7]} & 91.3 \textcolor{black!60}{[87.8,95.2]} \\
\qwenflash & 1.085 \textcolor{black!60}{[1.065,1.111]} & 1.160 & 15.227 / 20.295 & 0.0 \textcolor{black!60}{[0.0,0.0]} & 0.0 \textcolor{black!60}{[0.0,0.0]} \\
\qwenjson & 0.217 \textcolor{black!60}{[0.215,0.220]} & 0.232 & 0.001 / 0.002 & 100.0 \textcolor{black!60}{[100.0,100.0]} & \cellcolor{ebCoral!35}\textbf{100.0} \textcolor{black!60}{[100.0,100.0]} \\
\bottomrule
\end{tabular}}
\setlength{\tabcolsep}{0pt}\sbox0{\familytab}
\dimen0=\linewidth\advance\dimen0 by -1pt\advance\dimen0 by -\wd0\divide\dimen0 by 12
\ifdim\dimen0<1.5pt \dimen0=1.5pt\fi
\setlength{\tabcolsep}{\dimen0}\familytab
\end{table*}

\begin{figure*}[!htbp]\centering
\includegraphics[width=7in]{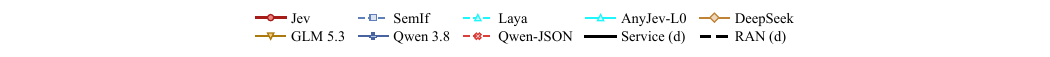}\\[-2pt]
\subfloat[Service completion]{\includegraphics[width=1.70in]{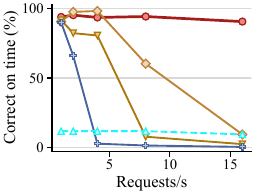}}\hfill
\subfloat[RAN slot load]{\includegraphics[width=1.70in]{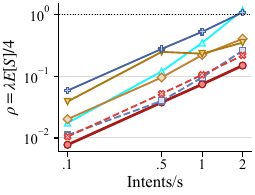}}\hfill
\subfloat[RAN queue tail]{\includegraphics[width=1.70in]{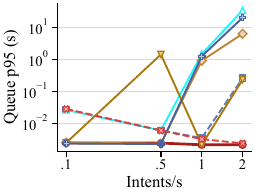}}\hfill
\subfloat[Native attainment]{\includegraphics[width=1.70in]{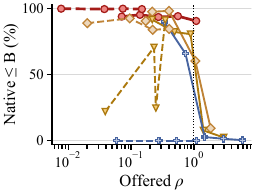}}
\caption{Native load traces borrowed and reanalysed from the Edge and RAN studies~\cite{li2026replacing,li2026intent}. Service admission reports correct on-time modeled service. RAN control reports a scheduled installable policy.}
\label{fig:rq-d}\end{figure*}

\begin{figure*}[!htbp]\centering
\includegraphics[width=7in]{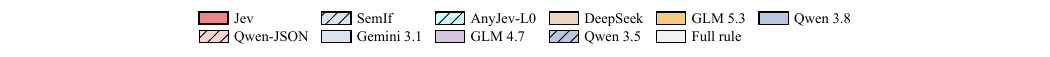}\\[-2pt]
\subfloat[RAN telemetry]{\includegraphics[width=1.70in]{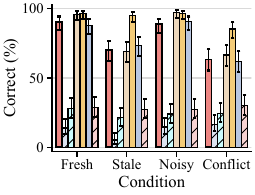}}\hfill
\subfloat[Placement refresh]{\includegraphics[width=1.70in]{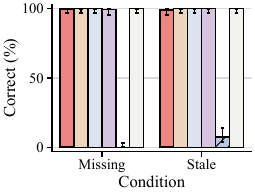}}\hfill
\subfloat[Version conflicts]{\includegraphics[width=1.70in]{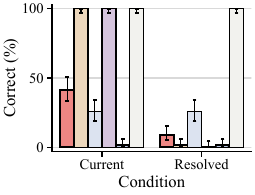}}\hfill
\subfloat[State representation]{\includegraphics[width=1.70in]{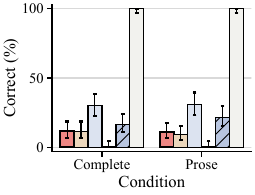}}
\caption{Observation-state correctness with 95\% Wilson intervals. Panel (a) reanalyses 150 RAN cases per condition~\cite{li2026intent}. Placement uses 120 cases per condition.}
\label{fig:rq-b}\end{figure*}
\begin{table*}[!htbp]
\centering\scriptsize
\caption{Observation-state correctness with 95\% Wilson intervals. The RAN rows are borrowed and reanalysed~\cite{li2026intent}. Placement rows are SoK-owned. Valid counts and return p50/p95 use contradictory telemetry or resolved conflict.}
\label{tab:rq-b-native}
\setlength{\tabcolsep}{3pt}
\renewcommand{\arraystretch}{0.94}
\setlength{\aboverulesep}{0pt}\setlength{\belowrulesep}{0pt}\setlength{\extrarowheight}{0pt}
\def\familytab{\begin{tabular}{lrrrrrr}
\toprule
\rowcolor{ebAmber!45}
\multicolumn{7}{l}{\textbf{\itshape RAN policy: fresh / stale / noisy / contradictory, 57 cells}} \\
\rowcolor{ebBlue}
Model & Fresh & Stale & Noisy & Contradictory & Valid / $n$ & \shortstack{Return (s) $\downarrow$\\p50 / p95} \\
\rowcolor{ebCoral!10}\jev & 90.7 \textcolor{black!60}{[84.9,94.4]} & 70.0 \textcolor{black!60}{[62.2,76.8]} & 88.7 \textcolor{black!60}{[82.6,92.8]} & 63.3 \textcolor{black!60}{[55.4,70.6]} & 150/150 & 0.292 / 0.408 \\
\semif & 14.0 \textcolor{black!60}{[9.3,20.5]} & 5.3 \textcolor{black!60}{[2.7,10.2]} & 14.7 \textcolor{black!60}{[9.9,21.2]} & 16.7 \textcolor{black!60}{[11.6,23.4]} & 150/150 & 0.601 / 0.713 \\
\anyjev & 28.0 \textcolor{black!60}{[21.4,35.7]} & 21.3 \textcolor{black!60}{[15.5,28.6]} & 24.0 \textcolor{black!60}{[17.9,31.4]} & 24.7 \textcolor{black!60}{[18.5,32.1]} & 150/150 & 0.492 / 0.504 \\
\deepseek & 96.0 \textcolor{black!60}{[91.5,98.2]} & 69.3 \textcolor{black!60}{[61.5,76.2]} & \cellcolor{ebCoral!35}\textbf{97.3} \textcolor{black!60}{[93.3,99.0]} & 66.7 \textcolor{black!60}{[58.8,73.7]} & 150/150 & 0.683 / 1.013 \\
\glm & \cellcolor{ebCoral!35}\textbf{96.7} \textcolor{black!60}{[92.4,98.6]} & \cellcolor{ebCoral!35}\textbf{95.3} \textcolor{black!60}{[90.7,97.7]} & 96.7 \textcolor{black!60}{[92.4,98.6]} & \cellcolor{ebCoral!35}\textbf{85.3} \textcolor{black!60}{[78.8,90.1]} & 137/150 & 1.851 / 2.975 \\
\qwenflash & 88.0 \textcolor{black!60}{[81.8,92.3]} & 73.3 \textcolor{black!60}{[65.7,79.8]} & 90.7 \textcolor{black!60}{[84.9,94.4]} & 62.0 \textcolor{black!60}{[54.0,69.4]} & 150/150 & 2.445 / 7.214 \\
\qwenjson & 28.7 \textcolor{black!60}{[22.0,36.4]} & 27.3 \textcolor{black!60}{[20.8,35.0]} & 27.3 \textcolor{black!60}{[20.8,35.0]} & 30.0 \textcolor{black!60}{[23.2,37.8]} & 150/150 & 0.478 / 0.486 \\
\midrule
\rowcolor{ebAmber!45}
\multicolumn{7}{l}{\textbf{\itshape Placement: missing / stale / conflict / resolved conflict}} \\
\rowcolor{ebBlue}
Model & Missing & Stale & Current conflict & Resolved conflict & Valid / $n$ & \shortstack{Return (s) $\downarrow$\\p50 / p95} \\
\rowcolor{ebCoral!10}\sokjev & \cellcolor{ebCoral!35}\textbf{100.0} \textcolor{black!60}{[96.9,100.0]} & 99.2 \textcolor{black!60}{[95.4,99.9]} & 41.7 \textcolor{black!60}{[33.2,50.6]} & 9.2 \textcolor{black!60}{[5.2,15.7]} & 120/120 & 0.374 / 0.590 \\
\deepseek & \cellcolor{ebCoral!35}\textbf{100.0} \textcolor{black!60}{[96.9,100.0]} & \cellcolor{ebCoral!35}\textbf{100.0} \textcolor{black!60}{[96.9,100.0]} & \cellcolor{ebCoral!35}\textbf{100.0} \textcolor{black!60}{[96.9,100.0]} & 1.7 \textcolor{black!60}{[0.5,5.9]} & 120/120 & 0.507 / 0.813 \\
\gemini & \cellcolor{ebCoral!35}\textbf{100.0} \textcolor{black!60}{[96.9,100.0]} & \cellcolor{ebCoral!35}\textbf{100.0} \textcolor{black!60}{[96.9,100.0]} & 25.8 \textcolor{black!60}{[18.8,34.3]} & 25.8 \textcolor{black!60}{[18.8,34.3]} & 120/120 & 1.015 / 1.276 \\
\sokglm & 99.2 \textcolor{black!60}{[95.4,99.9]} & \cellcolor{ebCoral!35}\textbf{100.0} \textcolor{black!60}{[96.9,100.0]} & \cellcolor{ebCoral!35}\textbf{100.0} \textcolor{black!60}{[96.9,100.0]} & 0.8 \textcolor{black!60}{[0.1,4.6]} & 120/120 & 0.861 / 6.439 \\
\sokqwen & 0.0 \textcolor{black!60}{[0.0,3.1]} & 7.5 \textcolor{black!60}{[4.0,13.6]} & 1.7 \textcolor{black!60}{[0.5,5.9]} & 1.7 \textcolor{black!60}{[0.5,5.9]} & 120/120 & 0.249 / 0.334 \\
Full rule & \cellcolor{ebCoral!35}\textbf{100.0} \textcolor{black!60}{[96.9,100.0]} & \cellcolor{ebCoral!35}\textbf{100.0} \textcolor{black!60}{[96.9,100.0]} & \cellcolor{ebCoral!35}\textbf{100.0} \textcolor{black!60}{[96.9,100.0]} & \cellcolor{ebCoral!35}\textbf{100.0} \textcolor{black!60}{[96.9,100.0]} & 120/120 & 0.000 / 0.000 \\
Per-stage rule & \cellcolor{ebCoral!35}\textbf{100.0} \textcolor{black!60}{[96.9,100.0]} & \cellcolor{ebCoral!35}\textbf{100.0} \textcolor{black!60}{[96.9,100.0]} & \cellcolor{ebCoral!35}\textbf{100.0} \textcolor{black!60}{[96.9,100.0]} & 50.0 \textcolor{black!60}{[41.2,58.8]} & 120/120 & 0.000 / 0.000 \\
Ungated & 0.0 \textcolor{black!60}{[0.0,3.1]} & 0.0 \textcolor{black!60}{[0.0,3.1]} & 0.0 \textcolor{black!60}{[0.0,3.1]} & \cellcolor{ebCoral!35}\textbf{100.0} \textcolor{black!60}{[96.9,100.0]} & 120/120 & 0.000 / 0.000 \\
First & \cellcolor{ebCoral!35}\textbf{100.0} \textcolor{black!60}{[96.9,100.0]} & \cellcolor{ebCoral!35}\textbf{100.0} \textcolor{black!60}{[96.9,100.0]} & \cellcolor{ebCoral!35}\textbf{100.0} \textcolor{black!60}{[96.9,100.0]} & 25.8 \textcolor{black!60}{[18.8,34.3]} & 120/120 & 0.000 / 0.000 \\
\bottomrule
\end{tabular}}
\setlength{\tabcolsep}{0pt}\sbox0{\familytab}
\dimen0=\linewidth\advance\dimen0 by -1pt\advance\dimen0 by -\wd0\divide\dimen0 by 14
\ifdim\dimen0<1.5pt \dimen0=1.5pt\fi
\setlength{\tabcolsep}{\dimen0}\familytab
\end{table*}

\begin{table*}[!htbp]
\centering\scriptsize
\caption{Contract-catalogue correctness (\%) with 95\% Wilson intervals. 48 cases per condition. Two new candidates form the replacement arm. Valid counts and return p50/p95 refer to absent coverage. N/A denotes an unmeasured condition. The rule baseline uses lexical matching.}
\label{tab:rq-c-contract}
\setlength{\tabcolsep}{3pt}
\renewcommand{\arraystretch}{0.94}
\setlength{\aboverulesep}{0pt}\setlength{\belowrulesep}{0pt}\setlength{\extrarowheight}{0pt}
\def\familytab{\begin{tabular}{lrrrrrr}
\toprule
\rowcolor{ebAmber!45}
\multicolumn{7}{l}{\textbf{\itshape Contract catalogue: stable / rekeyed / two new candidates / absent}} \\
\rowcolor{ebBlue}
Model & Stable & Rekeyed & Two new & Absent & Valid / $n$ & \shortstack{Return (s) $\downarrow$\\p50 / p95} \\
\rowcolor{ebCoral!10}\sokjev & \cellcolor{ebCoral!35}\textbf{100.0} \textcolor{black!60}{[92.6,100.0]} & \cellcolor{ebCoral!35}\textbf{100.0} \textcolor{black!60}{[92.6,100.0]} & \cellcolor{ebCoral!35}\textbf{100.0} \textcolor{black!60}{[92.6,100.0]} & 93.8 \textcolor{black!60}{[83.2,97.9]} & 48/48 & 0.318 / 0.511 \\
\deepseek & 97.9 \textcolor{black!60}{[89.1,99.6]} & 89.6 \textcolor{black!60}{[77.8,95.5]} & 77.1 \textcolor{black!60}{[63.5,86.7]} & 6.2 \textcolor{black!60}{[2.1,16.8]} & 48/48 & 0.413 / 2.696 \\
\gemini & \cellcolor{ebCoral!35}\textbf{100.0} \textcolor{black!60}{[92.6,100.0]} & \cellcolor{ebCoral!35}\textbf{100.0} \textcolor{black!60}{[92.6,100.0]} & \cellcolor{ebCoral!35}\textbf{100.0} \textcolor{black!60}{[92.6,100.0]} & 29.2 \textcolor{black!60}{[18.2,43.2]} & 48/48 & 0.934 / 1.600 \\
\sokglm & 56.2 \textcolor{black!60}{[42.3,69.3]} & 39.6 \textcolor{black!60}{[27.0,53.7]} & 2.1 \textcolor{black!60}{[0.4,10.9]} & 33.3 \textcolor{black!60}{[21.7,47.5]} & 48/48 & 0.825 / 2.943 \\
\sokqwen & 54.2 \textcolor{black!60}{[40.3,67.4]} & 54.2 \textcolor{black!60}{[40.3,67.4]} & 64.6 \textcolor{black!60}{[50.4,76.6]} & 2.1 \textcolor{black!60}{[0.4,10.9]} & 48/48 & 0.243 / 0.331 \\
Lexical rule & 50.0 \textcolor{black!60}{[36.4,63.6]} & 50.0 \textcolor{black!60}{[36.4,63.6]} & 50.0 \textcolor{black!60}{[36.4,63.6]} & \cellcolor{ebCoral!35}\textbf{100.0} \textcolor{black!60}{[92.6,100.0]} & 48/48 & 0.000 / 0.000 \\
MiniLM & 37.5 \textcolor{black!60}{[25.2,51.6]} & 37.5 \textcolor{black!60}{[25.2,51.6]} & 27.1 \textcolor{black!60}{[16.6,41.0]} & 50.0 \textcolor{black!60}{[36.4,63.6]} & 48/48 & 0.041 / 0.046 \\
First & 22.9 \textcolor{black!60}{[13.3,36.5]} & 29.2 \textcolor{black!60}{[18.2,43.2]} & N/A & 0.0 \textcolor{black!60}{[0.0,7.4]} & 48/48 & N/A / N/A \\
\bottomrule
\end{tabular}}
\setlength{\tabcolsep}{0pt}\sbox0{\familytab}
\dimen0=\linewidth\advance\dimen0 by -1pt\advance\dimen0 by -\wd0\divide\dimen0 by 14
\ifdim\dimen0<1.5pt \dimen0=1.5pt\fi
\setlength{\tabcolsep}{\dimen0}\familytab
\end{table*}

\begin{table*}[!htbp]
\centering\scriptsize
\caption{Public-gate minus self-check correctness (pp). Intervals are paired stratified-bootstrap 95\% estimates. $p_{\mathrm{Holm}}$ uses the 45-comparison workflow family. Validity and return time describe the public-gate arm. Instructions and refresh actions also change. Best correctness per condition is in bold.}
\label{tab:rq-e-paired}
\setlength{\tabcolsep}{3pt}
\renewcommand{\arraystretch}{0.94}
\setlength{\aboverulesep}{0pt}\setlength{\belowrulesep}{0pt}\setlength{\extrarowheight}{0pt}
\def\familytab{\begin{tabular}{lrrrrrr}
\toprule
\rowcolor{ebBlue}
Method & Self-check & Public gate & $\Delta$ pp [95\% CI] & $p_{\mathrm{Holm}}$ & Valid / $n$ & Return p50 / p95 (s) \\
\midrule
\rowcolor{ebAmber!45}
\multicolumn{7}{l}{\textbf{\itshape Independent quotas}} \\
\rowcolor{ebCoral!10}\sokjev & 29/120 & 84/120 & 45.8 \textcolor{black!60}{[37.5,54.2]} & $2.331\times10^{-15}$ & 120/120 & 0.358 / 0.604 \\
\deepseek & 19/120 & 60/120 & 34.2 \textcolor{black!60}{[25.0,43.3]} & $9.366\times10^{-9}$ & 120/120 & 0.509 / 1.175 \\
\gemini & 40/120 & 69/120 & 24.2 \textcolor{black!60}{[14.2,34.2]} & 0.0008467 & 120/120 & 1.009 / 1.241 \\
\sokglm & 1/120 & 71/120 & 58.3 \textcolor{black!60}{[49.2,66.7]} & $1.391\times10^{-18}$ & 119/120 & 0.843 / 3.553 \\
\sokqwen & 30/120 & 52/120 & 18.3 \textcolor{black!60}{[10.8,26.7]} & 0.003618 & 120/120 & 0.311 / 0.411 \\
Full rule & \cellcolor{ebCoral!35}\textbf{120/120} & \cellcolor{ebCoral!35}\textbf{120/120} & 0.0 \textcolor{black!60}{[0.0,0.0]} & 1 & 120/120 & 0.000 / 0.000 \\
Per-stage rule & \cellcolor{ebCoral!35}\textbf{120/120} & \cellcolor{ebCoral!35}\textbf{120/120} & 0.0 \textcolor{black!60}{[0.0,0.0]} & 1 & 120/120 & 0.000 / 0.000 \\
Ungated & \cellcolor{ebCoral!35}\textbf{120/120} & \cellcolor{ebCoral!35}\textbf{120/120} & 0.0 \textcolor{black!60}{[0.0,0.0]} & 1 & 120/120 & 0.000 / 0.000 \\
First & 28/120 & 28/120 & 0.0 \textcolor{black!60}{[0.0,0.0]} & 1 & 120/120 & 0.000 / 0.000 \\
\midrule
\rowcolor{ebAmber!45}
\multicolumn{7}{l}{\textbf{\itshape Shared quotas}} \\
\rowcolor{ebCoral!10}\sokjev & 14/120 & 42/120 & 23.3 \textcolor{black!60}{[15.8,30.8]} & $2.079\times10^{-6}$ & 120/120 & 0.353 / 0.715 \\
\deepseek & 14/120 & 33/120 & 15.8 \textcolor{black!60}{[8.3,23.3]} & 0.004852 & 120/120 & 0.501 / 1.087 \\
\gemini & 36/120 & 40/120 & 3.3 \textcolor{black!60}{[-5.8,12.5]} & 1 & 120/120 & 1.009 / 1.234 \\
\sokglm & 1/120 & 50/120 & 40.8 \textcolor{black!60}{[32.5,49.2]} & $1.457\times10^{-13}$ & 120/120 & 0.840 / 8.439 \\
\sokqwen & 20/120 & 39/120 & 15.8 \textcolor{black!60}{[8.3,23.3]} & 0.009012 & 120/120 & 0.319 / 0.389 \\
Full rule & \cellcolor{ebCoral!35}\textbf{120/120} & \cellcolor{ebCoral!35}\textbf{120/120} & 0.0 \textcolor{black!60}{[0.0,0.0]} & 1 & 120/120 & 0.000 / 0.000 \\
Per-stage rule & 60/120 & 60/120 & 0.0 \textcolor{black!60}{[0.0,0.0]} & 1 & 120/120 & 0.000 / 0.000 \\
Ungated & \cellcolor{ebCoral!35}\textbf{120/120} & \cellcolor{ebCoral!35}\textbf{120/120} & 0.0 \textcolor{black!60}{[0.0,0.0]} & 1 & 120/120 & 0.000 / 0.000 \\
First & 31/120 & 31/120 & 0.0 \textcolor{black!60}{[0.0,0.0]} & 1 & 120/120 & 0.000 / 0.000 \\
\midrule
\rowcolor{ebAmber!45}
\multicolumn{7}{l}{\textbf{\itshape Shared quotas + fault-domain separation}} \\
\rowcolor{ebCoral!10}\sokjev & 6/120 & 45/120 & 32.5 \textcolor{black!60}{[25.0,40.0]} & $1.455\times10^{-10}$ & 120/120 & 0.375 / 0.585 \\
\deepseek & 4/120 & 17/120 & 10.8 \textcolor{black!60}{[5.8,15.8]} & 0.007324 & 120/120 & 0.506 / 0.999 \\
\gemini & 42/120 & 61/120 & 15.8 \textcolor{black!60}{[5.8,25.8]} & 0.1404 & 120/120 & 1.021 / 1.276 \\
\sokglm & 0/120 & 34/120 & 28.3 \textcolor{black!60}{[20.8,35.8]} & $4.54\times10^{-9}$ & 120/120 & 0.808 / 3.075 \\
\sokqwen & 4/120 & 14/120 & 8.3 \textcolor{black!60}{[3.3,13.3]} & 0.1587 & 120/120 & 0.322 / 0.401 \\
Full rule & \cellcolor{ebCoral!35}\textbf{120/120} & \cellcolor{ebCoral!35}\textbf{120/120} & 0.0 \textcolor{black!60}{[0.0,0.0]} & 1 & 120/120 & 0.000 / 0.000 \\
Per-stage rule & \cellcolor{ebCoral!35}\textbf{120/120} & \cellcolor{ebCoral!35}\textbf{120/120} & 0.0 \textcolor{black!60}{[0.0,0.0]} & 1 & 120/120 & 0.000 / 0.000 \\
Ungated & \cellcolor{ebCoral!35}\textbf{120/120} & \cellcolor{ebCoral!35}\textbf{120/120} & 0.0 \textcolor{black!60}{[0.0,0.0]} & 1 & 120/120 & 0.000 / 0.000 \\
First & 32/120 & 32/120 & 0.0 \textcolor{black!60}{[0.0,0.0]} & 1 & 120/120 & 0.000 / 0.000 \\
\bottomrule
\end{tabular}}
\setlength{\tabcolsep}{0pt}\sbox0{\familytab}
\dimen0=\linewidth\advance\dimen0 by -1pt\advance\dimen0 by -\wd0\divide\dimen0 by 14
\ifdim\dimen0<1.5pt \dimen0=1.5pt\fi
\setlength{\tabcolsep}{\dimen0}\familytab
\end{table*}

\begin{table*}[!htbp]
\centering\scriptsize
\caption{Structural growth on three native axes. Contract and RAN rows are borrowed and reanalysed~\cite{li2026replacing,li2026intent}. Policy rows are SoK-owned. Correctness is in percent with 95\% Wilson intervals. Contract inputs differ, so their $\Delta$ is N/A.}
\label{tab:rq-a-native}
\setlength{\tabcolsep}{3pt}
\renewcommand{\arraystretch}{0.94}
\setlength{\aboverulesep}{0pt}\setlength{\belowrulesep}{0pt}\setlength{\extrarowheight}{0pt}
\def\familytab{\begin{tabular}{lrrrrrrr}
\toprule
\rowcolor{ebAmber!45}
\multicolumn{8}{l}{\textbf{\itshape Service contract: 4 / 6 / 8 fields, low constraint density}} \\
\rowcolor{ebBlue}
Model & 4 fields & 6 fields & 8 fields & N/A & \shortstack{$\Delta_{\mathrm{last-first}}$\\pp [CI]} & Valid / $n$ & \shortstack{Return (s) $\downarrow$\\p50 / p95} \\
\rowcolor{ebCoral!10}\jev & 92.3 \textcolor{black!60}{[88.8,94.8]} & 78.7 \textcolor{black!60}{[73.7,82.9]} & 57.7 \textcolor{black!60}{[52.0,63.1]} & N/A & N/A & 300/300 & 0.272 / 0.371 \\
\semif & 54.3 \textcolor{black!60}{[48.7,59.9]} & 40.0 \textcolor{black!60}{[34.6,45.6]} & 13.3 \textcolor{black!60}{[9.9,17.6]} & N/A & N/A & 300/300 & 0.133 / 0.194 \\
\laya & 25.3 \textcolor{black!60}{[20.7,30.5]} & 7.7 \textcolor{black!60}{[5.2,11.2]} & 1.0 \textcolor{black!60}{[0.3,2.9]} & N/A & N/A & 300/300 & 0.057 / 0.242 \\
\deepseek & 98.3 \textcolor{black!60}{[96.2,99.3]} & 95.7 \textcolor{black!60}{[92.7,97.5]} & 90.0 \textcolor{black!60}{[86.1,92.9]} & N/A & N/A & 300/300 & 0.460 / 0.591 \\
\glm & 89.7 \textcolor{black!60}{[85.7,92.6]} & 96.3 \textcolor{black!60}{[93.6,97.9]} & 80.3 \textcolor{black!60}{[75.5,84.4]} & N/A & N/A & 300/300 & 0.930 / 1.954 \\
\qwenflash & 98.3 \textcolor{black!60}{[96.2,99.3]} & 90.0 \textcolor{black!60}{[86.1,92.9]} & 82.3 \textcolor{black!60}{[77.6,86.2]} & N/A & N/A & 298/300 & 2.092 / 3.525 \\
\qwenjson & 53.0 \textcolor{black!60}{[47.4,58.6]} & 48.3 \textcolor{black!60}{[42.7,54.0]} & 30.0 \textcolor{black!60}{[25.1,35.4]} & N/A & N/A & 300/300 & 1.507 / 1.789 \\
Rule & 4.0 \textcolor{black!60}{[2.3,6.9]} & 0.3 \textcolor{black!60}{[0.1,1.9]} & 0.0 \textcolor{black!60}{[0.0,1.3]} & N/A & N/A & 300/300 & 0.000 / 0.000 \\
Oracle & \cellcolor{ebCoral!35}\textbf{100.0} \textcolor{black!60}{[98.7,100.0]} & \cellcolor{ebCoral!35}\textbf{100.0} \textcolor{black!60}{[98.7,100.0]} & \cellcolor{ebCoral!35}\textbf{100.0} \textcolor{black!60}{[98.7,100.0]} & N/A & N/A & 300/300 & 0.000 / 0.000 \\
\midrule
\rowcolor{ebAmber!45}
\multicolumn{8}{l}{\textbf{\itshape RAN policy: 3 / 7 / 21 / 57 cells, fresh, state-dependent target}} \\
\rowcolor{ebBlue}
Model & 3 cells & 7 cells & 21 cells & 57 cells & \shortstack{$\Delta_{\mathrm{last-first}}$\\pp [CI]} & Valid / $n$ & \shortstack{Return (s) $\downarrow$\\p50 / p95} \\
\rowcolor{ebCoral!10}\jev & \cellcolor{ebCoral!35}\textbf{100.0} \textcolor{black!60}{[97.5,100.0]} & 96.7 \textcolor{black!60}{[92.4,98.6]} & 90.0 \textcolor{black!60}{[84.2,93.8]} & 90.7 \textcolor{black!60}{[84.9,94.4]} & -9.3 \textcolor{black!60}{[-14.0,-5.3]} & 150/150 & 0.281 / 0.378 \\
\semif & 24.7 \textcolor{black!60}{[18.5,32.1]} & 12.7 \textcolor{black!60}{[8.3,18.9]} & 13.3 \textcolor{black!60}{[8.8,19.7]} & 14.0 \textcolor{black!60}{[9.3,20.5]} & -10.7 \textcolor{black!60}{[-18.0,-3.3]} & 150/150 & 0.432 / 0.524 \\
\anyjev & 59.3 \textcolor{black!60}{[51.3,66.9]} & 36.7 \textcolor{black!60}{[29.4,44.6]} & 28.0 \textcolor{black!60}{[21.4,35.7]} & 28.0 \textcolor{black!60}{[21.4,35.7]} & -31.3 \textcolor{black!60}{[-40.0,-22.0]} & 150/150 & 0.404 / 0.422 \\
\deepseek & 98.0 \textcolor{black!60}{[94.3,99.3]} & 98.7 \textcolor{black!60}{[95.3,99.6]} & 96.7 \textcolor{black!60}{[92.4,98.6]} & 96.0 \textcolor{black!60}{[91.5,98.2]} & -2.0 \textcolor{black!60}{[-4.7,0.0]} & 150/150 & 0.522 / 0.825 \\
\glm & 99.3 \textcolor{black!60}{[96.3,99.9]} & \cellcolor{ebCoral!35}\textbf{99.3} \textcolor{black!60}{[96.3,99.9]} & \cellcolor{ebCoral!35}\textbf{98.0} \textcolor{black!60}{[94.3,99.3]} & \cellcolor{ebCoral!35}\textbf{96.7} \textcolor{black!60}{[92.4,98.6]} & -2.7 \textcolor{black!60}{[-6.0,0.0]} & 150/150 & 1.234 / 2.248 \\
\qwenflash & 99.3 \textcolor{black!60}{[96.3,99.9]} & 97.3 \textcolor{black!60}{[93.3,99.0]} & 92.7 \textcolor{black!60}{[87.3,95.9]} & 88.0 \textcolor{black!60}{[81.8,92.3]} & -11.3 \textcolor{black!60}{[-16.7,-6.7]} & 150/150 & 2.213 / 3.053 \\
\qwenjson & 63.3 \textcolor{black!60}{[55.4,70.6]} & 47.3 \textcolor{black!60}{[39.5,55.3]} & 36.7 \textcolor{black!60}{[29.4,44.6]} & 28.7 \textcolor{black!60}{[22.0,36.4]} & -34.7 \textcolor{black!60}{[-43.3,-26.7]} & 150/150 & 0.432 / 0.440 \\
\midrule
\rowcolor{ebAmber!45}
\multicolumn{8}{l}{\textbf{\itshape Access policy: 4 / 16 / 64 distinct requirements}} \\
\rowcolor{ebBlue}
Model & 4 requirements & 16 requirements & 64 requirements & N/A & \shortstack{$\Delta_{\mathrm{last-first}}$\\pp [CI]} & Valid / $n$ & \shortstack{Return (s) $\downarrow$\\p50 / p95} \\
\rowcolor{ebCoral!10}\sokjev & 89.2 \textcolor{black!60}{[82.3,93.6]} & 80.8 \textcolor{black!60}{[72.9,86.9]} & 39.2 \textcolor{black!60}{[30.9,48.1]} & N/A & -50.0 \textcolor{black!60}{[-59.2,-40.8]} & 120/120 & 0.406 / 0.758 \\
\deepseek & 67.5 \textcolor{black!60}{[58.7,75.2]} & 52.5 \textcolor{black!60}{[43.6,61.2]} & 41.7 \textcolor{black!60}{[33.2,50.6]} & N/A & -25.8 \textcolor{black!60}{[-37.5,-14.2]} & 120/120 & 0.634 / 1.601 \\
\gemini & 71.7 \textcolor{black!60}{[63.0,79.0]} & 63.3 \textcolor{black!60}{[54.4,71.4]} & 65.8 \textcolor{black!60}{[57.0,73.7]} & N/A & -5.8 \textcolor{black!60}{[-16.7,5.0]} & 120/120 & 1.140 / 1.393 \\
\sokglm & 68.3 \textcolor{black!60}{[59.6,76.0]} & 52.5 \textcolor{black!60}{[43.6,61.2]} & 37.5 \textcolor{black!60}{[29.4,46.4]} & N/A & -30.8 \textcolor{black!60}{[-40.8,-20.8]} & 120/120 & 1.033 / 2.793 \\
\sokqwen & 18.3 \textcolor{black!60}{[12.4,26.2]} & 4.2 \textcolor{black!60}{[1.8,9.4]} & 1.7 \textcolor{black!60}{[0.5,5.9]} & N/A & -16.7 \textcolor{black!60}{[-23.3,-10.0]} & 120/120 & 0.601 / 0.688 \\
Full rule & \cellcolor{ebCoral!35}\textbf{100.0} \textcolor{black!60}{[96.9,100.0]} & \cellcolor{ebCoral!35}\textbf{100.0} \textcolor{black!60}{[96.9,100.0]} & \cellcolor{ebCoral!35}\textbf{100.0} \textcolor{black!60}{[96.9,100.0]} & N/A & 0.0 \textcolor{black!60}{[0.0,0.0]} & 120/120 & 0.000 / 0.000 \\
Cheapest & 77.5 \textcolor{black!60}{[69.2,84.1]} & 60.8 \textcolor{black!60}{[51.9,69.1]} & 37.5 \textcolor{black!60}{[29.4,46.4]} & N/A & -40.0 \textcolor{black!60}{[-49.2,-31.7]} & 120/120 & 0.000 / 0.000 \\
First & 30.8 \textcolor{black!60}{[23.3,39.6]} & 29.2 \textcolor{black!60}{[21.8,37.8]} & 25.0 \textcolor{black!60}{[18.1,33.4]} & N/A & -5.8 \textcolor{black!60}{[-15.0,3.3]} & 120/120 & 0.000 / 0.000 \\
\bottomrule
\end{tabular}}
\setlength{\tabcolsep}{0pt}\sbox0{\familytab}
\dimen0=\linewidth\advance\dimen0 by -1pt\advance\dimen0 by -\wd0\divide\dimen0 by 16
\ifdim\dimen0<1.5pt \dimen0=1.5pt\fi
\setlength{\tabcolsep}{\dimen0}\familytab
\end{table*}

\begin{table*}[!htbp]
\centering\scriptsize
\caption{Paired pure-length effects. Contract effects are borrowed and reanalysed from the Edge study~\cite{li2026replacing}. Policy effects are SoK-owned. Exact McNemar p-values use the complete 89-comparison structure family. The conditions retain the requirement set.}
\label{tab:rq-a-length}
\setlength{\tabcolsep}{3pt}
\renewcommand{\arraystretch}{0.94}
\setlength{\aboverulesep}{0pt}\setlength{\belowrulesep}{0pt}\setlength{\extrarowheight}{0pt}
\def\familytab{\begin{tabular}{lrrrrr}
\toprule
\rowcolor{ebBlue}
Method & Condition & n & $\Delta$ correctness pp [CI] & $p_{\mathrm{Holm}}$ & Paired $\Delta$ return (s) [CI] \\
\midrule
\rowcolor{ebAmber!45}
\multicolumn{6}{l}{\textbf{\itshape Service-contract padding: 16384 tokens minus base}} \\
\rowcolor{ebCoral!10}\jev & Padding & 300 & -2.7 \textcolor{black!60}{[-5.7,0.3]} & 1 & 0.149 \textcolor{black!60}{[0.135,0.157]} \\
\semif & Padding & 300 & -15.7 \textcolor{black!60}{[-21.3,-9.7]} & $4.31\times10^{-5}$ & 1.034 \textcolor{black!60}{[1.032,1.036]} \\
\laya & Padding & 300 & -9.3 \textcolor{black!60}{[-13.0,-6.0]} & $1.823\times10^{-5}$ & 0.082 \textcolor{black!60}{[0.081,0.082]} \\
\deepseek & Padding & 300 & -1.7 \textcolor{black!60}{[-3.3,-0.3]} & 1 & 0.259 \textcolor{black!60}{[0.249,0.276]} \\
\glm & Padding & 300 & -4.7 \textcolor{black!60}{[-7.3,-2.0]} & 0.07742 & 0.883 \textcolor{black!60}{[0.781,0.962]} \\
\qwenflash & Padding & 300 & 0.0 \textcolor{black!60}{[-2.3,2.0]} & 1 & 1.191 \textcolor{black!60}{[1.135,1.272]} \\
\qwenjson & Padding & 300 & -30.0 \textcolor{black!60}{[-35.3,-24.7]} & $2.117\times10^{-21}$ & 0.781 \textcolor{black!60}{[0.774,0.791]} \\
Rule & Padding & 300 & 0.0 \textcolor{black!60}{[0.0,0.0]} & 1 & 0.002 \textcolor{black!60}{[0.002,0.003]} \\
Oracle & Padding & 300 & 0.0 \textcolor{black!60}{[0.0,0.0]} & 1 & 0.000 \textcolor{black!60}{[0.000,0.000]} \\
\midrule
\rowcolor{ebAmber!45}
\multicolumn{6}{l}{\textbf{\itshape Access policy: irrelevant context or repetition minus base}} \\
\rowcolor{ebCoral!10}\sokjev & Irrelevant text & 120 & 0.8 \textcolor{black!60}{[-1.7,3.3]} & 1 & 0.044 \textcolor{black!60}{[0.022,0.084]} \\
\rowcolor{ebCoral!10}\sokjev & Repetition & 120 & 3.3 \textcolor{black!60}{[-0.8,8.3]} & 1 & 0.048 \textcolor{black!60}{[0.013,0.084]} \\
\deepseek & Irrelevant text & 120 & -5.0 \textcolor{black!60}{[-10.8,0.8]} & 1 & 0.030 \textcolor{black!60}{[-0.011,0.096]} \\
\deepseek & Repetition & 120 & -3.3 \textcolor{black!60}{[-10.8,4.2]} & 1 & 0.099 \textcolor{black!60}{[0.011,0.124]} \\
\gemini & Irrelevant text & 120 & 1.7 \textcolor{black!60}{[-6.7,9.2]} & 1 & 0.158 \textcolor{black!60}{[0.118,0.219]} \\
\gemini & Repetition & 120 & 1.7 \textcolor{black!60}{[-7.5,10.8]} & 1 & 0.172 \textcolor{black!60}{[0.148,0.221]} \\
\sokglm & Irrelevant text & 120 & -3.3 \textcolor{black!60}{[-8.3,0.8]} & 1 & 0.141 \textcolor{black!60}{[0.053,0.204]} \\
\sokglm & Repetition & 120 & -5.0 \textcolor{black!60}{[-11.7,1.7]} & 1 & 0.247 \textcolor{black!60}{[0.136,0.314]} \\
\sokqwen & Irrelevant text & 120 & -2.5 \textcolor{black!60}{[-8.3,3.3]} & 1 & 0.239 \textcolor{black!60}{[0.235,0.243]} \\
\sokqwen & Repetition & 120 & 8.3 \textcolor{black!60}{[0.8,15.8]} & 1 & 0.653 \textcolor{black!60}{[0.647,0.662]} \\
Full rule & Irrelevant text & 120 & 0.0 \textcolor{black!60}{[0.0,0.0]} & 1 & 0.000 \textcolor{black!60}{[0.000,0.000]} \\
Full rule & Repetition & 120 & 0.0 \textcolor{black!60}{[0.0,0.0]} & 1 & 0.000 \textcolor{black!60}{[0.000,0.000]} \\
Cheapest & Irrelevant text & 120 & 0.0 \textcolor{black!60}{[0.0,0.0]} & 1 & 0.000 \textcolor{black!60}{[0.000,0.000]} \\
Cheapest & Repetition & 120 & 0.0 \textcolor{black!60}{[0.0,0.0]} & 1 & 0.000 \textcolor{black!60}{[0.000,0.000]} \\
First & Irrelevant text & 120 & 0.0 \textcolor{black!60}{[0.0,0.0]} & 1 & 0.000 \textcolor{black!60}{[0.000,0.000]} \\
First & Repetition & 120 & 0.0 \textcolor{black!60}{[0.0,0.0]} & 1 & 0.000 \textcolor{black!60}{[0.000,0.000]} \\
\bottomrule
\end{tabular}}
\setlength{\tabcolsep}{0pt}\sbox0{\familytab}
\dimen0=\linewidth\advance\dimen0 by -1pt\advance\dimen0 by -\wd0\divide\dimen0 by 12
\ifdim\dimen0<1.5pt \dimen0=1.5pt\fi
\setlength{\tabcolsep}{\dimen0}\familytab
\end{table*}

\FloatBarrier
\section{Three-Axis Literature Map}\label{app:literature}
Table~\ref{tab:e0-systematization} maps the 139 families to their decision interfaces, execution paths and check responsibilities. Each row combines the reviewed tasks and versions of a family. Checks are qualified by component, test or formal model.
\begingroup\footnotesize
\setlength{\tabcolsep}{3pt}
\setlength{\aboverulesep}{0pt}\setlength{\belowrulesep}{0pt}
\renewcommand{\arraystretch}{1.10}
\setlength{\LTcapwidth}{\textwidth}
\newlength{\soklitwidth}\setlength{\soklitwidth}{\dimexpr\textwidth-1pt-8\tabcolsep\relax}
\begin{longtable}{>{\raggedright\arraybackslash}p{.09\soklitwidth}>{\centering\arraybackslash}p{.07\soklitwidth}>{\raggedright\arraybackslash}p{.36\soklitwidth}>{\raggedright\arraybackslash}p{.48\soklitwidth}}
\caption{Three-axis literature map of 139 paper families, ordered by the lowest reference number in each row. S: selection; G: generation; C: deterministic computation; U: interface unspecified. Multiple letters retain multiple interfaces. Check labels distinguish observation state, feasibility or related checks, and candidate coverage. Unlisted check types have unresolved ownership. NE marks rows with all three unresolved.}\label{tab:e0-systematization}\\
\toprule
\rowcolor{ebBlue} Work & Interface & Task and execution path & Check responsibility \\
\midrule\endfirsthead
\caption[]{Three-axis literature map (continued). S: selection; G: generation; C: computation; U: unspecified. Unlisted checks and NE denote unresolved ownership.}\\
\toprule
\rowcolor{ebBlue} Work & Interface & Task and execution path & Check responsibility \\
\midrule\endhead
\bottomrule\endfoot
\cite{li2026replacing} & S/G/C & Intent interpretation to edge admission/execution & \emph{State:} Scheduler reads queues/workers; \emph{Feasibility:} Scheduler: predicted feasibility; \emph{Coverage:} Scheduler: reject if no eligible node \\
\addlinespace[1pt]
\cite{wei2025inta} & G & Retrieved manuals to multivendor command-line output & \emph{Feasibility:} INTA: syntax matching; LLM: semantic equivalence \\
\addlinespace[1pt]
\cite{miyaoka2025chatdriven,e0sourceS13} & S/G/C & Intent parameters to placement/routing solver & \emph{Feasibility:} Integer linear program: resource/placement constraints; \emph{Coverage:} Optimizer: joint request-subset admission \\
\addlinespace[1pt]
\cite{wang2025network,e0arxiv250701997} & G & Tool-driven anomaly localization and diagnosis & \textit{NE} \\
\addlinespace[1pt]
\cite{manias2024intentbased} & S/G & Intent extraction; proposed core placement & \textit{NE} \\
\addlinespace[1pt]
\cite{nam2026llmbased} & G & Deployment-workflow generation and twin testing & \textit{NE} \\
\addlinespace[1pt]
\cite{wang2024netconfeval,e0arxiv230906342} & G & Task-specific policy, API and routing generation & \emph{Feasibility:} LLM diagnosis; task-specific tests/verifiers \\
\addlinespace[1pt]
\cite{el-hassany2018netcomplete} & C & Routing-sketch completion before activation & \emph{Feasibility:} NetComplete: sketch and routing constraints \\
\addlinespace[1pt]
\cite{hossain2025netintent} & G/C & Flow generation, installation and drift repair & \emph{Feasibility:} LLM: conflict detection; controller: resolution \\
\addlinespace[1pt]
\cite{clark2026jev} & S/G & Agent diagnosis with ranked tests and evidence gates & \emph{State:} Planner reads fresh snapshot; review: current failure; \emph{Feasibility:} Jev review: cause, restoration and durability \\
\addlinespace[1pt]
\cite{li2026fast} & S/G/C & Versioned intent contracts to edge placement & \emph{State:} Controller versions; scheduler reads queues; \emph{Feasibility:} Execution layer: permissions; scheduler: placement \\
\addlinespace[1pt]
\cite{asadli2026evaluating} & G & Offline configuration repair with feedback & \emph{Feasibility:} Batfish/Config2Spec: final scoring; optional agent feedback \\
\addlinespace[1pt]
\cite{angi2025llnet} & G/C & Intent JSON to confirmed switch rules & \emph{Feasibility:} Network Manager: installed-intent conflicts; \emph{Coverage:} Intent Converter: devices/scenarios \\
\addlinespace[1pt]
\cite{beckett2016dont} & C & Policy compilation to BGP configurations & \emph{Feasibility:} Propane: permitted paths/preferences \\
\addlinespace[1pt]
\cite{brodimas2025intentbased} & S/G & Agent plans to cluster/service deployment & \textit{NE} \\
\addlinespace[1pt]
\cite{chen2022softwaredefined} & S/C & Offline device-model and parameter mapping & \emph{Feasibility:} NAssim: command-line hierarchy and device-model tests \\
\addlinespace[1pt]
\cite{dinh2025endtoend} & G & Business intent to RAN/core service configuration & \textit{NE} \\
\addlinespace[1pt]
\cite{dzeparoska2024intent} & S/G & Intent fulfillment and drift-triggered repair & \emph{Feasibility:} LLM validator: policy-tree review \\
\addlinespace[1pt]
\cite{el-hassany2017networkwide} & C & Solver-based network configuration synthesis & \emph{Feasibility:} SyNET: modeled forwarding constraints; \emph{Coverage:} Solver: solution if one exists in model \\
\addlinespace[1pt]
\cite{islam2026intent} & G & State-grounded service-objective construction & \textit{NE} \\
\addlinespace[1pt]
\cite{jacobs2021hey} & G/S/C & Confirmed intent to compiler/controller & \emph{Feasibility:} Exploratory classifier: intent contradictions \\
\addlinespace[1pt]
\cite{jha2025itbench} & G & Network-fault diagnosis and mitigation in ITBench & \textit{NE} \\
\addlinespace[1pt]
\cite{li2025jaunt} & S & Per-request tool selection with latency prediction & \textit{NE} \\
\addlinespace[1pt]
\cite{e0arxiv260822644,manias2024semantic} & S/G & Static route selection and dynamic parameters & \textit{NE} \\
\addlinespace[1pt]
\cite{mcnamara2023nlp} & S & Registered workflow/model to slice or network function virtualization (NFV) action & \textit{NE} \\
\addlinespace[1pt]
\cite{mondal2023what} & G & Configuration translation/generation with feedback & \emph{Feasibility:} Campion: equivalence; Batfish: routing properties \\
\addlinespace[1pt]
\cite{njah2023intentbased} & G & Intent entity extraction; proposed SDN activation & \textit{NE} \\
\addlinespace[1pt]
\cite{ramanathan2023practical} & C & BGP compilation, staging and activation & \emph{Feasibility:} Aura: switch-emulation policy tests \\
\addlinespace[1pt]
\cite{schneider2021snowcap} & C & Safe update ordering with convergence waits & \emph{Feasibility:} Snowcap: temporal safety over modeled states \\
\addlinespace[1pt]
\cite{tian2019safely} & C & Access-control-list verification, repair and migration & \emph{Feasibility:} Jinjing: formal reachability/placement constraints \\
\addlinespace[1pt]
\cite{tseng2026faultbench} & G & Tool-driven diagnosis and proposed remediation & \emph{State:} Agent tests ticket premises against observations \\
\addlinespace[1pt]
\cite{e0sourceS02,tu2025intentbased} & G/C & Confirmed intent translation to NFV deployment & \emph{Feasibility:} NFV-Intent/NI-MANO: capabilities and resource flavors \\
\addlinespace[1pt]
\cite{wu2024netllm} & S & Learned bitrate and cluster-action selection & \textit{NE} \\
\addlinespace[1pt]
\cite{zhou2025netarena} & G & Agent planning/repair with environment feedback & \textit{NE} \\
\addlinespace[1pt]
\cite{zhu2026omniplan} & G/S/C & Intent preferences to network-planning experts & \emph{Feasibility:} Mixed-integer programs: constraints; fast experts: heuristics \\
\addlinespace[1pt]
\cite{e0arxiv180702205} & C & Policies to SDN paths and conflict advice & \emph{Feasibility:} OSDF: rule-based conflict classification \\
\addlinespace[1pt]
\cite{e0arxiv190403228} & G/C & Confirmed voice intent to OpenFlow rules & \textit{NE} \\
\addlinespace[1pt]
\cite{e0arxiv191000967} & C & Behavior-guided search for P4 programs & \emph{Feasibility:} Search fitness: supplied examples/rules \\
\addlinespace[1pt]
\cite{e0arxiv200805509} & G/C & Confirmed intent to NFV service-chain deployment & \emph{Feasibility:} Intent Deployer: configuration-consistency check \\
\addlinespace[1pt]
\cite{e0arxiv210505179} & C & Conceptual access-policy parsing and refinement & \emph{Feasibility:} Proposed configurator: authorization conflicts \\
\addlinespace[1pt]
\cite{e0arxiv210505594} & U & Conceptual smart-grid intent to network slices & \emph{Feasibility:} Proposed orchestrator: SLA/profile checks \\
\addlinespace[1pt]
\cite{e0arxiv220503932} & C & Controlled intent to simulated public-safety service & \emph{Feasibility:} Proposed intent validator: resource/conflict checks \\
\addlinespace[1pt]
\cite{e0arxiv221211874} & C & Optical feasibility to lightpath deployment & \emph{Feasibility:} Lightpath computation and routing/spectrum assignment: optical and wavelength constraints \\
\addlinespace[1pt]
\cite{e0arxiv230208544} & C & Knowledge-based intent to simulated services & \emph{Feasibility:} Network-layer validator: resource/model checks \\
\addlinespace[1pt]
\cite{e0arxiv230409711} & C & Internet Protocol (IP)/optical intent compilation in simulation & \textit{NE} \\
\addlinespace[1pt]
\cite{e0arxiv230409916} & C & Constrained vehicular-edge service mapping & \emph{Feasibility:} Priority-aware installation/location-aware mapping heuristics: node/link constraints \\
\addlinespace[1pt]
\cite{e0arxiv230702757} & G & Distributed transmit-power proposals & \textit{NE} \\
\addlinespace[1pt]
\cite{e0arxiv240108341} & C & Modeled RAN tilt-conflict resolution & \emph{Feasibility:} Bargaining: feasible utility sets \\
\addlinespace[1pt]
\cite{e0arxiv240202768} & S & Device-intent symbols to quality-of-service (QoS) slices & \textit{NE} \\
\addlinespace[1pt]
\cite{e0arxiv240309369} & G & Configuration generation and vendor translation & \textit{NE} \\
\addlinespace[1pt]
\cite{e0arxiv240415091} & S/C & Time-series monitoring of intent drift & \textit{NE} \\
\addlinespace[1pt]
\cite{e0arxiv240507850} & G/S & Entity/knowledge-graph completion to intent templates & \emph{Feasibility:} KG2E classifier: triple plausibility \\
\addlinespace[1pt]
\cite{e0arxiv240606059} & G/S/C & Intent validity to hierarchical RAN control & \emph{Feasibility:} Prediction and QoS-drift validity rules \\
\addlinespace[1pt]
\cite{e0arxiv240717767} & S & Parent-bandit selection of joint service priorities & \emph{Feasibility:} Learned parent policy: modeled resource conflicts \\
\addlinespace[1pt]
\cite{e0arxiv240813298} & S/G & Intent to iterative verified configuration & \emph{Feasibility:} Batfish: behavior checks and repair feedback \\
\addlinespace[1pt]
\cite{e0arxiv241114283} & G & Retrieved configuration to error reports & \emph{Feasibility:} LLM: syntax/dependency diagnosis \\
\addlinespace[1pt]
\cite{e0arxiv241116232} & S & Wireless-mesh action recommendation & \textit{NE} \\
\addlinespace[1pt]
\cite{e0arxiv250311933} & G/S & Intent/model selection to service and xApp creation & \textit{NE} \\
\addlinespace[1pt]
\cite{e0arxiv250509339} & G & Retrieved context to structured network intent & \textit{NE} \\
\addlinespace[1pt]
\cite{e0arxiv250518389} & G/C & Lua scheduler construction before slot execution & \textit{NE} \\
\addlinespace[1pt]
\cite{e0arxiv250604594} & S/C & Wi-Fi channel initialization followed by search & \textit{NE} \\
\addlinespace[1pt]
\cite{e0arxiv250712061} & S/U & Proposed intent choice; implemented enforcement & \emph{Coverage:} Ontology: operable candidate techniques \\
\addlinespace[1pt]
\cite{e0arxiv250714230} & S/G & History-based transmit-power strategy over O1 & \textit{NE} \\
\addlinespace[1pt]
\cite{e0arxiv250921026} & G/S & Intent translation to learned throughput control & \emph{Feasibility:} Translator: basic prior-intent conflict check \\
\addlinespace[1pt]
\cite{e0arxiv251021894} & G/C & Fact graph and generated network queries & \textit{NE} \\
\addlinespace[1pt]
\cite{e0arxiv251105133} & S & Malicious-intent classification at ingress & \textit{NE} \\
\addlinespace[1pt]
\cite{e0arxiv251121084} & S/G & 5G request classification to command generation & \textit{NE} \\
\addlinespace[1pt]
\cite{e0arxiv251220627} & G/S/C & Strategies to predicted deployment approval & \emph{Feasibility:} Federated evaluators: predicted deployability \\
\addlinespace[1pt]
\cite{lira2025network} & S/G & Iterative configuration generation with verification & \emph{Feasibility:} Assigned verifier; implementation unspecified \\
\addlinespace[1pt]
\cite{e0arxiv251220275} & G/C & Language-planned actions through graph admission & \emph{State:} Shape constraints and graph queries reject stale entities; \emph{Feasibility:} G-SPEC: next-state policy guards \\
\addlinespace[1pt]
\cite{e0arxiv260106640} & S/G & RAN/core recommendations to slice provisioning & \emph{Feasibility:} Specialists/orchestrator: model consistency checks \\
\addlinespace[1pt]
\cite{e0arxiv260112744} & G/C & Optimization-code generation and SDN demonstration & \emph{Feasibility:} Gurobi: generated-program constraints \\
\addlinespace[1pt]
\cite{e0arxiv260113114} & G/C & Planned, timed slice-rate changes via tools & \emph{Feasibility:} Feasibility/session tools: action constraints \\
\addlinespace[1pt]
\cite{e0arxiv260213672} & S/C & Risk-based drift alerts and attribution & \textit{NE} \\
\addlinespace[1pt]
\cite{e0arxiv260214283} & S/C & Multi-intent risk alerts and root-cause attribution & \textit{NE} \\
\addlinespace[1pt]
\cite{e0arxiv260306647} & G/S & Agent plans/code to emulated-network deployment & \emph{Feasibility:} Senior model: policy comparison; emulation: tests \\
\addlinespace[1pt]
\cite{e0arxiv260301663} & G/C & Intent contract to rApp policy and xApp control & \emph{Feasibility:} Agent: intent; rApp: history-based feasibility estimate \\
\addlinespace[1pt]
\cite{e0arxiv260323772} & G/S/U & Policy activation; proposed ranked assurance repair & \textit{NE} \\
\addlinespace[1pt]
\cite{e0arxiv260409678} & G & Interactive routing commands to converged state & \textit{NE} \\
\addlinespace[1pt]
\cite{e0arxiv260620564} & G/S & Multivendor command-line generation and model judging & \emph{Feasibility:} Model-judge panel: benchmark semantic scoring \\
\addlinespace[1pt]
\cite{e0arxiv260502911} & S/G & Expert/weight selection to resource allocation & \textit{NE} \\
\addlinespace[1pt]
\cite{e0arxiv260406856} & S/G & Security intent completion to domain agents & \emph{Coverage:} Orchestrator: support check and alert \\
\addlinespace[1pt]
\cite{e0arxiv260407264} & G/C/S & Offline intent compilation to certified low-Earth-orbit routing & \emph{Feasibility:} Validator/certifier: supported hard-constraint fragments; \emph{Coverage:} Certifier: fragment-bound accept/reject; unresolved abstain \\
\addlinespace[1pt]
\cite{e0arxiv260422513} & S/G & Configuration-context selection and repair & \textit{NE} \\
\addlinespace[1pt]
\cite{e0arxiv260427359} & C & Resource Description Framework intent representation/ontology validation & \textit{NE} \\
\addlinespace[1pt]
\cite{parra-ullauri2026rolebased} & G/C & Business intent to governed service orders & \emph{Feasibility:} Assigned tools: feasibility; supervisor: admission \\
\addlinespace[1pt]
\cite{e0arxiv260527428} & G/C & Event-driven parameter updates to task routing & \emph{State:} Feedback model uses completed, available records; \emph{Coverage:} Router: device/risk filter; routes with penalties if all risky \\
\addlinespace[1pt]
\cite{e0arxiv260600417} & G & Core-network plans to user-plane-function/user reassignment & \emph{Feasibility:} Executor agent: plan-feasibility assessment \\
\addlinespace[1pt]
\cite{e0arxiv260601222} & G & 5G base-station configuration/repair to approved startup & \emph{Feasibility:} Verification agent: review; operator: startup approval \\
\addlinespace[1pt]
\cite{e0arxiv260605076} & C & Offline port-policy compliance and drift monitoring & \textit{NE} \\
\addlinespace[1pt]
\cite{e0arxiv260708809} & S/C & Intent profiles to O1 assurance counters & \textit{NE} \\
\addlinespace[1pt]
\cite{e0arxiv260728826} & S & Distilled policy for isolation/restoration actions & \textit{NE} \\
\addlinespace[1pt]
\cite{e0arxiv260800908} & C/S & Candidate screening to network deployment & \emph{Feasibility:} Executor: exact residual check and repair/fallback; \emph{Coverage:} Guard: no executable generated candidate; fallback \\
\addlinespace[1pt]
\cite{e0arxiv260823179} & G & Interactive command-line changes scored in emulation & \textit{NE} \\
\addlinespace[1pt]
\cite{e0arxiv260904486} & G/C & Command generation with uncertainty-based review & \textit{NE} \\
\addlinespace[1pt]
\cite{e0arxiv260912144} & S & Security classification of intent windows & \textit{NE} \\
\addlinespace[1pt]
\cite{e0arxiv260921103} & G/S/C & Intent translation to tool-based policy inspection & \emph{State:} State manager: atomic, versioned snapshots; \emph{Feasibility:} Tools: network predicates; LLM: final verdict \\
\addlinespace[1pt]
\cite{e0arxiv260925356} & S/G & Telecom fault classification with reasoning & \textit{NE} \\
\addlinespace[1pt]
\cite{e0arxiv260927386} & G/C & Numerical algorithm search to path allocation & \emph{Feasibility:} Trusted network functions: feasibility constraints \\
\addlinespace[1pt]
\cite{e0arxiv260931397} & G/C & Intent compilation to corrected traffic-shaping rules & \emph{Feasibility:} Fixed-template module: policy and tc constraints \\
\addlinespace[1pt]
\cite{e0arxiv180303106} & C & Multilayer SDN service setup and teardown & \textit{NE} \\
\addlinespace[1pt]
\cite{e0arxiv200807162} & U & Conceptual domain/capability/service orchestration & \textit{NE} \\
\addlinespace[1pt]
\cite{e0arxiv241210107} & S/G & Wireless tool selection and allocation generation & \textit{NE} \\
\addlinespace[1pt]
\cite{e0arxiv250216198} & S/G & Semantic objectives to sequential resource actions & \emph{State:} Controller updates state after each allocation; \emph{Feasibility:} Executor: represented-state feasibility mask \\
\addlinespace[1pt]
\cite{e0arxiv250708164} & G/S & Network queries and service subscriptions & \textit{NE} \\
\addlinespace[1pt]
\cite{e0arxiv260316458} & G/S/C & Semantic rewards to reinforcement-learning placement/resource solving & \textit{NE} \\
\addlinespace[1pt]
\cite{e0arxiv260321251} & G/S & Tool-assisted wireless allocation and QoS decisions & \emph{Feasibility:} Agent: predicted-throughput QoS verdict \\
\addlinespace[1pt]
\cite{martins2026intentdriven} & G/S/C & Catalog discovery to service-profile decomposition & \emph{Feasibility:} Capability predicates: extracted requirements; \emph{Coverage:} Catalog layer: supplied-profile match; greedy coverage \\
\addlinespace[1pt]
\cite{e0arxiv190905528} & G/S & Troubleshooting dialogue acts and responses & \textit{NE} \\
\addlinespace[1pt]
\cite{e0arxiv240608305} & S/G/C & Network-health diagnosis and proposed mitigation & \emph{Feasibility:} LLM prompts: evidence and action checks \\
\addlinespace[1pt]
\cite{e0arxiv240706085} & S/C & Packet reconstruction to failure/localization labels & \textit{NE} \\
\addlinespace[1pt]
\cite{e0arxiv250116842} & G/C/S & Network knowledge to diagnostic plans/reports & \emph{Feasibility:} Z3 component: rules; model: diagnostic report \\
\addlinespace[1pt]
\cite{e0arxiv251100651} & S/G & Delegated telecom diagnosis and action planning & \textit{NE} \\
\addlinespace[1pt]
\cite{e0arxiv260405400} & G/S & Runbook, API and diagnostic-branch assistance & \textit{NE} \\
\addlinespace[1pt]
\cite{e0arxiv260413353} & S/G & Clarification to cross-domain diagnostic queries & \textit{NE} \\
\addlinespace[1pt]
\cite{e0arxiv260504530} & S/C & Symptom probes to skill-based fault localization & \emph{State:} Diagnostic agent cross-checks IP/layer evidence \\
\addlinespace[1pt]
\cite{e0arxiv260600582} & S/G & Multimodal faults to root-cause/propagation chains & \textit{NE} \\
\addlinespace[1pt]
\cite{e0sourceS01} & G & Multidomain intent translation to service descriptors & \emph{Feasibility:} Proposed handlers: conflict/resource checks \\
\addlinespace[1pt]
\cite{e0sourceS03} & S/C & Parameter/command mapping to target command-line trees & \emph{Feasibility:} Constraint-satisfaction solver: translation and view consistency \\
\addlinespace[1pt]
\cite{e0sourceS04} & G & Graph-grounded routing-parameter updates & \textit{NE} \\
\addlinespace[1pt]
\cite{e0sourceS05} & G/S/C & Dialogue-based intent to Open Network Operating System installation & \textit{NE} \\
\addlinespace[1pt]
\cite{e0sourceS06} & G & Command generation and probes in emulation & \textit{NE} \\
\addlinespace[1pt]
\cite{e0sourceS08} & G/S & Production alert diagnosis with operator assistance & \textit{NE} \\
\addlinespace[1pt]
\cite{e0sourceS09} & G/S/C & Dialogue-selected workflows to health reports & \textit{NE} \\
\addlinespace[1pt]
\cite{e0sourceS10} & G/S & Workflow/domain-specific-language generation with diagnostic tools & \emph{Feasibility:} Parsers, dry runs and invariant validators \\
\addlinespace[1pt]
\cite{e0sourceS12} & G/C/S & Alarm propagation graphs to root-cause diagnosis & \textit{NE} \\
\addlinespace[1pt]
\cite{e0sourceS15} & S & LLM-assisted learning for O-RAN slice allocation & \textit{NE} \\
\addlinespace[1pt]
\cite{asif2025ribn} & G/S/C & Intent translation to learned service-chain placement & \emph{Feasibility:} Similarity/k-nearest-neighbour classifier: intent contradictions \\
\addlinespace[1pt]
\cite{e0sourceS17} & G/S/C & Example/template selection to formal synthesis & \emph{Feasibility:} CEGS: formal checks of encoded intent, topology and templates \\
\addlinespace[1pt]
\cite{e0sourceS18} & G/S & Clarified intent to guide-based mitigation actions & \textit{NE} \\
\addlinespace[1pt]
\cite{e0sourceS19} & G/S/U & Conceptual incident diagnosis and mitigation & \textit{NE} \\
\addlinespace[1pt]
\cite{e0sourceS20} & S/G & Intrusion classification with explanatory alerts & \textit{NE} \\
\addlinespace[1pt]
\cite{e0sourceS21} & S & Federated traffic/log anomaly classification & \textit{NE} \\
\addlinespace[1pt]
\cite{doro2022orchestran} & C/S & Request-driven model placement and xApp dispatch & \emph{Feasibility:} Binary integer linear program: resources, timescales, input-data reachability; \emph{Coverage:} Optimizer: request admission from catalog models \\
\addlinespace[1pt]
\cite{e0arxiv211107724} & C/S/U & Intent policies; ML-hardware performance estimation & \emph{Feasibility:} Policy configurator: conflict check; \emph{Coverage:} Policy configurator: technique availability; both unevaluated \\
\addlinespace[1pt]
\cite{e0arxiv250922834} & G/C & Offline intent to planned optical network design & \emph{Feasibility:} Grammar parser: structure; planner: deployment constraints; \emph{Coverage:} Planner: reports infeasible requirements \\
\addlinespace[1pt]
\cite{e0arxiv260427084} & G/C & Offline causal priors to RAN power-control configuration & \emph{State:} Bayesian network conditions on current measurements; incremental updates \\
\addlinespace[1pt]
\cite{e0arxiv260816422} & G & Offline emulator-scenario and attack-agent generation & \emph{Feasibility:} Validator: schema/topology; NASim load-and-run test \\
\addlinespace[1pt]
\cite{e0arxiv250913901} & C & Declarative intents to GitOps/Nephio deployment & \emph{State:} GitOps operators: desired/runtime state drift \\
\addlinespace[1pt]
\end{longtable}
\endgroup

\section{Search Strategy and Evidence Definitions}\label{app:e0-protocol}
\subsection{Search and Eligibility}
The arXiv API was queried on 28 September 2026 with the following title-and-abstract searches, covering 1 January 2016--28 September 2026. Q1, Q2 and Q3 returned 115, 28 and 16 hits, all of which were retrieved. Duplicates within arXiv results were removed by identifier. Cross-source matches used exact normalized titles. Related reports were grouped by work while preserving their distinct tasks and versions.
\smallskip\noindent\textbf{Q1 (115 hits).}
\begingroup\footnotesize\ttfamily\raggedright
((ti:"network configuration" OR abs:"network configuration") OR (ti:"router configuration" OR abs:"router configuration") OR (ti:"intent based networking" OR abs:"intent based networking") OR (ti:"network intent" OR abs:"network intent")) AND ((ti:"language model" OR abs:"language model") OR (ti:"language models" OR abs:"language models") OR (ti:"LLM" OR abs:"LLM") OR (ti:"LLMs" OR abs:"LLMs") OR (ti:"natural language" OR abs:"natural language") OR (ti:"intent based" OR abs:"intent based") OR (ti:"intent driven" OR abs:"intent driven")) AND submittedDate:[201601010000 TO 202609282359]
\par\endgroup
\smallskip\noindent\textbf{Q2 (28 hits).}
\begingroup\footnotesize\ttfamily\raggedright
((ti:"network orchestration" OR abs:"network orchestration") OR (ti:"service orchestration" OR abs:"service orchestration") OR (ti:"network function deployment" OR abs:"network function deployment") OR (ti:"intent driven resource allocation" OR abs:"intent driven resource allocation")) AND ((ti:"language model" OR abs:"language model") OR (ti:"language models" OR abs:"language models") OR (ti:"LLM" OR abs:"LLM") OR (ti:"LLMs" OR abs:"LLMs") OR (ti:"natural language" OR abs:"natural language") OR (ti:"intent based" OR abs:"intent based") OR (ti:"intent driven" OR abs:"intent driven")) AND submittedDate:[201601010000 TO 202609282359]
\par\endgroup
\smallskip\noindent\textbf{Q3 (16 hits).}
\begingroup\footnotesize\ttfamily\raggedright
((ti:"network troubleshooting" OR abs:"network troubleshooting") OR (ti:"network diagnosis" OR abs:"network diagnosis") OR (ti:"network fault diagnosis" OR abs:"network fault diagnosis") OR (ti:"network fault localization" OR abs:"network fault localization")) AND ((ti:"language model" OR abs:"language model") OR (ti:"language models" OR abs:"language models") OR (ti:"LLM" OR abs:"LLM") OR (ti:"LLMs" OR abs:"LLMs") OR (ti:"natural language" OR abs:"natural language") OR (ti:"intent based" OR abs:"intent based") OR (ti:"intent driven" OR abs:"intent driven")) AND submittedDate:[201601010000 TO 202609282359]
\par\endgroup
\smallskip\noindent\textbf{Publisher-page discovery and citation tracing.}
Targeted web searches used the following queries to locate individual IEEE Xplore and ACM Digital Library pages.
\par\smallskip
\noindent{\footnotesize\ttfamily "network configuration" ("LLM" OR "language models"); domains: ieeexplore.ieee.org, dl.acm.org\par}
\noindent{\footnotesize\ttfamily "intent" "orchestration" "language models"; domains: ieeexplore.ieee.org, dl.acm.org\par}
\noindent{\footnotesize\ttfamily "network troubleshooting" ("LLM" OR "language models"); domains: ieeexplore.ieee.org, dl.acm.org\par}
\noindent{\footnotesize\ttfamily site:dl.acm.org "network troubleshooting" "language models"\par}
\noindent{\footnotesize\ttfamily site:dl.acm.org "network configuration" "language models"\par}
\noindent{\footnotesize\ttfamily site:dl.acm.org "intent" "service orchestration"\par}
One backward-reference round examined 152 references from INTA (arXiv:2501.08760), Chat-Driven Optimal Management (arXiv:2512.24614) and A Network Arena for Benchmarking AI Agents on Network Troubleshooting (arXiv:2512.16381). These seeds represent configuration, orchestration and diagnosis. The 21 additional records combine targeted discovery, this citation round and one previously uncited source.

Eligibility requires an identifiable network configuration, orchestration or diagnostic decision. Conceptual designs and studies without latency measurements remain eligible when their decision task is explicit. Mixed-domain studies contribute their network tasks. Title/abstract screening assigned 79 records to background and excluded 26 as out of scope. Relevant full-text review assigned six further records to background. Independent researcher review excluded one work concerned only with CPU-consumption prediction. A blind second screening of all 111 set-aside records found seven eligible under this criterion, and these enter the corpus as additional families ($\S$\ref{sec:review-verification}). Figure~\ref{fig:e0-screening} gives the resulting record, report and family counts.
\subsection{Coverage Audit}\label{app:coverage-audit}
The OpenAlex search applied a title-and-abstract filter for publications dated from 1 January 2016 to 28 September 2026. Three queries paired a task concept with an interface concept. The configuration concept was ``network configuration'', ``router configuration'', ``intent based networking'' or ``network intent''. The orchestration concept was ``network orchestration'', ``service orchestration'', ``network function deployment'' or ``intent driven resource allocation''. The diagnosis concept was ``network troubleshooting'', ``network diagnosis'', ``network fault diagnosis'' or ``network fault localization''. Each was combined by AND with ``language model'', ``language models'', LLM, LLMs, ``natural language'', ``intent based'' or ``intent driven''.

The sampling frame is the 594 records that passed title-and-abstract screening. A fixed pattern over title and abstract assigned 267 records to the language-model stratum when they mention LLM, language model, GPT, generative AI, GenAI, agentic, agent, natural language, NLP or SLM. The other 327 form the earlier intent-networking stratum. Fifty records were drawn from each stratum with seed 42 before any access check. Full text was sought from open-access locations, then by arXiv identifier or a unique arXiv title match, then through a library request.

In the language-model stratum, 49 draws were coded and 45 are eligible, with 15 explicit loop claims, 3 matched claims, 2 p95 reports and 1 deadline report. In the earlier stratum, 49 were coded, 32 are eligible and 2 remain unresolved, with 10, 2, 2 and 1 positives. Each coded draw carries the weight of its stratum frame size over the coded draws in that stratum. A weighted rate divides weighted positives by weighted eligible draws. Its interval comes from 2,000 bootstrap resamples of coded draws within each stratum, with seed 42. This estimate assumes that access and coding are missing at random within each stratum. The bounds in $\S$\ref{sec:review-verification} drop that assumption by counting the two unretrieved draws and every unresolved code as negative or as positive.

\subsection{Evidence Definitions}
Table~\ref{tab:e0-codebook} defines the 18 reporting indicators. ``Not reported'' means that the relevant methods, evaluation and appendices supply no explicit report. ``Unclear'' means that the available description does not resolve the category, and N/A denotes an inapplicable measurement or operation. A positive indicator satisfies the stated criterion. Explicit absence, missing reporting and inapplicability remain separate categories.

A coding unit is a work--task--decision step at a specified measurement boundary. Different implementations or versions can contribute linked scopes of the same step. A shared aggregate measurement belongs to its measured scope. Inference location concerns decision execution, and code/data indicators concern public pointers supplied by the study.

\begingroup\footnotesize\setlength{\tabcolsep}{4pt}
\setlength{\aboverulesep}{0pt}\setlength{\belowrulesep}{0pt}
\renewcommand{\arraystretch}{1.10}
\setlength{\LTcapwidth}{\textwidth}
\newlength{\sokdefinitionwidth}
\setlength{\sokdefinitionwidth}{\dimexpr\textwidth-1pt-6\tabcolsep\relax}
\begin{longtable}{>{\raggedright\arraybackslash}p{.20\sokdefinitionwidth}>{\raggedright\arraybackslash}p{.47\sokdefinitionwidth}>{\raggedright\arraybackslash}p{.33\sokdefinitionwidth}}
\caption{Definitions and response categories for the 18 evidence fields. The criterion specifies positive coverage in Table~\ref{tab:e0-audit}. Timing endpoint and execution location permit multiple selections. All other fields are single-choice.}\label{tab:e0-codebook}\\
\toprule\rowcolor{ebBlue} Field & Positive reporting criterion & Response categories \\
\midrule\endfirsthead
\caption[]{Evidence definitions (continued).}\\
\toprule\rowcolor{ebBlue} Field & Positive reporting criterion & Response categories \\
\midrule\endhead\bottomrule\endfoot
\rowcolor{ebAmber!45}\multicolumn{3}{l}{\textbf{\itshape Timing, workload, and comparison}} \\*
Timing endpoint & Explicit end event for a measured interval: model return, action dispatch, confirmed network effect, verified service result, or another named event. & Model; dispatch; network; service; other; not reported; unclear; N/A. \\ \addlinespace[3pt]
Metric denominator & Explicit population entering the metric: all attempts, successful requests, another subset, or different denominators for different metrics. & All; successful; other; mixed; unclear; N/A. \\ \addlinespace[3pt]
Tail latency & A latency quantile at p95 or higher. & Reported; not reported; unclear; N/A. \\ \addlinespace[3pt]
Deadline attainment & The fraction or count attaining an explicit deadline. & Reported; not reported; unclear; N/A. \\ \addlinespace[3pt]
Input load & Operational arrival rate, traffic load, concurrency or queue input, as defined below. & Reported; not reported; unclear; N/A. \\ \addlinespace[3pt]
Queue / waiting & Explicit queueing or waiting evidence for the decision or service path. & Reported; not reported; unclear; N/A. \\ \addlinespace[3pt]
Stability evidence & Timing or operating-state behavior over operational time or load, as defined below. & Reported; not reported; unclear; N/A. \\ \addlinespace[3pt]
Non-learning comparison & A rule, solver or traditional controller evaluated as an independent comparator. & Reported; not reported; unclear; N/A. \\ \addlinespace[3pt]
\rowcolor{ebAmber!45}\multicolumn{3}{l}{\textbf{\itshape Control-loop claim and deployment}} \\*
Explicit loop claim & An explicit control-loop class or time-budget claim. & RT/fast; near-RT; non-RT; other budget or multiple loops; not stated; unclear; N/A. \\ \addlinespace[3pt]
Supported loop claim & A measurement supports the claimed budget at the corresponding execution boundary. & Matched; mismatch; insufficient evidence; N/A. \\ \addlinespace[3pt]
Execution location & An explicit location of inference or decision computation. & Hosted API; local GPU; edge device; local CPU; other; unclear; N/A. \\ \addlinespace[3pt]
Network round trip & The reported decision latency explicitly includes its network communication. & Included; excluded; unclear; N/A. \\ \addlinespace[3pt]
\rowcolor{ebAmber!45}\multicolumn{3}{l}{\textbf{\itshape Validation responsibilities}} \\*
Format check & Evaluation of output syntax or structural validity. & Reported; explicitly unchecked; not reported; unclear; N/A. \\ \addlinespace[3pt]
Semantic check & Evaluation of task meaning or decision correctness. & Reported; explicitly unchecked; not reported; unclear; N/A. \\ \addlinespace[3pt]
Final-state check & Evaluation of the resulting network or service state. & Reported; explicitly unchecked; not reported; unclear; N/A. \\ \addlinespace[3pt]
Three gates distinguished & Format, semantic and final-state checking are explicitly distinguished within the scope. & All three; partial; combined; unclear; N/A. \\ \addlinespace[3pt]
\rowcolor{ebAmber!45}\multicolumn{3}{l}{\textbf{\itshape Artifact pointers}} \\*
Public code pointer & A public code address or resource is given in the paper or its supplement. & Public link; on request/private; explicitly not public; not reported; unclear; N/A. \\ \addlinespace[3pt]
Public data pointer & A public data address or resource is given in the paper or its supplement. & Public link; on request/private; explicitly not public; not reported; unclear; N/A. \\ \addlinespace[3pt]
\end{longtable}\endgroup
\subsection{Load and Stability Definitions}\label{app:e0-c1}
During adjudication, the researchers clarified two ambiguous field definitions after comparing their independent answers. Operational load includes request/service arrivals, concurrency, traffic offered to the system and queue input, with the decision or service boundary identified. Evidence can be measured, simulated or analytical. Training batch size, prompt length, topology size and single-instance computational complexity describe different quantities.

Stability evidence concerns timing or operating-state behavior over operational time or load, including queue/control oscillation and delay under varying traffic. Both stable and unstable behavior qualify. Fixed-condition run-to-run dispersion, a single-request runtime curve and training convergence are distinct from this indicator. The clarification applies to final reporting coverage. The pre-adjudication agreement in Table~\ref{tab:e0-agreement} uses the earlier independent responses.

\section{Coding Distributions and Agreement}\label{app:agreement}
Table~\ref{tab:e0-categories} gives the category counts over the 405 scopes coded by the two researchers. The seven families added by the second screening contribute the other 24 scopes ($\S$\ref{sec:review-verification}). Table~\ref{tab:e0-agreement-options} supplies the option-level reliability of the two multi-select fields, and Table~\ref{tab:e0-agreement} gives the field-level results. Table~\ref{tab:tx-reliability} reports the reproducibility of the three-axis systematization.

The companion files provide scope-level coding records and extended descriptions of the literature map for data reuse. Experimental companion tables contain 497 task-condition cells, 239 matched contrasts and 151 online summaries. They also contain 580 timing strata, 83 native load cells and 131 physical-path summaries. The six-case and 24-case policy, repair, forwarding, placement and network records remain in the machine-readable supplement. The execution-path table contains 176 domain-condition rows, including 24 empirical timing strata at each load. JSON-encoded columns store intervals and native metadata. Missing values are coded as unavailable. Measurement metadata identify the inputs, deployment versions and collection periods for each result.
\begingroup\footnotesize
\setlength{\tabcolsep}{4pt}
\setlength{\aboverulesep}{0pt}\setlength{\belowrulesep}{0pt}
\renewcommand{\arraystretch}{1.10}
\begin{table}[!htbp]
\centering
\caption{Pre-adjudication coding agreement: 231 initial pairs and 174 independent rechecks after scope alignment. A is exact agreement (\%), and $\kappa$ and AC1 use the original codebook categories. Multi-select A is exact-set agreement, and its option-wise coefficients appear in Table~\ref{tab:e0-agreement-options}. Combined pools these two pre-adjudication strata.}
\label{tab:e0-agreement}
\def\familytab{\begin{tabular}{lrrrrrrrrr}
\toprule
\rowcolor{ebBlue} Field & \multicolumn{3}{c}{\shortstack{Initial comparable\\$n=231$}} & \multicolumn{3}{c}{\shortstack{Aligned recheck\\$n=174$}} & \multicolumn{3}{c}{\shortstack{Combined\\$n=405$}} \\
\rowcolor{ebBlue} & A (\%) & $\kappa$ & AC1 & A (\%) & $\kappa$ & AC1 & A (\%) & $\kappa$ & AC1 \\
\midrule
\rowcolor{ebAmber!45}\multicolumn{10}{l}{\textbf{\itshape Timing, workload, and comparison}} \\*
Named timing endpoint & 93.5 & N/A & N/A & 87.9 & N/A & N/A & 91.1 & N/A & N/A \\
Specified metric denominator & 91.8 & 0.839 & 0.908 & 89.7 & 0.835 & 0.882 & 90.9 & 0.838 & 0.897 \\
p95 or higher & 100.0 & 1.000 & 1.000 & 99.4 & 0.935 & 0.994 & 99.8 & 0.965 & 0.997 \\
Deadline attainment & 99.1 & 0.663 & 0.991 & 97.1 & 0.674 & 0.970 & 98.3 & 0.674 & 0.982 \\
Input load & 26.0 & 0.062 & 0.071 & 30.5 & 0.108 & 0.115 & 27.9 & 0.083 & 0.088 \\
Queue / waiting & 37.7 & 0.006 & 0.268 & 41.4 & 0.083 & 0.300 & 39.3 & 0.041 & 0.281 \\
Stability evidence & 26.4 & 0.021 & 0.099 & 36.2 & 0.144 & 0.203 & 30.6 & 0.075 & 0.143 \\
Non-learning comparison & 92.2 & 0.736 & 0.914 & 90.8 & 0.725 & 0.896 & 91.6 & 0.731 & 0.906 \\
\rowcolor{ebAmber!45}\multicolumn{10}{l}{\textbf{\itshape Control-loop claim and deployment}} \\*
Explicit loop claim & 86.1 & 0.692 & 0.850 & 81.6 & 0.649 & 0.799 & 84.2 & 0.673 & 0.828 \\
Supported loop claim & 97.8 & 0.000 & 0.978 & 98.9 & 0.851 & 0.988 & 98.3 & 0.625 & 0.982 \\
Named execution location & 83.5 & N/A & N/A & 77.6 & N/A & N/A & 81.0 & N/A & N/A \\
Network round trip included & 45.0 & 0.187 & 0.314 & 39.1 & 0.118 & 0.232 & 42.5 & 0.169 & 0.276 \\
\rowcolor{ebAmber!45}\multicolumn{10}{l}{\textbf{\itshape Validation responsibilities}} \\*
Format check reported & 82.7 & 0.645 & 0.803 & 84.5 & 0.661 & 0.825 & 83.5 & 0.652 & 0.812 \\
Semantic check reported & 85.7 & 0.294 & 0.850 & 85.6 & 0.464 & 0.846 & 85.7 & 0.380 & 0.848 \\
Final-state check reported & 37.2 & 0.213 & 0.245 & 48.9 & 0.340 & 0.380 & 42.2 & 0.269 & 0.302 \\
Three gates distinguished & 65.8 & 0.183 & 0.619 & 73.6 & 0.365 & 0.706 & 69.1 & 0.261 & 0.656 \\
\rowcolor{ebAmber!45}\multicolumn{10}{l}{\textbf{\itshape Artifact pointers}} \\*
Public code pointer & 96.5 & 0.925 & 0.962 & 92.5 & 0.846 & 0.917 & 94.8 & 0.891 & 0.943 \\
Public data pointer & 86.6 & 0.728 & 0.851 & 90.8 & 0.817 & 0.898 & 88.4 & 0.768 & 0.871 \\
\bottomrule
\end{tabular}}
\setlength{\tabcolsep}{0pt}\sbox0{\familytab}
\dimen0=\linewidth\advance\dimen0 by -1pt\advance\dimen0 by -\wd0\divide\dimen0 by 20
\ifdim\dimen0<1.5pt \dimen0=1.5pt\fi
\setlength{\tabcolsep}{\dimen0}
\sokfitcolumns{\familytab}{20}
\familytab
\end{table}
\endgroup

\begin{table}[!htbp]
\centering\footnotesize
\setlength{\tabcolsep}{3pt}
\renewcommand{\arraystretch}{1.08}
\caption{Reproducibility of the three-axis systematization over the 129 families outside the 10-family pilot. Two model coders (GPT and Claude) recoded each family blind from full text under a frozen codebook. A third blind pass mapped the published cells into the same fields. A is observed agreement. Rows without $\kappa$ report exact set agreement. The last two columns give each coder's agreement with the published entries ($\kappa$, or exact agreement in percent), over families whose published cells settle the field (path 103, receiving roles 125). Owner and mechanism comparisons with fewer than 20 shared checks are omitted.}\label{tab:tx-reliability}
\begin{tabular}{@{}p{0.34\columnwidth}rrrrrr@{}}
\toprule
\rowcolor{ebBlue} & & \multicolumn{3}{c}{Coder vs coder} & \multicolumn{2}{c}{Coder vs table} \\
\rowcolor{ebBlue} Field & $n$ & A (\%) & $\kappa$ & AC1 & GPT & Claude \\
\midrule
\rowcolor{ebAmber!45}\multicolumn{7}{l}{\textbf{\itshape Decision interface}} \\*
Selection (S) & 129 & 76.7 & 0.50 & 0.58 & 0.39 & 0.51 \\
Generation (G) & 129 & 86.8 & 0.72 & 0.75 & 0.76 & 0.72 \\
Computation (C) & 129 & 82.9 & 0.67 & 0.66 & 0.57 & 0.69 \\
Exact interface set & 129 & 53.5 & -- & -- & 46.5\% & 53.5\% \\
\rowcolor{ebAmber!45}\multicolumn{7}{l}{\textbf{\itshape Execution path}} \\*
Offline / request-driven / both & 129 & 88.4 & 0.69 & 0.87 & 0.47 & 0.54 \\
Exact receiving-role set & 129 & 58.9 & -- & -- & 43.2\% & 55.2\% \\
\rowcolor{ebAmber!45}\multicolumn{7}{l}{\textbf{\itshape Check present}} \\*
Observation & 129 & 81.4 & 0.62 & 0.64 & 0.13 & 0.12 \\
Feasibility & 129 & 89.1 & 0.73 & 0.82 & 0.42 & 0.42 \\
Coverage & 129 & 86.8 & 0.72 & 0.75 & 0.10 & 0.17 \\
\rowcolor{ebAmber!45}\multicolumn{7}{l}{\textbf{\itshape Check owner and mechanism, where both coders find the check}} \\*
Observation owner & 40 & 67.5 & -- & -- & -- & -- \\
Observation mechanism & 40 & 70.0 & -- & -- & -- & -- \\
Feasibility owner & 87 & 66.7 & -- & -- & 43.4\% & 47.2\% \\
Feasibility mechanism & 87 & 81.6 & -- & -- & 63.2\% & 64.3\% \\
Coverage owner & 43 & 65.1 & -- & -- & -- & -- \\
Coverage mechanism & 43 & 81.4 & -- & -- & -- & -- \\
\bottomrule
\end{tabular}
\end{table}

\FloatBarrier
\begin{table}[!htbp]
\centering\footnotesize
\setlength{\tabcolsep}{4pt}
\setlength{\aboverulesep}{0pt}\setlength{\belowrulesep}{0pt}
\renewcommand{\arraystretch}{1.10}
\newlength{\sokoptionwidth}
\setlength{\sokoptionwidth}{\dimexpr\textwidth-1pt-12\tabcolsep\relax}
\caption{Pre-adjudication option agreement for the two multi-select fields over all 405 paired scopes (231 initial pairs and 174 independent rechecks after scope alignment). Counts give the number of selections by each researcher. A counts both joint selections and joint non-selections. N/A denotes undefined $\kappa$ for a constant category.}\label{tab:e0-agreement-options}
\begin{tabular}{>{\raggedright\arraybackslash}p{.30\sokoptionwidth}>{\raggedleft\arraybackslash}p{.18\sokoptionwidth}>{\raggedleft\arraybackslash}p{.18\sokoptionwidth}>{\raggedleft\arraybackslash}p{.12\sokoptionwidth}>{\raggedleft\arraybackslash}p{.11\sokoptionwidth}>{\raggedleft\arraybackslash}p{.11\sokoptionwidth}}
\toprule
\rowcolor{ebBlue} Option & \shortstack{Researcher 1\\selected} & \shortstack{Researcher 2\\selected} & A (\%) & $\kappa$ & AC1 \\
\midrule
\rowcolor{ebAmber!45}\multicolumn{6}{l}{\textbf{\itshape Timing endpoint}} \\*
Model return & 38 & 41 & 97.3 & 0.846 & 0.967 \\
Dispatch & 2 & 6 & 98.5 & 0.244 & 0.985 \\
Network effect & 11 & 12 & 99.3 & 0.866 & 0.992 \\
Service result & 23 & 22 & 97.8 & 0.788 & 0.975 \\
Other & 92 & 81 & 94.3 & 0.831 & 0.914 \\
Not reported & 251 & 252 & 98.8 & 0.974 & 0.977 \\
Unclear & 7 & 5 & 98.0 & 0.324 & 0.980 \\
N/A & 0 & 0 & 100.0 & N/A & 1.000 \\
\rowcolor{ebAmber!45}\multicolumn{6}{l}{\textbf{\itshape Execution location}} \\*
Hosted API & 147 & 128 & 93.3 & 0.852 & 0.879 \\
Local GPU & 62 & 67 & 97.3 & 0.899 & 0.963 \\
Edge device & 0 & 0 & 100.0 & N/A & 1.000 \\
Local CPU & 21 & 22 & 98.8 & 0.877 & 0.986 \\
Other & 39 & 24 & 91.4 & 0.400 & 0.899 \\
Unclear & 103 & 115 & 92.6 & 0.812 & 0.878 \\
N/A & 94 & 89 & 95.8 & 0.880 & 0.935 \\
\bottomrule
\end{tabular}
\end{table}

\begin{table}[!htbp]
\centering\footnotesize
\setlength{\tabcolsep}{4pt}
\setlength{\aboverulesep}{0pt}\setlength{\belowrulesep}{0pt}
\renewcommand{\arraystretch}{1.10}
\newlength{\sokcategorywidth}
\setlength{\sokcategorywidth}{\dimexpr\textwidth-1pt-4\tabcolsep\relax}
\caption{Adjudicated category counts across the 405 scopes coded by the two researchers. Each single-choice row sums to 405. Timing endpoint and execution location allow multiple selections. N/A denotes inapplicability, and unlisted response categories have zero count. Table~\ref{tab:e0-audit} reports positive coverage and its family-clustered intervals.}\label{tab:e0-categories}
\begin{tabular}{>{\raggedright\arraybackslash}p{.27\sokcategorywidth}>{\raggedright\arraybackslash}p{.73\sokcategorywidth}}
\toprule
\rowcolor{ebBlue} Field & Response category: number of scopes \\
\midrule
\rowcolor{ebAmber!45}\multicolumn{2}{l}{\textbf{\itshape Timing, workload, and comparison}} \\*
Named timing endpoint & Model return:~39; Dispatch:~6; Network effect:~12; Service result:~23; Other:~86; Not reported:~253; Unclear:~2 \\
\addlinespace[3pt]
Specified metric denominator & All attempts:~48; Successful:~5; Other:~16; Mixed:~7; Unclear:~76; N/A:~253 \\
\addlinespace[3pt]
p95 or higher & Reported:~14; Not reported:~390; Unclear:~1 \\
\addlinespace[3pt]
Deadline attainment & Reported:~7; Not reported:~397; Unclear:~1 \\
\addlinespace[3pt]
Input load & Reported:~81; Not reported:~324 \\
\addlinespace[3pt]
Queue / waiting & Reported:~11; Not reported:~394 \\
\addlinespace[3pt]
Stability evidence & Reported:~44; Not reported:~361 \\
\addlinespace[3pt]
Non-learning comparison & Reported:~64; Not reported:~337; Unclear:~4 \\
\addlinespace[3pt]
\rowcolor{ebAmber!45}\multicolumn{2}{l}{\textbf{\itshape Control-loop claim and deployment}} \\*
Explicit loop claim & RT/fast:~54; Near-RT:~11; Non-RT:~36; Other:~19; Not stated:~285 \\
\addlinespace[3pt]
Supported loop claim & Matched:~7; Mismatch:~1; Insufficient:~397 \\
\addlinespace[3pt]
Named execution location & Hosted API:~143; Local GPU:~66; Local CPU:~21; Other:~29; Unclear:~105; N/A:~95 \\
\addlinespace[3pt]
Network round trip included & Included:~35; Excluded:~5; Unclear:~91; N/A:~274 \\
\addlinespace[3pt]
\rowcolor{ebAmber!45}\multicolumn{2}{l}{\textbf{\itshape Validation responsibilities}} \\*
Format check reported & Reported:~135; Not reported:~266; N/A:~4 \\
\addlinespace[3pt]
Semantic check reported & Reported:~344; Explicitly unchecked:~3; Not reported:~58 \\
\addlinespace[3pt]
Final-state check reported & Reported:~75; Explicitly unchecked:~31; Not reported:~96; Unclear:~8; N/A:~195 \\
\addlinespace[3pt]
Three gates distinguished & All three:~12; Partial:~262; Combined:~130; Unclear:~1 \\
\addlinespace[3pt]
\rowcolor{ebAmber!45}\multicolumn{2}{l}{\textbf{\itshape Artifact pointers}} \\*
Public code pointer & Public link:~96; On request/private:~3; Not public:~4; Not reported:~284; Unclear:~18 \\
\addlinespace[3pt]
Public data pointer & Public link:~139; On request/private:~4; Not public:~1; Not reported:~248; Unclear:~13 \\
\addlinespace[3pt]
\bottomrule
\end{tabular}
\end{table}

\end{document}